\documentclass[12pt]{article}
\usepackage[english]{babel}
\RequirePackage[colorlinks]{hyperref}
\usepackage{setspace}
\usepackage{booktabs}
\usepackage{graphicx}
\usepackage{rotating}
\usepackage{hyperref}
\usepackage{pdfsync}
\usepackage[margin=2cm,a4paper]{geometry}
\usepackage{amsmath,amsfonts,bm,amssymb} 
\usepackage{amsthm}
\usepackage{soul}
\usepackage{setspace}
\usepackage{multirow}
\usepackage{afterpage}
\newcommand{\be}{\begin{equation}}
\newcommand{\ee}{\end{equation}}
\newcommand{\bes}{\begin{equation*}}
\newcommand{\ees}{\end{equation*}}
\newcommand{\bd}{\begin{displaymath}}
\newcommand{\ed}{\end{displaymath}}
\newcommand{\e}{\mathbb{E}}
\newcommand{\q}{\mathbb{Q}}

\newcommand{\eonia}{{}}

\newcommand{\FRA}{\mathrm{FRA}}

\graphicspath{{./figures/}}

\begin{document}

%\title{HJM for risk management in the multi-curve framework}
\title{Multi-curve HJM modelling for risk management}

\author{
Chiara Sabelli$^{a}$\footnote{Corresponding author: chiara.sabelli@sns.it, phone +39 050 509259.}~, Michele Pioppi$^{b}$\footnote{The views, thoughts and opinions expressed in this paper are those of the author in his
individual capacity and should not be attributed to UniCredit S.p.A. or to the author as a 
representative or employee of UniCredit S.p.A..}~, Luca Sitzia$^{b\dag}$~and Giacomo Bormetti$^{a,c}$~\footnote{Author contributions: CS and GB designed research; CS and GB performed research; CS, MP, LS and GB analysed data; CS and GB wrote the paper.}}

\date{September 9, 2015}
\maketitle
\begin{center}
  $^{a}$~\emph{Scuola Normale Superiore, Piazza dei Cavalieri 7, Pisa, 56126, Italy}\\
  $^{b}$~\emph{UniCredit S.p.A., Piazza Gae Aulenti, Milano, 20154, Italy}\\
  $^{c}$~\emph{QUANTLab\hspace{2pt}\footnote{www.quantlab.it}, via Pietrasantina 123, Pisa, 56122, Italy}\\
\end{center}

\normalsize

\vspace{0.3cm}

\smallskip

\begin{abstract}
  We present a HJM approach to the projection of multiple yield curves developed to capture the volatility content of historical term structures for risk management purposes. Since we observe the empirical data at daily frequency and only for a finite number of time-to-maturity buckets, we propose a modelling framework which is inherently discrete. In particular, we show how to approximate the HJM continuous time description of the multi-curve dynamics by a Vector Autoregressive process of order one. The resulting dynamics lends itself to a feasible estimation of the model volatility-correlation structure and market risk-premia. Then, resorting to the Principal Component Analysis we further simplify the dynamics reducing the number of covariance components.  
Applying the constant volatility version of our model on a sample of curves from the Euro area, we demonstrate its forecasting ability through an out-of-sample test.
%We test our approach on a sample of curves from the Euro area starting from March 2004 and we carry out different numerical experiments. In particular, we perform scenario generation and out-of-sample tests which assess, finally, the reliability of our methodology.
\end{abstract}


\newpage

%%%%%%%%%%%%%%%%%%%%%%%%%%%%%%%%%%%%%%%%%%%%%%%%%%%%%%%%%%%%%%%%%%%%%%%%%%%%%%%%%%%%%%%%%%%%%%%%%%%%%%%%%%%%%%%%%%%%%%%%%%%%%%%%%%%%%%%%%%%%%%%%%%
\section{Introduction}
%
%%%%%%%%%%%%%%%%%%%%%%%%%%%%%%%%%%%%%%%%%%%%%%%%%%%%%%%%%%%%%%%%%%%%%%%%%%%%%%%%%%%%%%%%%%%%%%%%%%%%%%%%%%%%%%%%%%%%%%%%%%%%%%%%%%%%%%%%%%%%%%%%%%

Future yield curve scenarios are a necessary ingredient for many activities which are carried out in the financial risk management area.
Indeed the price of several securities depends on the future values of interest rate term structures. As a consequence, expected exposures of portfolios made of such instruments and in turn risk measures like VaR and Expected Shortfall are tightly related to the future distributions of underlying interest rates. Financial institutions thus need to forecast confidence intervals for yields with various times-to-maturity, ranging from few days up to 30 years, for a series of fixed dates in the future. 

The main issue when building a model to describe the evolution of the yield curve is the large number of degrees of freedom.
The dynamics which drives the term structure is inherently multivariate and thus one needs to model both the volatility of the individual yields and the correlation among yields with different maturities. For this reason many of the approaches that have been proposed in the literature reduce substantially the dimension of the curve by relying on factor models. The choice of the factors is either based on well known dimensional reduction techniques, such as Principal Component Analysis (PCA) or Factor Analysis~\cite{litterman_scheinkman1991,knez1994,jamshidian_zhu1996,rebonato1998,scherer2002,driessen2003}, or driven by economic intuition.  The latter approach is the one firstly proposed by Nelson and Siegel~\cite{nelson1987} and developed later on by Diebold and Li~\cite{diebold_li2006}. It describes the yield curve in terms of three dynamical factors - level, slope and curvature - combined using factor loadings. The three factors evolve following a Vector Autoregressive dynamics estimated on the historical time series of yields, whereas the structure of the loading functions is fixed. Related works on factor models are~\cite{bliss1997,dai2000,dejong1999,dejong2000,duffee2002}, but nearly none of them consider forecasting directly. More recently an enriched dynamics has been proposed for the three factor models \`a la Nelson and Siegel, in order to account for  regime changes induced by central banks interventions~\cite{bernadell2005,Coroneo2011393}. A hidden Markov chain is introduced to mimic the evolution of macroeconomic variables, such as GDP and CPI growth, which influence the factor dynamics. A systematic study of regime switching factor models which include long memory effects and heteroskedasticity has been carried out in~\cite{monfort2007,gourieroux2013}.

The characterisation of the factor dynamics, either for factors coming from unrestricted analysis or selected on the basis of economic intuition, is usually performed parametrically exploiting the information content of the empirical data. A different approach is the non-parametric filtered historical simulation, proposed by Barone-Adesi et al. in~\cite{barone1999,barone2002}. This approach preliminarily computes the standardized residuals in a yield curve model with state dependent conditional means and volatilities. Then, residuals are bootstrapped to generate out-of-sample scenarios for yields with different maturities. This method does not rely on any distributional assumption and it is thus capable to account for a quite broad variety of historical patterns. A recent application of filtered historical simulation can be found in~\cite{audrino2004}, where the authors compute standardised residuals using the Functional Gradient Descent method.

A quite different, yet related, stream of literature dealing with future yield curve forecasting employs classical modelling approaches, such as short rate affine models or instantaneous forward rate models. In these frameworks, absence of arbitrage is automatically embedded in the stochastic differential equations which govern the evolution. Absence of arbitrage may indeed represent a desirable property to obtain reliable P\&L distributions for interest rate dependent portfolios, as explained in~\cite{teichmann_wuthrich2013}. In affine models, such as~\cite{Vasicek1977}, the yield curve is related to the short rate $r(t)$ by an exponential relation, $P(t,T)=\exp\{A(t,T)-B(t,T)\,r(t)\}$, where the functional form of $A(t,T)$ and $B(t,T)$ is fixed to ensure absence of arbitrage. In the second class of models, firstly introduced by Heath Jarrow and Morton (HJM)~\cite{heath1992}, the term structure is related to the instantaneous forward curve by $P(t,T)=\exp\{-\int_t^Tdu\,f(t,u)\}$. The absence of arbitrage here constraints the drift coefficient of $f(t,T)$ in the risk-neutral measure in terms of its volatility functions. To obtain a model capable to describe the evolution of the yield curve under the objective probability measure, one has to add a drift correction proportional to the market price of risk. The problem of measuring the price of risk implied by historical data is in general quite challenging. Attempts to build a viable estimation procedure have been discussed in~\cite{sahalia2010}, where the authors consider a multi factor dynamics for $r(t)$ and a number of different parametrisations for the market price of risk. In this respect, one advantage of HJM models is that they are less sensitive to a misspecification of the price of risk. As explained in~\cite{teichmann_wuthrich2013}, an error on its estimation affects the evolution of forward rates comparatively less than the evolution of the short rate. A recent attempt to generate interest rate scenarios in a HJM framework can be found in~\cite{teichmann_wuthrich2013}. In this paper we propose a feasible Maximum Likelihood methodology which provides a statistically significant estimate of the market risk-premia and also returns its historical evolution. As we describe in Section~\ref{sec:results} the interest rate market -- especially in the time window centered around the credit crisis -- implicitly quotes strongly positive risk-premia. This implies a strong negative correction to the drift of the HJM model which pushes future rate curves downward with respect to the observed forward curve.

\begin{figure}[h]
  \centering
  \includegraphics[width=0.9\textwidth]{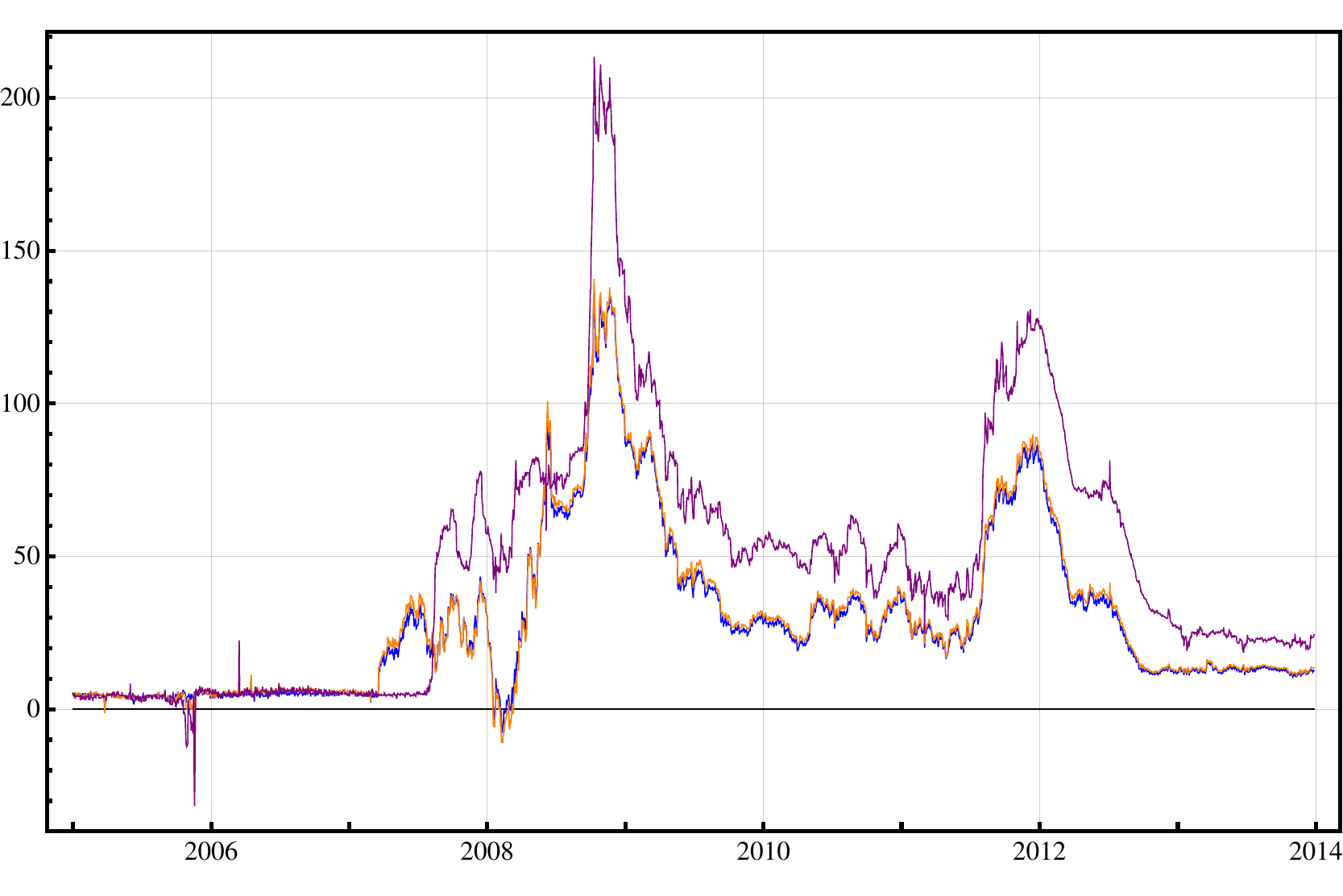}
  \caption{Differences in basis points between the two year continuously compounded yield computed from the curves with tenor $\Delta=$ 3M, 6M, and 1Y and the EONIA two year rate. (Blue $\to$ EUR3M, Orange $\to$ EUR6M, Purple $\to$ EUR1Y.)}
  \label{fig:hts_multicurve}
\end{figure}

All the approaches we have discussed so far describe the evolution of a single yield curve. The liquidity/credit crisis of 2007-2008 has strongly changed the interest rate landscape leading to a multiple yield curve scenario~\cite{henrard2007}. In old financial markets quotes of Forward Rate Agreements (FRA) and Zero Coupon Bond (ZCB) prices were related by simple no-arbitrage rules. Under the pressure of both the liquidity issues and the evidence that no counterparties could be still considered as risk-free entities, the post-crisis money market appears as a place where each forward rate seems to act as a different asset. As a consequence, a set of yield curves, instead of a single one, is today necessary to accommodate for the prices of interest rate derivatives quoted on the market: The Euro OverNight Index Average (EONIA) curve~\footnote{See Section~\ref{subsec:var} for a precise definition of the EONIA curve.}, i.e. the yield term structure for overnight borrowing in the Euro area, and the EUR3M, EUR6M, and EUR1Y. While the former is commonly assumed to be the best proxy for risk-free rates, the last three are sensitive to the credit and liquidity risks associated to longer tenors. For an introductory discussion on this topic see~\cite{morini2009,mercurio2009}. In Figure~\ref{fig:hts_multicurve} we show the rise of the difference between the two year continuously compounded yield computed from the curves with tenor $\Delta=$ 3M, 6M, and 1Y and the EONIA ZC yield with the same time-to-maturity.
\begin{center}
  ~\\
FIGURE~\ref{fig:hts_multicurve} SHOULD BE HERE.
  ~\\
\end{center}
In literature, several authors have proposed different approaches to extend classical interest rate models to the new multiple yield curve scenario. All these attempts share the pricing perspective even though they capture different aspects: Libor Market Models~\cite{mercurio2009,mercurio2010,mercurio2010b}, HJM modelling~\cite{pallavicini2010,fujii2011,crepey2012,crepey2013,pallavicini2014,cuchiero2014}, multiplicative spreads~\cite{henrard2007,henrard2010,henrard2013}, foreign currency approach~\cite{bianchetti2010}, SABR model extensions~\cite{bianchetti2011}, and short rate model extensions~\cite{kijima2009,kenyon2010,grasselli2014,morino2014}. For a comprehensive review on the subject we refer to the book by Henrard~\cite{henrard2014}. As far as we know, none of these models has been studied with the aim of producing reliable forecasts of confidence intervals for all the yield curves simultaneously.

%In this article we take as a starting point the multiple curve extension of the HJM framework developed in~\cite{pallavicini2014}.
%The advantage of this approach is that it is flexible enough to be rephrased in a discrete setting both for the time variable and the times-to-maturity which characterise each yield curve. This is a crucial feature which allows to design a viable estimation procedure based on time series analysis. We thus conceive our model in order to incorporate the information coming from the historical time series and thus produce reliable future scenarios for financial risk management purposes.

In this article we take as a starting point the multiple curve extension of the HJM framework proposed in~\cite{pallavicini2014}.
We develop a general Vector Autoregressive representation of such modelling framework by rephrasing it in a discrete setting both for the time variable
and the times-to-maturity. As dynamical objects we take the instantaneous forward term structure, to describe the risk-free curve, and FRA rates, to describe the longer tenor curves. We thus derive a joint Vector Autoregressive process of order one (VAR(1)) which accounts for the evolution both of the EONIA and the longer tenor curves. This representation is tailored for a viable Maximum Likelihood estimation of the volatility-correlation structure and market risk-premia on the time series of yields, and is thus capable to generate density distributions for the future values of the yields belonging to different term structures. To the best of our knowledge, this is the first attempt to forecast yield confidence levels in the novel multiple curve environment inclusive of market risk-premia and thus fully consistent with historical observations. 

Our setting is also well suited to the application of the PCA. We show how to reduce the number of Brownian shocks associated to each time-to-maturity bucket retaining only the minimum number of principal components which suffice to explain at least 95\% of the total variance. More in detail, 
we chose constant (in time) volatilities and estimate the model on the available historical data for the EONIA and the EUR3M term structures.
Starting from a high dimensional object, namely we need 22 buckets to describe two yield curves, we select a smaller number, i.e. always lower than 9, of principal components which account for most of the correlation. After a significant reduction of the model dimension, we devise a numerical procedure for the yield curve projection and, then, we perform an out-of-sample test. To assess the forecasting power of our specific model, we consider the unconditional coverage test, described in~\cite{kupiec1995,christoffersen1998}. For a coverage probability equal to 95\% and short forecasting horizons, we obtain very satisfactory results. Some inadequacies of the approach arise for more extreme coverage probabilities and longer horizons, especially for the shortest times-to-maturity of the EONIA curve. This effect can be attributed in part to the fact that the monetary policy of the European Central Bank (ECB) largely determines curve dynamics. The interventions of the ECB affect the level of the overnight rate in terms of discrete shifts, which a model driven by Brownian shocks can hardly reproduce. In order to explicitly incorporate the ECB policy one has to move to a modelling framework specifically designed to capture discrete movements in the target rate and shifts from tightening to easing regimes~\cite{renne2014}. Moreover, as far as the out-of-sample performance is concerned, \cite{renne2014,kimorphanides2012} prove that the use of survey-based forecasts of the ECB policy rate enhances the out-of-sample forecasting ability of future rates.

The paper is organised as follows. Section~\ref{sec:model} presents the Heath-Jarrow-Morton modelling framework and details the derivation of the first order Vector AutoRegressive (VAR(1)) representation of the equations which govern the dynamics. We also discuss the numerical approach used for the generation of future scenarios and the methodology employed for the estimation of the model parameters. Section~\ref{sec:results} presents our application to a data sample of yield curves from the Euro area and shows the results of our back-testing procedure. Finally, in Section~\ref{sec:conclusions}, we conclude and draw future perspectives.

%%%%%%%%%%%%%%%%%%%%%%%%%%%%%%%%%%%%%%%%%%%%%%%%%%%%%%%%%%%%%%%%%%%%%%%%%%%%%%%%%%%%%%%%%%%%%%%%%%%%%%%%%%%%%%%%%%%%%%%%%%%%%%%%%%%%%%%%%%%%%%%%%%
\section{The Model}
\label{sec:model}
%
%%%%%%%%%%%%%%%%%%%%%%%%%%%%%%%%%%%%%%%%%%%%%%%%%%%%%%%%%%%%%%%%%%%%%%%%%%%%%%%%%%%%%%%%%%%%%%%%%%%%%%%%%%%%%%%%%%%%%%%%%%%%%%%%%%%%%%%%%%%%%%%%%%

In this section we describe our model for the multiple yield curve environment. Preliminarily, we review the standard formulation of the Heath Jarrow Morton setting, and we specify the modelling in order to effectively describe the covariance structure of the historical time series.

We denote by $Y(t,x)$ the (continuously compounded) yield observed at time $t$ with time-to-maturity $x$, whose relation with the price $P(t,t+x)$ of a ZCB is given by 
\be
  P(t,t+x)=\exp\{-x\,Y(t,x)\}\,,
\ee
where $T=t+x$ is the maturity of the contract. Since it is common to report the historical yield term structures in terms of time-to-maturity buckets, in our description $x$ plays a central role. The most prominent quantity in the HJM framework is the instantaneous forward rate $f(t,x)$ defined as
\be
\label{eq:ffromP}
  f(t,x):=-\partial_x\ln{P(t,t+x)}\,,
\ee
or, in equivalent terms as an explicit function of $Y(t,x)$, as
\be
\label{eq:ffromY}
  f(t,x)=Y(t,x)+x\,\partial_x Y(t,x)\,.
\ee

%%%%%%%%%%%%%%%%%%%%%%%%%%%%%%%%%%%%%%%%%%%%%%%%%%%%%%%%%%%%%%%%%%%%%%%%%%%%%%%%%%%%%%%%%%%%%%%%%%%%%%%%%%%%%%%%%%%%%%%%%%%%%%%%%%%%%%%%%%%%%%%%%%
\subsection{Vector Autoregressive representation of the HJM framework}
\label{subsec:var}
%
%%%%%%%%%%%%%%%%%%%%%%%%%%%%%%%%%%%%%%%%%%%%%%%%%%%%%%%%%%%%%%%%%%%%%%%%%%%%%%%%%%%%%%%%%%%%%%%%%%%%%%%%%%%%%%%%%%%%%%%%%%%%%%%%%%%%%%%%%%%%%%%%%%

The starting point of our model is the Heath Jarrow Morton framework, where the dynamics of the yield curve is rephrased in term of the instantaneous forward rate dynamics. Under the risk-neutral measure the stochastic differential equation (SDE) driving the evolution of $f(t,x)$ reads as follows

\be
\label{eq:hjm}
df(t,x)=\left[\bm{\sigma}_f(t,x)\cdot\int_0^xdu\,\bm{\sigma}_f(t,u)+\partial_x\,f(t,x)\right]\,dt+\bm{\sigma}_f(t,x)\cdot d\bm{W}^0(t),
\ee
where $\bm{\sigma}_f(t,x)$ and  $\bm{W}^0(t)$ are $N$ dimensional vectors of volatility functions and independent Brownian motions, respectively~\footnote{The symbol $\cdot$ stands for the usual scalar product.}. 
The drift term is made up of two components. The first component corresponds to the HJM drift condition ensuring the absence of arbitrage, and it is completely determined after the specification of the volatility vectors $\bm{\sigma}_f(t,x)$. The second term is a differential correction originally introduced by Musiela~\cite{musiela1993,brace_musiela1994,bjork2004} which accounts for the description of the forward rate dynamics in terms of the time-to-maturity $x$. As far as the diffusion coefficient is concerned, for the moment we put no restrictions on the volatility vectors and they can be deterministic, local (i.e. $\bm{\sigma}_f(t,x)$ are functions of $f(t,x)$), or even depend on some additional stochastic process.

Academic and specialised literature provides several extensions of this modelling framework to the multiple yield curve environment, as we mentioned in the introduction.
%see for instance~\cite{mercurio2009,mercurio2010,mercurio2010b,pallavicini2010,fujii2011,crepey2012,crepey2013,pallavicini2014,henrard2007,henrard2010,henrard2013,bianchetti2010,bianchetti2011,kijima2009,kenyon2010,grasselli2014,morino2014}
In our paper we refer in particular to the work of Moreni and Pallavicini~\cite{pallavicini2014}.

The rate for overnight borrowing in the Euro area is the Euro OverNight Index Average (EONIA). It represents the underlying for Overnight Indexed Swaps (OIS). An OIS is a swap contract exchanging fixed \textit{versus} floating, where the floating rate is computed as the geometric average of EONIA rates. OIS rates are commonly assumed to be the best proxy for risk-free rates and the fact that they are usually employed as collateral rates in collateralised transactions has lead to OIS discounting. By relying on bootstrapping techniques (e.g. see\cite{ametrano2009}) we can obtain from OIS rates the term structure of OIS ZCB prices $x\mapsto P(t,t+x)$. Then, employing relation~(\ref{eq:ffromP}), we compute the term structure of the EONIA instantaneous forward rates. In addition, we consider fixed income instruments, such as FRA, swaps, caps/floors, and swaptions, whose underlying Libor (Euribor) rates are sensitive to tenors longer than the overnight one. Before the financial crisis Libor rates associated to different tenors were related by simple no-arbitrage relations. In the post-crisis interest rate market, this is no longer the case and a specific yield curve is constructed from market instruments whose underlying rate depends on a specific tenor. Among all financial instruments, FRA's, interest rate swaps, and basis swaps represent the most liquid interest rate linear derivatives, whose simple structure naturally lends itself to a bootstrap approach~(see, again,~\cite{ametrano2009}). For each tenor $\Delta=$3M, 6M, 1Y, we construct three \textit{risky} curves, i.e. the EUR three month, six month, and one year curves, and we denote with $Y_{\Delta}(t,x)$ the associated yields.
In the extension of the HJM modelling framework to the multiple yield curve environment that we are considering, the evolution of the risk-free curve is described in terms of instantaneous forward rates, whose dynamics corresponds to Equation~(\ref{eq:hjm}),
while the evolution of longer tenor curves is provided in terms of FRA par rates. By definition, the time $t$ FRA rate with tenor $\Delta$ and time-to-maturity $x$ is given by
\be
\FRA_{\Delta}(t,x):=\e^{\q^{t+x}}_t\left[F_\Delta(t+x-\Delta;t+x-\Delta,t+x)\right],
\ee
where $\q^{t+x}$ is the martingale measure whose numeraire $P(t,t+x)$ is the EONIA bond price, while $F_\Delta(t+x-\Delta;t+x-\Delta,t+x)$
is the Libor rate with tenor $\Delta$ which applies for unsecured deposit rates over the period $[t+x-\Delta,t+x]$.
Since both the instantaneous forward rates and the FRA rates are martingale under the same terminal measure $\q^{t+x}$, we express the joint dynamics as
\begin{eqnarray}
\label{eq:hjmmultif}
  &&df(t,x)=\partial_x\,f(t,x)\,dt+\bm{\sigma}_f(t,x)\cdot d\bm{W}^{t+x}(t)\,,\\
\label{eq:hjmmultifra}
  &&d\FRA_\Delta(t,x)=\partial_x\,\FRA_\Delta(t,x)\,dt+\bm{\sigma}_\Delta^\FRA(t,x)\cdot d\bm{W}^{t+x}(t)\,,
\end{eqnarray}
where $\bm{W}^{t+x}$ is an $N$ dimensional Brownian motion under $\q^{t+x}$~\footnote{As before, the drift terms appear because we parametrise the rate dynamics in terms of the time-to-maturity.}.

With the aim of describing the historical evolution of the yield curves we need to adjust the risk-neutral dynamics including the contribution associated to the market price of risk. As discussed in standard textbooks, see for instance~\cite{bjork2004}, this amounts to the addition in Equation~(\ref{eq:hjmmultif}) and~(\ref{eq:hjmmultifra}) of a drift correction equal to $-\bm{\lambda}\cdot\bm{\sigma}_f(t,x)\,dt$ and $-\bm{\lambda}\cdot\bm{\sigma}_{\Delta}^\FRA(t,x)\,dt$, respectively. The $N$ entries of the vector $\bm{\lambda}$ correspond to non-negative risk premia associated to the instantaneous forward and FRA rate volatilities. Since historical data sets are typically constituted by observations of a finite number of time-to-maturity buckets collected at daily or weekly frequency, we approximate the continuous time real-world dynamics by means of a discrete time process. We preliminarily detail our approach for the EONIA dynamics, and then extend it to the EUR3M, EUR6M, and EUR1Y curves. We denote with $\bm{s}$ the finite set of $K$ time-to-maturity buckets which describes the empirical EONIA term structure. Thus, Equation~(\ref{eq:hjmmultif}) is relevant only for those $x$ belonging to $\bm{s}$ and time $t$ corresponding to a discrete grid. Employing the Euler discretisation scheme we rewrite the SDE~(\ref{eq:hjmmultif}) as a set of $K$ equations
\begin{eqnarray}
  f(t_{k+1},s_i)-f(t_{k},s_i)\!&=&\!\left[\bm{\sigma}_f(t_k,s_i)\cdot\int_0^{s_i}du\,\bm{\sigma}_{f,\text{int}}(t_k,u)-\bm{\lambda}\cdot\bm{\sigma}_f(t_k,s_i)+\partial_x\,f_{\text{int}}(t_k,x)\Big|_{s_i}\right]\,\Delta t\notag\\
  \label{eq:hjmdiscrete}
  &&+\bm{\sigma}_f(t_k,s_i)\cdot \bm{\varepsilon}(t_{k+1})\,\sqrt{\Delta t}\,,
\end{eqnarray}
for $i=1,\ldots, K$ with $\bm{\varepsilon}(t_{k+1})\sim\mathcal{N}(\bm{0},I_N)$, $I_N$ being the $N\times N$ identity matrix.
It is important to observe that in order to compute the first order derivate of the instantaneous forward rate curve and integrate the volatility functions, we need to define interpolated versions of both quantities, $f_{\text{int}}(t_k,x)$ and $\bm{\sigma}_{f,\text{int}}(t_k,x)$. As suggested in~\cite{hagan2006}, a conventional choice corresponds to the use of the Bessel cubic spline method. 
The spline representation allows to write the derivative term in Equation~(\ref{eq:hjmdiscrete}) in matrix form (see Appendix~\ref{app:spline})
\be
\label{eq:derf}
\partial_x f_{\text{int}}(t_k,x)\Big|_{x=s_i}=\left[M_f(\bm{s})\,\bm{f}(t_k)\right]_i\,,
\ee
where we have introduced the vector of instantaneous forward rates
\bes
\bm{f}(t_k)=\left[f(t_k,s_1),\,f(t_k,s_2),\ldots,f(t_k,s_K)\right]^\intercal\,,
\ees
and $M_f(\bm{s})$ is a tridiagonal $K\times K$ matrix which depends only on the buckets vector $\bm{s}$~\footnote{We drop the dependence of $A_i$, $B_i$ and $C_i$ on the vector $\bm{s}$ of maturity buckets for ease of notation.}
\be
\label{eq:splinematrix}
M_f(\bm{s})=\left[
  \begin{array}{ccccccc}
    A_1 & B_1 & C_1 & 0 &0&\dots & 0\\ 
    A_2 & B_2 & C_2 & 0 &0&\dots & 0 \\ 
    0&A_3&B_3& C_3 &0& \dots &0\\
    \vdots&&&&&&\vdots\\
    0&0&\dots &0&A_{K_\eonia-1}&B_{K_\eonia-1}& C_{K_\eonia-1}\\
    0&0&\dots &0&A_{K_\eonia}&B_{K_\eonia}& C_{K_\eonia}\\
  \end{array}\right].
\ee
The same happens with the integral of the volatility functions
\be
\label{eq:intsigma}
\bm{\sigma}_f(t_k,s_i)\cdot\int_0^{s_i}du\,\bm{\sigma}_{f,\text{int}}(t_k,u)=\sum_{\substack{h=1}}^{K}\left[P_f(\bm{s})\right]_{ih}\,\bm{\sigma}_f(t_k,s_i)\cdot \bm{\sigma}_f(t_k,s_h)=\left[P_f(\bm{s})\,\Sigma_f(t_k)\,\Sigma_f^\intercal(t_k)\right]_{ii}\,,
\ee
where $P_f(\bm{s})$ is a $K \times K$ matrix of the form~\footnote{We drop the dependence of $\left[P_f\right]_{ij}$ on the vector $\bm{s}$ of maturity buckets for ease of notation.} (see Appendix~\ref{app:spline})
\be
\label{eq:pf}
P_f(\bm{s})=\left[\begin{array}{ccccccc}
\left[P_f\right]_{11} & 0 & 0 & 0 &0&\dots & 0\\ 
\left[P_f\right]_{21} & \left[P_f\right]_{22} & \left[P_f\right]_{23} & 0 &0&\dots & 0 \\ 
\left[P_f\right]_{31} & \left[P_f\right]_{32} & \left[P_f\right]_{33} & \left[P_f\right]_{34} &0& \dots &0\\
\\
\vdots&&&&&&\vdots\\
\\
\left[P_f\right]_{K\,1}& \left[P_f\right]_{K\,1} & \left[P_f\right]_{K\,2} & \dots &  \dots &  \dots  & \left[P_f\right]_{K\,K}\\
\end{array}\right]\,,\\
\ee
and $\Sigma_f(t_k)$ is a $K\times N$ matrix of volatilities
\be
\label{eq:volmatrix}
\Sigma_f(t_k)=\left[\bm{\sigma}_f(t_k,s_1),\,\bm{\sigma}_f(t_k,s_2),\ldots,\bm{\sigma}_f(t_k,s_{K})\right]^\intercal\,.
\ee
Then, the system~(\ref{eq:hjmdiscrete}) can be represented as a VAR(1) process
\be
\label{eq:hjm3}
\bm{f}(t_{k+1})=\left(I_{K}+M_f(\bm{s})\,\Delta t\right)\,\bm{f}(t_k)+\bm{\mu}_f(t_k)\,\Delta t+\Sigma_f(t_k)\, \bm{\varepsilon}(t_{k+1})\,\sqrt{\Delta t}\,,
\ee
where the drift term $\bm{\mu}_f(t_k)$ is defined as
\be
\label{eq:muf}
\bm{\mu}_f(t_k)=\text{diag}\left[P_f(\bm{s})\,\Sigma_f(t_k)\,\Sigma_f^\intercal(t_k)\right]-\Sigma_f(t_k)\,\bm{\lambda}.
\ee
The same procedure can be applied to each of the infinite-dimensional SDE's which drive the evolution of the FRA term structures associated to the EUR3M, EUR6M and EUR1Y curves, i.e. Equation~(\ref{eq:hjmmultifra}), and reduce them to finite-dimensional systems.
We employ the Bessel cubic spline interpolation method on $FRA_\Delta(t,x)$, keeping in mind that for each curve we have in principle a different vector of time-to-maturity buckets, denoted by $\bm{s}_{\Delta}$.
We approximate the derivative term which appears in the drift term of Equation~(\ref{eq:hjmmultifra}) as
\be
\label{eq:derfra}
\partial_x\, FRA_{\Delta~\text{int}}(t_k,x)\Big|_{x=s_{\Delta,i}}=\left[M_\Delta(\bm{s}_{\Delta})\,\bm{\tilde{F}}_\Delta(t_k)\right]_i\,,
\ee
where
$\bm{\tilde{F}}_\Delta(t_k)$ is the $K_\Delta$ dimensional vector of FRA rates
\be
\bm{\tilde{F}}_\Delta(t_k)=\left[FRA_\Delta(t_k,s_{\Delta,1}),\, FRA(t_k,s_{\Delta,2}),\ldots, FRA(t_k,s_{\Delta,K_\Delta})\right]^\intercal,
\ee

and $M_\Delta(\bm{s}_\Delta)$ is a $K_\Delta\times K_\Delta$ matrix which depends only on $\bm{s}_\Delta$ and has the same form of $M_f(\bm{s})$, see Equation~(\ref{eq:splinematrix}).
As one can see in Equation~(\ref{eq:hjmmultifra}), beside the derivative component the FRA drift contains the integral term, which we can work out as we did for the instantaneous forward curve
\bes
\bm{\sigma}_{\Delta}(t,s_{\Delta,i})\cdot\int_0^{s_{\Delta,i}}du\,\bm{\sigma}_{f,\text{int}}(t,u)=\sum_{\substack{h=1}}^{K_\Delta}\left[P_\Delta(\bm{s},\bm{s}_\Delta)\right]_{ih}\,\bm{\sigma}_\Delta(t_k,s_{\Delta,i})\cdot \bm{\sigma}_f(t_k,s_h)=\left[P_\Delta(\bm{s},\bm{s}_\Delta)\,\Sigma_f(t_k)\,\Sigma_\Delta^\intercal(t_k)\right]_{ii}
\ees
where $P_\Delta(\bm{s},\bm{s}_\Delta)$ is a $K_\Delta\times K$ matrix which depends on the vectors $\bm{s}$ and $\bm{s}_\Delta$. It is the analog of $P_f(\bm{s})$ defined in Equation~(\ref{eq:pf}). Here $\Sigma_\Delta(t_k)$ is a $K_\Delta\times N$ matrix containing the FRA rates volatilities
\be
\label{eq:volmatrixS}
\Sigma_\Delta(t_k)=\left[\bm{\sigma}_\Delta(t_k,s_{\Delta,1}),\ldots,\bm{\sigma}_\Delta(t_k,s_{\Delta,K_\Delta})\right]^\intercal.
\ee

Finally, the joint dynamics of the risk-free instantaneous forward curve and the FRA curves under the real world measure reads as follows
\be
\label{eq:multihjmfra}
\begin{split}
  &\bm{f}(t_{k+1})=\left(I_{K}+M_f(\bm{s})\,\Delta t\right)\,\bm{f}(t_k)+\bm{\mu}_f(t_k)\,\Delta t+\Sigma_f(t_k)\, \bm{\varepsilon}(t_{k+1})\,\sqrt{\Delta t},\\
  &\bm{\tilde{F}}_\Delta(t_{k+1})=\left(I_{K_\Delta}+M_{\Delta}(\bm{s}_\Delta)\,\Delta t\right)\,\bm{\tilde{F}}_\Delta(t_k)+\bm{\mu}_\Delta(t_k)\,\Delta t+\Sigma_\Delta(t_k)\, \bm{\varepsilon}(t_{k+1})\,\sqrt{\Delta t}\,,
\end{split}
\ee
for $\Delta=\text{3M, 6M, 1Y}$, where the drift coefficient $\bm{\mu}_\Delta(t_k)$ of the FRA rates has the form
\be
\label{eq:mufra}
\bm{\mu}_\Delta(t_k)=\text{diag}\left[P_\Delta(\bm{s},\bm{s}_\Delta)\,\Sigma_f(t_k)\,\Sigma^\intercal_{\Delta}(t_k)\right]-\Sigma_\Delta(t_k)\,\bm{\lambda}.
\ee

Equation~(\ref{eq:multihjmfra}) is one of the main contribution of this paper. 
It corresponds to a VAR(1) representation of the HJM modelling framework extended to describe multiple yield curves by including the FRA rate dynamics.
It is conceived to describe the historical evolution of the EONIA curve through the instantaneous forward rates and of the higher tenor curves by means of the FRA term structures. The model is completely specified after choosing the volatility matrices $\Sigma_f(t)$ and $\Sigma_\Delta(t)$. 
As we will detail in the next sections, we consider the case of constant (in time) volatility functions, both for the EONIA and the EUR3M, EUR6M and EUR1Y curves.

\subsection{Constant volatility model}
In this section we specify the form of the volatility matrices $\Sigma_f(t)$ and $\Sigma_\Delta(t)$. We consider constant volatility functions both for the instantaneous forward curve and the FRA term structures, i.e.
\bes
\begin{split}
&\Sigma_f(t_k)\equiv\Sigma_f\,,\\
&\Sigma_\Delta(t_k)\equiv\Sigma_\Delta\,.\\
\end{split}
\ees
The major advantages of this choice are that it allows to devise a simple approach to estimation and to reduce the dimensionality of the problem by means of the PCA~\cite{anderson1963,flury1997}. Indeed, among the $N$ Brownian drivers entering the yield and FRA dynamics, PCA permits to select a subset of componets with dimension $F<<N$ which still explains a large fraction of the observed variance.

Preliminarily, we define the following vectors which can be computed directly from the available historical time series
\be
\label{eq:shiftedvectorsfra}
\begin{split}
&\bm{y}_f(t_{k+1})=\bm{f}(t_{k+1})-\left(I_{K}+M_f(\bm{s})\,\Delta t\right)\,\bm{f}(t_k)\,,\\
&\bm{y}_{\Delta}(t_{k+1})=\bm{\tilde{F}}_\Delta(t_{k+1})-\left(I_{K_\Delta}+M_\Delta(\bm{s}_\Delta)\,\Delta t\right)\,\bm{\tilde{F}}_\Delta(t_k), \qquad \Delta=3M, 6M, 1Y.
\end{split}
\ee
The advantage of introducing such vectors lies in the fact that now the Equations in~(\ref{eq:multihjmfra}) can be expressed as covariance stationary processes
\be
\label{eq:ydynamics}
\begin{split}
&\bm{y}_f(t_{k})=\bm{\mu}_f\,\Delta t+\Sigma_f\, \bm{\varepsilon}(t_{k})\,\sqrt{\Delta t}\,,\\
&\bm{y}_\Delta(t_{k})=\bm{\mu}_\Delta\,\Delta t+\Sigma_\Delta\, \bm{\varepsilon}(t_{k})\,\sqrt{\Delta t},\qquad \Delta=\text{3M, 6M, 1Y}.
\end{split}
\ee
where $\bm{\mu}_f$ and $\bm{\mu}_\Delta$ correspond to the quantities defined in Equations~(\ref{eq:muf}) and~(\ref{eq:mufra}), respectively, dropping the dependence on time.
%\be
%\label{eq:squarevolmodel}
%\begin{split}
%&\bm{\mu}_f=\text{diag}\left[P_f\,\Sigma_f\,\Sigma_f^\intercal\right]\\
%&\bm{\mu}_\Delta=\text{diag}\left[P_\Delta\,\Sigma_f\,\Sigma^\intercal_{\Delta}-\frac{1}{2}\,\Sigma_\Delta\,\Sigma^\intercal_{\Delta}\right],\qquad \Delta=\text{3M, 6M, 1Y}.
%\end{split}
%\ee

% \xxx{Then, from Equation~(\ref{eq:ydynamics}) it readily follows that
% \bes
% \e\left[\left(\bm{y}_f-\e\left[\bm{y}_f\right]\right)\otimes\left(\bm{y}_f-\e\left[\bm{y}_f\right]\right)\right]=\Sigma_f\,\Sigma^\intercal_f\,\Delta t\,,
% \ees
% and analogously for $\bm{y}_\Delta$.}

Now we can embed the four vectors of Equation~(\ref{eq:shiftedvectorsfra}) in a single vector with $D=K+K_{\text{3M}}+K_{\text{6M}}+K_{\text{1Y}}$ components
\bes
\bm{y}(t_k)=\left[\bm{y}^\intercal_f(t_k),\,\bm{y}^\intercal_{\text{3M}}(t_k),\,\bm{y}^\intercal_{\text{6M}}(t_k),\bm{y}^\intercal_{\text{1Y}}(t_k)\right]^\intercal\,,
\ees
which also has covariance stationary dynamics
\be
\label{eq:deary}
\bm{y}(t_{k})=\bm{\mu}\,\Delta t+\Sigma\, \bm{\varepsilon}(t_{k})\,\sqrt{\Delta t},
\ee
where $\bm{\mu}$ is a $D$ dimensional drift term
\bes
\bm{\mu}=\left[\bm{\mu}^\intercal_f,\,\bm{\mu}^\intercal_{\text{3M}},\,\bm{\mu}^\intercal_{\text{6M}},\bm{\mu}^\intercal_{\text{1Y}}\right]^\intercal\,
\ees
and $\Sigma$ is a $D\times N$ volatility matrix
\bes
\Sigma=\left[\Sigma^\intercal_f,\,\Sigma^\intercal_{\text{3M}},\,\Sigma^\intercal_{\text{6M}},\Sigma^\intercal_{\text{1Y}}\right]^\intercal.
\ees
If we now assume that $N=D$, a convenient parametrization for the volatility matrix $\Sigma$ is the following one:
\be
\label{eq:factorvol}
\Sigma=\Omega\, R,
\ee
where $\Omega$ is a diagonal matrix containing the volatilities of the components of $\bm{y}$
\bes
\Omega=\left[\begin{array}{cccc}
\omega_1 & 0 &\dots & 0\\ 
0 & \omega_2 &\dots & 0 \\ 
\vdots&&\ddots&\vdots\\
0&\dots &0&\omega_{D}\\
\end{array}\right],
\ees
whereas $R$ is the lower triangular Cholesky decomposition of the correlation matrix $\Gamma$
\bes
R\,R^\intercal=\Gamma.
\ees
With this choice the covariance matrix of the vector $\bm{y}$ reads
\bd
\text{Cov}\left[\bm{y}(t_k)\right]=\Sigma\,\Sigma^\intercal=\Omega\,\Gamma\,\Omega.
\ed
We also introduce the vector $\bm{\omega}$ containing the volatilities $\omega_1,\ldots,\omega_D$
\bes
\bm{\omega}=\left[\omega_1,\ldots,\omega_D\right]^\intercal,
\ees
in such a way that the matrix $\Omega$ can be rewritten synthetically as $\text{diag}\left[\bm{\omega}\right]$.
If we now insert Equation~(\ref{eq:factorvol}) inside the expression for $\bm{\mu}$ we obtain
\be
\label{eq:driftpar}
\bm{\mu}=\bm{\omega}\circ\left(P\circ \Gamma\right)\,\bm{\omega}-\bm{\omega}\circ R\,\bm{\lambda},
\ee
where $\circ$ denotes the Hadamard product between two matrices, i.e. $\left(A\circ B\right)_{ij}=A_{ij}\,B_{ij}$, or equivalently between two vectors, i.e. $\left(\bm{a}\circ \bm{b}\right)_{i}=a_i\,b_i$, whereas the $P$ matrix is a $D\times D$ matrix defined in terms of the matrices $P_f$ and $P_\Delta$
\bes
P=\left[\begin{array}{ccccc}P_f & 0 & 0 &\dots &0 \\ P_{\text{3M}} & 0 & 0 &\dots &0 \\ P_{\text{6M}} & 0 & 0 &\dots & 0 \\ P_{\text{1Y}}&0 &0 &\dots&0\end{array} \right].
\ees
Finally it is useful to introduce the vector $\bm{\eta}(t_k)$
\bes
\bm{\eta}(t_k)=R\,\bm{\varepsilon}(t_k),
\ees
in order to write the generating process for $\bm{y}(t_k)$ as follows
\be
\label{eq:multihjm2}
\bm{y}(t_k)=\bm{\mu}\,\Delta t+\bm{\omega} \circ \bm{\eta}(t_k)\,\sqrt{\Delta t},
\ee
with $\bm{\eta(t_k)}\sim\mathcal{N}(\bm{0},\Gamma)$.
From Equation~(\ref{eq:factorvol}), one can easily see that the covariance matrix is parametrized in terms of $D$ volatilities and $D(D-1)/2$ correlation coefficients, whereas Equation~(\ref{eq:driftpar}) states that $D$ additional components of the market price of risk vector $\bm{\lambda}$ are needed to determine the drift vector.

\subsection{Simulation of future scenarios}
\subsubsection{Gaussian diffusive model}
\label{sec:gaussian}
Equation~(\ref{eq:multihjm2}) allows two different implementations of the forecasting procedure. The first possibility is to carry out a step by step Monte Carlo simulation sampling iteratively from a multivariate normal distribution whose mean and covariance structure can be computed from the drift and diffusion coefficients. The second one is computationally more convenient and relies on the following equation, which is an immediate consequence of the relation~(\ref{eq:multihjm2}) 
\be
\label{eq:intinstfwd}
\bm{f}(t_{k+1})=(I_{K}+M_f(\bm{s})\,\Delta t)^k\, \bm{f}(t_1)+\bm{\mu}_f\,k\,\Delta t+\sum_{\substack{h=0}}^{k-1}(I+M_f(\bm{s})\,\Delta t)^h\,\Sigma_f\,\bm{\varepsilon}(t_{k-h+1})\sqrt{\Delta t}.
\ee
Thus, the instantaneous forward rate vector at time $t_{k+1}$ conditionally on the value of $\bm{f}(t_1)$ is normally distributed with mean vector and covariance matrix given by
\begin{eqnarray}
%&&\bm{f}(t_{k+1})\sim\mathcal{N}(\bm{\mu}_{f}(t_{k+1}),\Sigma_f(t_{k+1}))\\
&&\e\left[\bm{f}(t_{k+1})|\bm{f}(t_1)\right]=\bm{\mu}_f\,k\,\Delta t+(I_{K}+M_f(\bm{s})\,\Delta t)^k\, \bm{f}(t_1)\,,\nonumber\\
&&\text{Cov}\left[\bm{f}(t_{k+1})\right]=\Delta t\,\sum_{\substack{h=0}}^{k-1}(I_{K}+M_f(\bm{s})\,\Delta t)^h \,\Sigma_f\,\Sigma^\intercal_f\,\left((I_{K}+M_f(\bm{s})\,\Delta t)^h\right)^{T}.\nonumber
\end{eqnarray}
A similar result holds also for the FRA vectors and allows to perform Monte Carlo simulation sampling directly over a long time horizon. Moreover, starting from Equation~(\ref{eq:intinstfwd}) we can compute the distribution of the ZC yield vector $\bm{Y}(t_{k+1})$ since
\bes
Y_i(t_{k})=\frac{1}{s_i}\int_0^{s_i}du\, f_{int}(t_k,u)=\frac{1}{s_i}\sum_{\substack{h=1}}^{K_\Delta}\left[P_f(\bm{s})\right]_{ih}\,f(t_k,s_h)=\frac{1}{s_i}\left[P_f(\bm{s})\bm{f}(t_k)\right]_{i}\,.
\ees
From the expression above it is evident that also the ZC yields are multivariate normally distributed random variables at each point in time, and we can explicitly compute the associated conditional mean and covariance matrix.

The present model corresponds essentially to a Gaussian dynamics for the EONIA instantaneous forward rates and for the EUR3M, EUR6M, and EUR1Y FRA curves. As a consequence, it allows for negative rates, a feature which is not ruled out by data. Given the extremely low level of the interest rates at the shortest maturities, especially at the time of writing, it can represent a valuable characteristic of our model. The $M_f(\bm{s})$ matrix, which appears in both expressions of mean and covariance, plays a role in the time evolution of the forward rates variance. Its effects depend crucially on whether the term structure is flat (little or no effect), upward sloping (the variance grows faster than in a simple diffusion model) or downward sloping (slows down the variance growth with respect to a purely diffusive dynamics) at a specific time-to-maturity bucket. Similar conclusions can be drawn for the FRA curves. 

\subsubsection{Bootstrap of innovation vectors}
\label{sec:bootstrap}
The assumption of normally distributed disturbances may result inadequate under same circumstances, especially in presence of market turmoils. Moreover, the analysis of the residuals computed after estimation exhibits evidences of excess of kurtosis and heteroskedastic behaviour of rate volatilities. In order to partially capture these effects we consider a complementary strategy to project interest rate term structures into the future. We  resort to the bootstrap technique and sample with replacement the vectors $\bm{\eta}(t_k)$ from the historical time series~\cite{efron1979}. We expect that this slight modification of the forecasting procedure improves the capability of the model to capture the tail behaviour of rate distribution with respect to the multivariate normal case. It is important to stress that our set up does not correct for the possible serial correlation of residuals or heteroskedasticity of volatility time series. The results we present in the next section are in line with these expectations.

\subsection{Estimation of model parameters}\label{sec:estimation}
As explained in the previous sections, from the yield curve  historical time series we need to estimate the $D\times D$ covariance matrix $\Sigma$, parametrized in terms of $D$ volatilities, $D(D-1)/2$ correlation coefficients, and $D$ components of the market price of risk vector $\bm{\lambda}$.
We devise an estimation procedure which consists in an iterative search of the maximum of the likelihood function, computed by fixing all the parameters of the model except one. 

We introduce the vector of parameters $\bm{\theta}$
\bd
\bm{\theta}=\{\lambda_1,\ldots,\lambda_D,\omega_1,\ldots,\omega_D\}
\ed
neglecting for the moment the $D(D-1)/2$ correlation coefficients $\Gamma_{12},\Gamma_{13},\ldots,\Gamma_{D-1\,D}$. This choice is motivated by the fact that the optimization algorithm will treat the vector $\bm{\theta}$ and the correlation coefficients in different ways during estimation.

The log-likelihood function of the series $\{\bm{y}(t_k)\}_{k=1}^L$ changed by sign reads
\bd
\begin{split}
&\mathcal{L}\left(\bm{y}(t_L),\ldots,\bm{y}(t_1)\Big|\bm{\theta},\Gamma\right)\\
&\hspace{50pt}=\frac{L\,D}{2}\,\ln\left(2\pi\right)+ \frac{L}{2} \ln\left(\det\left(\Gamma\right)\right)+ \frac{L}{2}\sum_{\substack{i=1}}^D\,\ln\left(\omega_{i}\right)+ \frac{1}{2} \sum_{\substack{k=1}}^{L}\bm{\eta}(t_k)\cdot\Gamma^{-1}\bm{\eta}(t_k).
\end{split}
\ed
In what follows we will denote by $\bm{\theta}^{(n)}$ and $\Gamma^{(n)}$ the values of the parameters at step $n$ of the algorithm. To start the calibration we initalise the market price of risk vector to zero, whereas volatilities and correlations are set equal to their Pearson estimates. At each subsequent step all the components of $\bm{\theta}$ except one, say $\theta_i$, are fixed to the value obtained at the previous step
\bd
\begin{cases}
&\theta_i\quad\text{treated as a variable},\\
&\theta_j=\theta^{(n-1)}_j,\quad \forall j\neq i.
\end{cases}
\ed
The correlation matrix is chosen as
\bd
\Gamma=\rho^{(n-1)},
\ed
where the matrix $\rho^{(n-1)}$ is estimated at the step $n-1$ and will be defined in a while. With this choice for $\bm{\theta}$ and $\Gamma$ the vector of residuals can be easily computed as
\bd
\tilde{\eta}_j(t_k;\theta_i)=\frac{y_j (t_k)-\mu_j\,\Delta t}{\omega_j\,\sqrt{\Delta t}},\quad j=1,\ldots,D,\quad k=1,\ldots,L.
\ed
The associated likelihood $\mathcal{L}$ is treated as a function only of $\theta_i$, then we define $\theta^\star_i$ as the value which minimises it
\bd
\theta^\star_i = \text{argmin}_{\substack{\theta_i}}\,\mathcal{L}\left(\bm{y}(t_L),\ldots,\bm{y}(t_1)\Big|\theta^{(n-1)}_1,\ldots,\theta^{(n-1)}_{i-1},\theta_i,\theta^{(n-1)}_{i+1},\ldots,\theta^{(n-1)}_{2D},\rho^{(n-1)}\right).
\ed
Finally the value of $\theta_i$ is updated $\theta^{(n)}_i=\theta^\star_i$. Using the value $\bm{\theta}^{(n)}$ one computes the vector of residuals
\bd
\eta^{(n)}_i(t_k)=\frac{y_i(t_k)-\mu^{(n)}_i\,\Delta t}{\omega^{(n)}_i\,\sqrt{\Delta t}},\quad i=1,\ldots,D,
\ed
and then the sample mean of its outer product
\bd
Q^{(n)}_{ij}=\frac{1}{L}\sum_{\substack{k=1}}^L\eta^{(n)}_i(t_k)\,\eta^{(n)}_j(t_k).
\ed
The correlation matrix to be used at each step is then computed as follows
\bd
\rho^{(n)}_{ij}=\frac{Q^{(n)}_{ij}}{\sqrt{Q^{(n)}_{ii}}\,\sqrt{Q^{(n)}_{jj}}}.
\ed
We repeat the procedure until
\bd
\begin{split}
&\left|\theta^{(n)}_i-\theta^{(n-1)}_i\right|<\gamma\, \left|\theta^{(n-1)}_i\right|,\quad \forall\, i=1,\ldots,2D,\\
&\left|\rho^{(n)}_{ij}-\rho^{(n-1)}_{ij}\right|<\gamma\, \left|\rho^{(n-1)}_{ij}\right|,\quad \forall\, i=1,\ldots,D,\quad j=1,\ldots,i,\\
\end{split}
\ed
with $\gamma=10^{-4}$.
We denote the final value of the parameters as
\bes
\begin{split}
&\hat{\bm{\theta}}=\{\hat{\lambda}_1,\ldots,\hat{\lambda}_D,\hat{\omega}_1\ldots,\hat{\omega}_D\}\\
&\hat{\Gamma}_{12},\hat{\Gamma}_{13},\ldots,\hat{\Gamma}_{D-1\,D}.
\end{split}
\ees
In the empirical analysis we make an additional assumption on the market price of risk vector, by guessing that its component are step-wise constant. In particular we introduce two components for the EONIA curve, one accounting for the shortest maturities and the other for the medium-term and long-term ones, and one component for each longer tenor curve. Specifically, in the case of four curves the $\bm{\lambda}$ vector will be defined as follows
\bd
\bm{\lambda}=\left[
\underbrace{\lambda_s,\ldots,\lambda_s}_{K_s},\:
\underbrace{\lambda_l,\ldots,\lambda_l}_{K-K_s},\:
\underbrace{\lambda_{\text{3M}},\ldots,\lambda_{\text{3M}}}_{K_{\text{3M}}},\:
\underbrace{\lambda_{\text{6M}},\ldots,\lambda_{\text{6M}}}_{K_\text{6M}},\:
\underbrace{\lambda_{\text{1Y}},\ldots,\lambda_{\text{1Y}}}_{K_\text{1Y}},
\right]^\intercal
\ed
Our choice is motivated by the idea that the market is more sensitive to and prices differently the risk associated with the short-term and long-term components of the yield curves. This assumption is also empirically supported by a preliminary analysis which shows that the relative improvement in the likelihood function due to the inclusion of extra risk factors is negligible. 

The errors on the estimated parameter are computed with the bootstrap technique. Once we have obtained the values $\hat{\bm{\theta}}$ and $\hat{\Gamma}$,
we sample the associated series of residuals $\{\hat{\bm{\eta}}(t_k)\}_{k=1}^L$ to build $N_\text{b}=500$ time series of adjusted returns with the same length of the historical one. We repeat estimation on each bootstrap copy and obtain a bootstrap sample of parameters $\{\hat{\bm{\theta}}_{i},\hat{\Gamma}_{i}\}_{i=1}^{N_\text{b}}$. Conditionally on the non existence of a significant bias between the historical estimate and the bootstrap mean, we take the standard deviation of the $N_\text{b}$ estimates as a proxy for the standard error of the parameters.

We have extensively tested the estimation procedure presented above on synthetic time series computed via Monte Carlo. We have considered several different scenarios with high and low volatilities, and with relatively high and low level of the market price of risk. For all scenarios the estimate provided by the iterative algorithm was in statistical agreement with the true parameter value and statistically different from zero.

In the next section we present the empirical results. We first carry out the estimation of the model on the historical time series at our disposal, both in a single and multiple yield curve framework. Then we move to the analysis of the forecasting ability for multiple yield curves and test the model through an out-of-sample exercise.

%%%%%%%%%%%%%%%%%%%%%%%%%%%%%%%%%%%%%%%%%%%%%%%%%%%%%%%%%%%%%%%%%%%%%%%%%%%%%%%%%%%%%%%%%%%%%%%%%%%%%%%%%%%%%%%%%%%%%%%%%%%%%%%%%%%%%%%%%%%%%%%%%%
\section{Empirical results}
\label{sec:results}
%
%%%%%%%%%%%%%%%%%%%%%%%%%%%%%%%%%%%%%%%%%%%%%%%%%%%%%%%%%%%%%%%%%%%%%%%%%%%%%%%%%%%%%%%%%%%%%%%%%%%%%%%%%%%%%%%%%%%%%%%%%%%%%%%%%%%%%%%%%%%%%%%%%%
\subsection{Data set}
The data set at our disposal consists of daily time series of four ZC yield curves from the Euro area: The EONIA curve, the EUR3M curve, the EUR6M curve, and the EUR1Y curve~\footnote{Market quotes are taken from Bloomberg and Reuters.}.  For computational reasons we limit our analysis to two curves, the EONIA and the EUR3M, but the case with four curves straightforwardly follows the same line of reasoning. We report in Table~\ref{tab:dataset2dates} the starting and ending dates of the two time series.
\begin{center}
  ~\\
  TABLE~\ref{tab:dataset2dates} SHOULD BE HERE\\
  ~\\
\end{center}
Each curve is made of a finite number of time-to-maturity buckets
\begin{itemize}
  \item $s^\prime_i \mapsto Y(t_k,s^\prime_i)$\\ 
  ZC yields of the EONIA curve for $i=1,\ldots,K^\prime_{\eonia}$;
  \item $s^\prime_{\text{3M},i}\mapsto Y_{\text{3M}} (t_k,s^\prime_{\text{3M},i})$\\
  ZC yields of the EUR3M curve, for $i=1,\ldots,K^\prime_{\text{3M}}$.
\end{itemize}
The two vectors $\bm{s}^\prime$, and $\bm{s}^\prime_{\text{3M}}$ are reported in Table~\ref{tab:dataset2mat}.
\begin{center}
  ~\\
  TABLE~\ref{tab:dataset2mat} SHOULD BE HERE\\
  ~\\
\end{center}
The instantaneous forward term structure can be obtained from the yields by means of Equation~(\ref{eq:ffromY}).
As for the longer tenor curve, we compute the $\FRA_{\text{3M}}$ rates following equation~(\ref{eq:forward}) where $P(t,t+x)$ is substituted by $P_{\text{3M}}(t,x):=\exp\left\{-x\,Y_{\text{3M}}(t,x)\right\}$
\be
\label{eq:forward}
\FRA_{\text{3M}}(t,x):=\frac{1}{\Delta}\left(\frac{P_{\text{3M}}(t,t+x-\Delta)}{P_{\text{3M}}(t,t+x)}-1\right),
\ee
with $\Delta =$ three months.
From the expression above it is evident that the quantity $\FRA_{\text{3M}}(t,x)$ is defined only for $x\geq \Delta$.

In order to speed up computation, we select a limited number of times to maturity, which are collected in the vectors $\bm{s}$ and $\bm{s}_{\text{3M}}$ reported in Table~\ref{tab:dataset2matspread}. 
\begin{center}
  ~\\
  TABLE~\ref{tab:dataset2matspread} SHOULD BE HERE\\
  ~\\
\end{center}

%%%%%%%%%%%%%%%%%%%%%%%%%%%%%%%%%%%%%%%%%%%%%%%%%%%%%%%%%%%%%%%%%%%%%%%%%%%%%%%%%%%%%%%%%%%%%%%%%%%%%%%%%%%%%%%%%%%%%%%%%%%%%%%%%%%%%%%%%%%%%%%%%%
\subsection{Estimation results}
In this section we show the results of the procedure described in Section~\ref{sec:estimation}. We start from the single curve case, i.e. we limit our analysis to the risk-free curve described by the EONIA instantaneous forward term structure. In this setting the vector $\bm{y}$ reduces to the $K=12$ dimensional vector $\bm{y}_f$. We thus need to estimate $66$ correlation coefficient, $12$ volatilities and the market price of risk vector. We describe the vector $\bm{\lambda}$ in terms of two components, $\lambda_s$ and $\lambda_l$:
\bd
\bm{\lambda}=\left[\underbrace{\lambda_s,\lambda_s}_{K_s=2},\:\underbrace{\lambda_l,\ldots,\lambda_l}_{K-K_s=10}\right]^\intercal.
\ed
The first component is associated to the two shortest buckets of the EONIA curve, the one month and two month instantaneous rates, whereas the second one refers to the time-to-maturity buckets ranging from three months up to 30 years.

Then we move to the multiple yield curve environment. 
Since $K=12$, and $K_{\text{3M}}=10$, the dimension $D$ of the vector $\bm{y}$ is equal to $22$. Thus the number of parameters to be estimated amounts to 231 correlation coefficients, 22 volatilities and 3 components of the market price of risk vector. Indeed we introduce only one additional component in the vector $\bm{\lambda}$ which accounts for the entire $FRA_{\text{3M}}$ curve
\be
\label{eq:riskpremia}
\bm{\lambda}=\left[\underbrace{\lambda_s,\lambda_s}_{K_s=2},\:\underbrace{\lambda_l,\ldots,\lambda_l}_{K-K_s=10},\:\underbrace{\lambda_{\text{3M}},\ldots,\lambda_{\text{3M}}}_{K_{\text{3M}}=10}\right]^\intercal.
\ee
It is worth to observe that as $\lambda_l$ refers to EONIA rates with time-to-maturity at least three months, consistently $\lambda_{\text{3M}}$ is associated to FRA rates with time-to-maturity longer than three months.

\subsubsection{Principal Components Analysis}
The optimization we carry out generates an estimate of the volatility matrix $\hat{\Sigma}$, as stated by Equation~(\ref{eq:factorvol}). One can thus build the $D\times D$ covariance matrix of $\bm{y}$~\footnote{$\text{Cov}\left[\bm{y}\right]=\hat{C}\,\Delta t$.}
\be
\hat{C}=\hat{\Sigma}\,\hat{\Sigma}^\intercal,
\ee
and perform the Principal Component Analysis (PCA). PCA is a well-known dimensional reduction technique which allows to identify the linear combination of the vector components which carries the largest fraction of total volatility. Since $\hat{C}$ is symmetric and semi-positive definite, it can be diagonalised with an orthogonal matrix $\hat{O}$, so that we obtain
\bes
\hat{C}=\hat{O}\,\text{diag}\left[\bm{\gamma}\right]\,\hat{O}^\intercal,
\ees
where $\bm{\gamma}=\left[\gamma_1,\ldots,\gamma_D\right]^\intercal$ contains the non-negative eigenvalues of $\hat{C}$ and the columns of $\hat{O}$ are its eigenvectors. The PCA suggests that if we keep only the $F$ largest eigenvalues, neglecting the smaller $K-F$ ones, we preserve a fraction of the total variance $\phi(F)$ equal to
\bd
\phi(F)=\frac{\sum_{\substack{i=1}}^F\gamma_i}{\sum_{\substack{i=1}}^D\gamma_i}.
\ed
In order to fix the number of principal components to be retained in the analysis, we chose a threshold value for this quantity, e.g. $\phi(F)\geq 95\%$.
We can thus define a sort of {\it modified volatility functions} as
\be
\label{eq:modvol}
\bm{w}_{m}=\sqrt{\gamma_m}\,\left[\hat{O}_{1m},\ldots, \hat{O}_{Dm}\right]^\intercal,\qquad m=1,\ldots,F,
\ee
which basically are the first $F$ eigenvectors of $\hat{C}$ rescaled by the relative eigenvalues.
\subsubsection{Single curve}

\begin{figure}[h]
  \centering
  \includegraphics[width=0.9\textwidth]{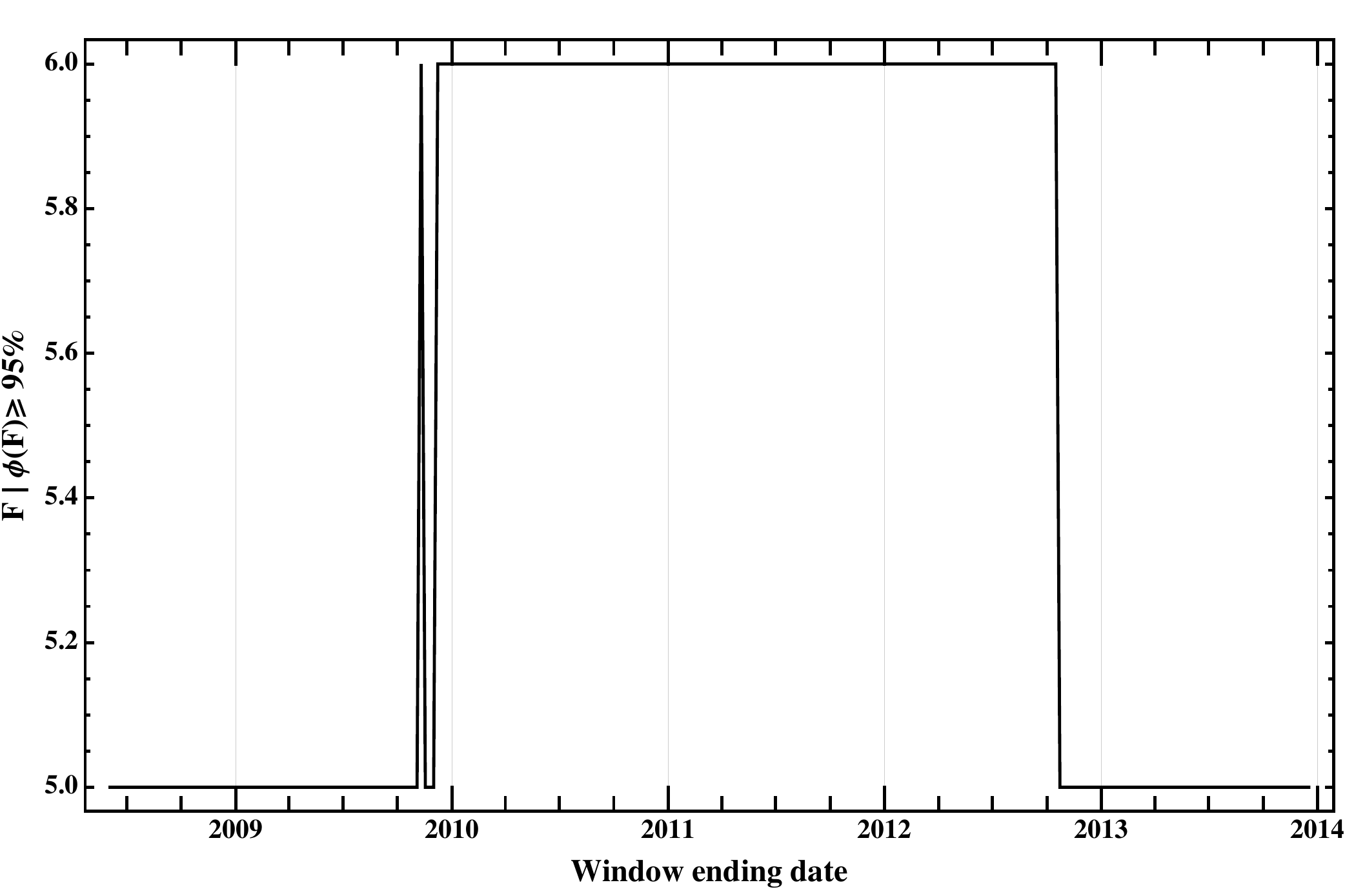}
  \caption{Minimum number of principal components of the EONIA covariance matrix $\hat{C}$ to be retained in order to preserve a fraction of the original variance larger than $95\%$. The results are obtained considering $K=12$ buckets for the EONIA curve. Each point refers to three years of data with weekly sampling frequency. The date reported on the $x$ axis corresponds to the ending date of the three year period, which is then moved five days ahead to obtain the consecutive point.}
  \label{fig:minnumofpc3ywlambda995}
\end{figure}

As a first step in the estimation procedure, we perform a rolling analysis on the time series at our disposal. We start from the first day of the series and consider three years of weekly spaced realizations of the vector $\bm{y}_f$. As described in Section~\ref{sec:estimation}, we perform the iterative optimization and obtain the maximum likelihood estimation of the model parameters. Then, we move one week ahead and repeat the same procedure. Finally, we obtain a sample of $290$ historical covariance matrices on which we perform the PCA and compute the minimum number of principal components needed to reproduce at least 95\% of the historical variance.
The plot in Figure~\ref{fig:minnumofpc3ywlambda995} shows how this minimum number changes across time.
The date reported on the $x$ axis corresponds to the ending date of the three year period of data used to compute each point in the plot.
\begin{center}
  ~\\
  FIGURE~\ref{fig:minnumofpc3ywlambda995} SHOULD BE HERE\\
  ~\\
\end{center}
As one can see from Figure~\ref{fig:minnumofpc3ywlambda995}, during the credit crisis the number of principal components needed to account for a consistent fraction of the historical volatility rises up from 5 to 6, and cools down to 5 afterwards.
This result is in contrast to what typically happens in the equity market. During periods characterised by declining markets, the number of principal components which describes the return covariance in the equity sector diminishes and indicates that the correlation among different assets has increased.
Here we observe the opposite trend: During the credit crisis the cross-correlation among rates sensitive to different tenors decreases. This effect might be due to the ECB interventions on the monetary policy, which strongly affects the short part of the curve during periods of economic downturn, whereas the long end of the term structure evolves almost unaffected by those interventions. As a consequence, this translates in a substantial lack of correlation among the two ends of the curve. Certainly, this empirical evidence denotes the segmentation of the term structure, and might indicate that market operators look at the different components of the curve as distinct investment opportunities.

\begin{figure}[h]
  \centering
  \includegraphics[width=0.9\textwidth]{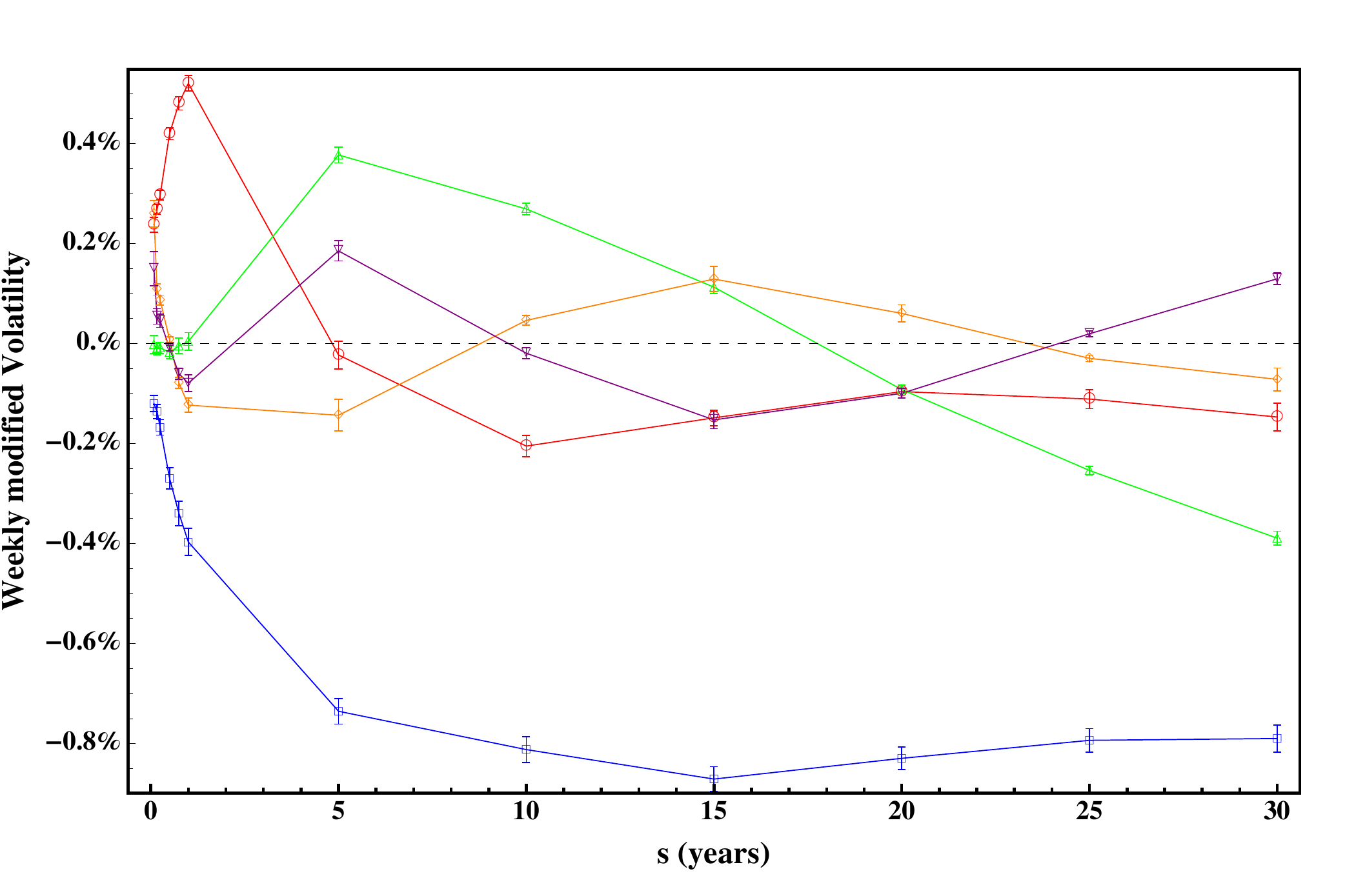}
  \caption{First five modified modified volatility functions for the EONIA term structure in a single curve framework. On the $x$ axis we report the times-to-maturity on a yearly basis. The points correspond to $K=12$ buckets. Blue$\to\bm{w}^{(f)}_1$, Red$\to\bm{w}^{(f)}_2$, Green$\to\bm{w}^{(f)}_3$, Orange$\to\bm{w}^{(f)}_4$, Purple$\to\bm{w}^{(f)}_5$. Data refer to the three year period January 5 2010 - January 5 2013 and are sampled with weekly frequency.}
  \label{fig:eoniamodvolwema}
\end{figure}

We fix now a specific estimation window, which starts on January 5 2010 and ends on the same date in 2013. Keeping $F=5$ principal components, we are able to preserve $96.74\%$ of the total historical volatility. In Figure~\ref{fig:eoniamodvolwema} we plot the five {\it modified volatility functions} $\bm{w}^{(f)}_1,\ldots,\bm{w}^{(f)}_5$, with the associated standard errors \footnote{For the computation of the statistical errors affecting the principal components we refer to~\cite{anderson1963,flury1997}.}. 
%The x axis reports the times-to-maturity on a yearly basis and the points correspond to the finite grid of $K$ buckets. 
The modified volatility functions represent the columns of the matrix $\hat{O}$ which diagonalize the covariance matrix, scaled by their eigenvalue, see Equation~(\ref{eq:modvol}).
\begin{center}
  ~\\
  FIGURE~\ref{fig:eoniamodvolwema} SHOULD BE HERE\\
  ~\\
\end{center}
The first factor is characterised by a flat structure over the long part of the curve. Then it declines to zero for small times-to-maturity, but does not change its sign. The second factor switches sign around the five year bucket, thus accounts for the difference among the long and short components of the curve. The humped shape of the third factor accounts for the convexity of the curve. Thus, the first three components have been usually associated with the level, slope, and convexity of the term structure. These evidences trace back to~\cite{litterman_scheinkman1991}. Since then, the PCA has been largely used in interest rate applications, e.g. in~\cite{jamshidian_zhu1996,rebonato1998,diebold_li2006}, whereas, from the modelling side, the link with the approach described in~\cite{nelson1987} have spurred a stream of research about factor models, see again~\cite{diebold_li2006} or for recent achievements~\cite{bernadell2005,Coroneo2011393} and references therein. However, Figure~\ref{fig:eoniamodvolwema} shows that these days a sufficiently large level of the total variance can be captured only including higher order components and a clear interpretation of such components is lacking. As we will see in the next section where we perform the PCA directly on the multiple yield term structures, the number and shape of the principal components is similar to the single curve case.

\subsubsection{Multiple yield curves}
\begin{figure}
\centering
\includegraphics[width=0.9\textwidth]{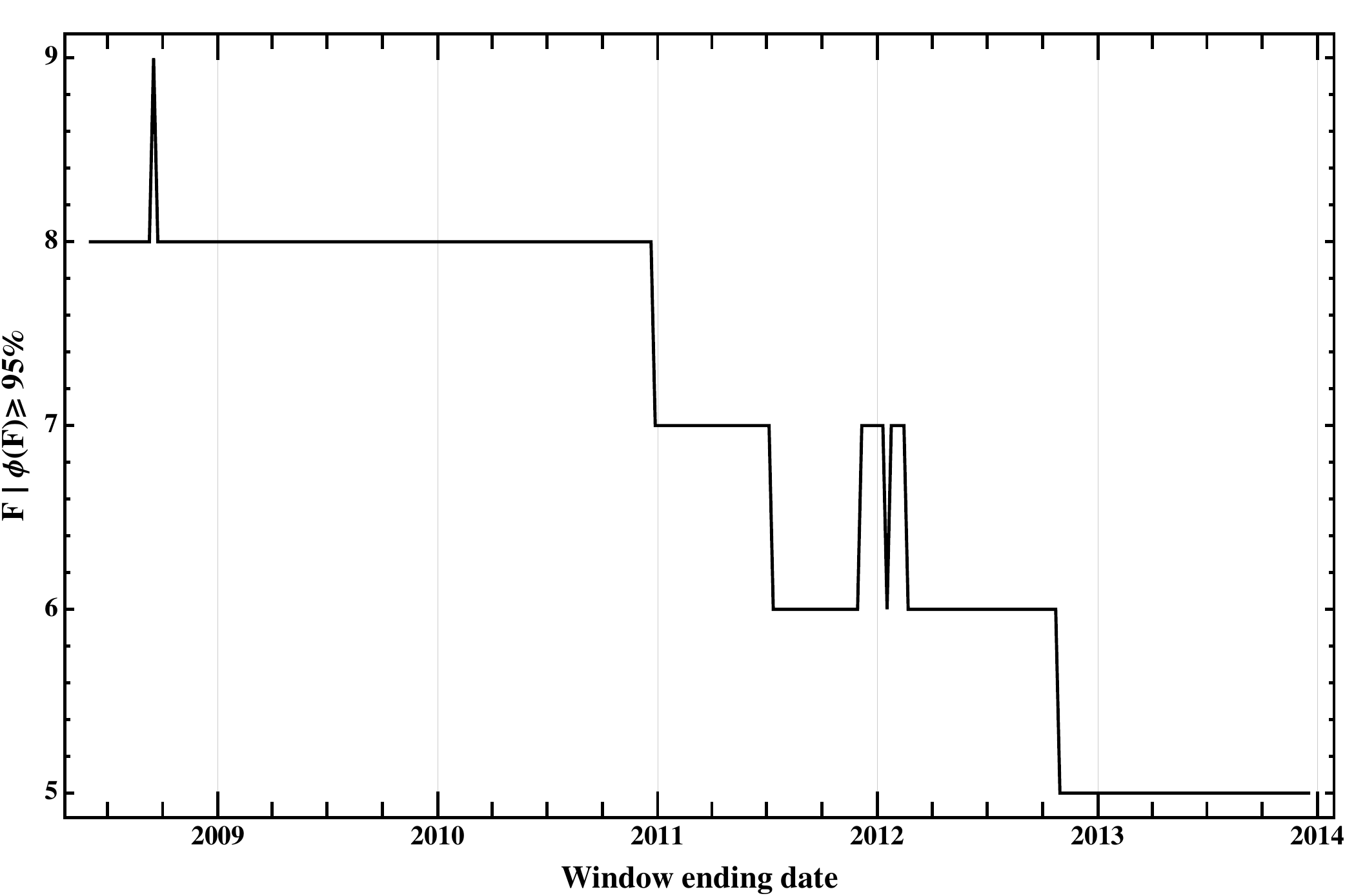}
\caption{Minimum number of principal components of the covariance matrix $\hat{C}$ to be retained in order to preserve a fraction of the original variance larger than $95\%$. The results are obtained considering $K=12$ buckets for the EONIA curve and $K_{\text{3M}}=10$ buckets for the EUR3M term structure, making a total of $D=22$ components. Each point refers to three years of data with weekly sampling frequency. The date reported on the $x$ axis corresponds to the ending date of the three year period, which is then moved five days ahead to obtain the consecutive point.}
\label{fig:multiminnumofpc3ywlambda995}
\end{figure}

In order to fix the number of principal components in the multiple yield curve case, we perform the same analysis described for the single curve framework. In Figure~\ref{fig:multiminnumofpc3ywlambda995} we report the minimum number of principal components needed to capture at least the $95\%$ of the total variance present in the historical data for the four curves, i.e. the EONIA curve $f(t,x)$ and the additional FRA curve $FRA_\text{3M}(t,x)$.
\begin{center}
  ~\\
  FIGURE~\ref{fig:multiminnumofpc3ywlambda995} SHOULD BE HERE\\
  ~\\
\end{center}
During the credit crisis the minimum number of principal components is stable around 8, then it gradually declines starting from 2011 and at the beginning of 2013 it cools down at 5. 

\begin{figure}
  \begin{minipage}[t]{\textwidth}
    \centering
    \includegraphics[width=0.75\textwidth]{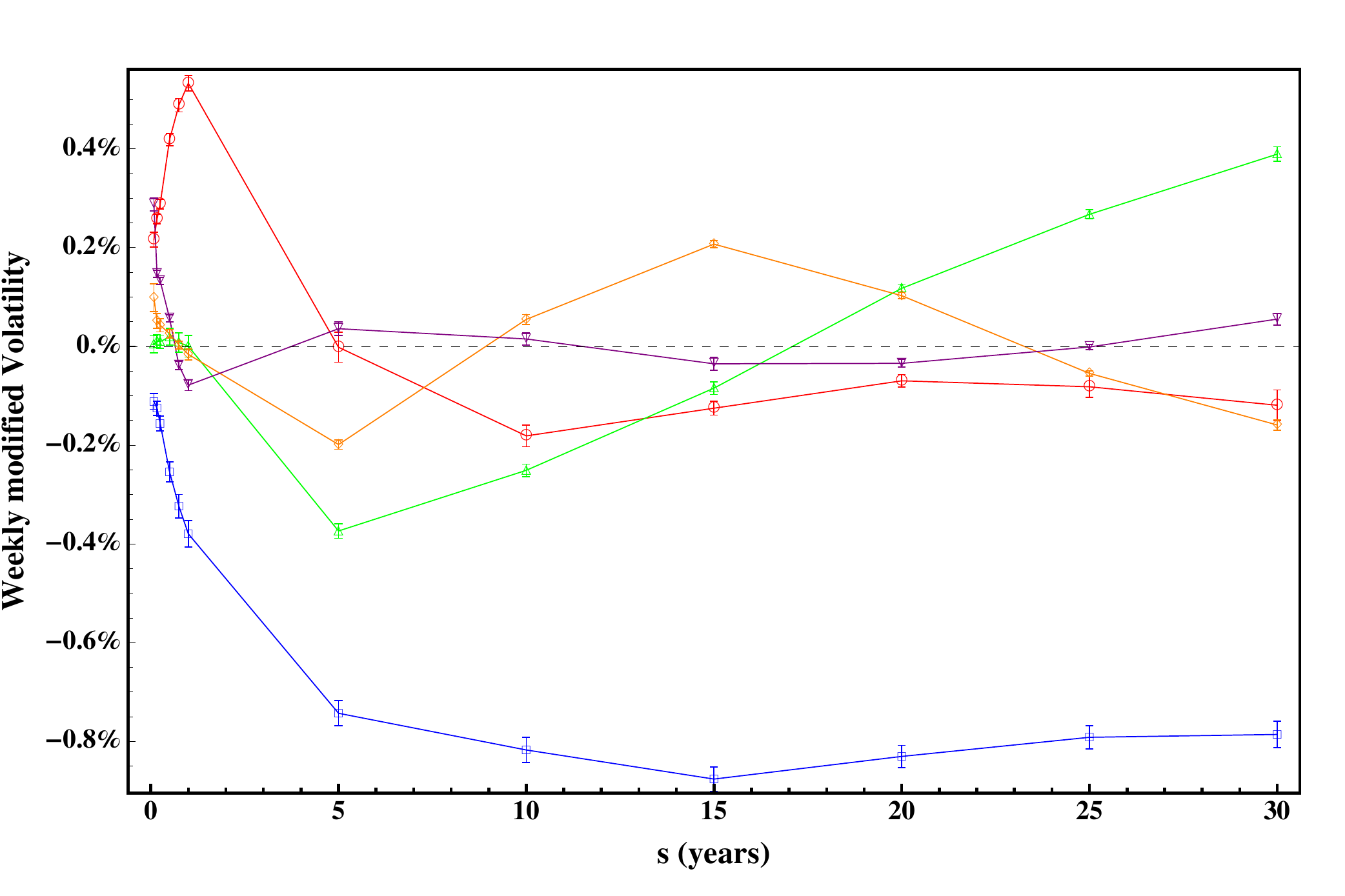}
  \end{minipage}
  \hspace{1truecm} % To get a little bit of space between the figures
  \begin{minipage}[t]{\textwidth}
    \centering \includegraphics[width=0.75\textwidth]{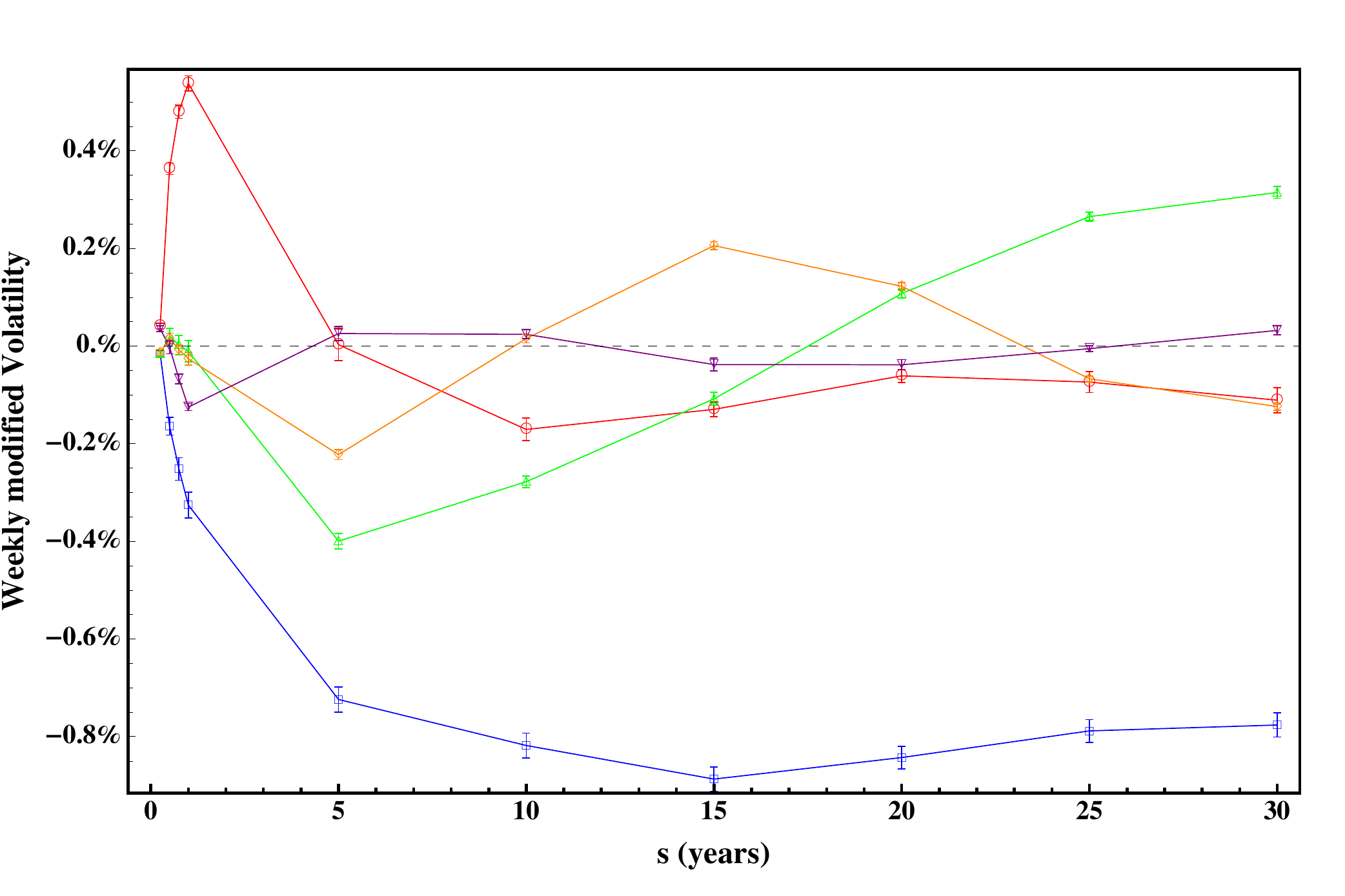}
  \end{minipage}
  \caption{First five modified volatility functions for the EONIA ($K=12$) and the EUR3M ($K_{\text{3M}}=10$) curves in a multiple curve framework. On the $x$ axis we report the times-to-maturity on a yearly basis. Left panel: $\bm{w}^{(f)}_{m}$ with $m=1,\ldots,5$. Righ panel: $\bm{w}^{(\text{3M})}_m$ with $m=1,\ldots,5$. Colours as in Figure~\ref{fig:eoniamodvolwema}.  Data refer to the three year period January 5 2010 - January 5 2013 and are sampled with weekly frequency.}
  \label{fig:modvolwema}
\end{figure}

We then fix the usual window ranging from January 5 2010 to January 5 2013, and compute the associated modified volatility functions.  
In the two panels of Figure~\ref{fig:modvolwema} we show the five modified volatility functions needed to account for 95.67\% of the total historical variance. For a better visualisation of the functions, we split each principal component in two parts, the first referring to the EONIA curve and the second one to the EUR3M curve
%\be
%\begin{split}
%&\bm{w}^{(f)}_m=\left[V_{1m},\ldots,V_{K\,m}\right]^\intercal\\
%&\bm{w}^{(\text{3M})}_m=\left[V_{K+1\,m},\ldots,V_{K+K_{\text{3M}}\,m}\right]^\intercal\\
%&\bm{w}^{(\text{6M})}_m=\left[V_{K+K_{\text{3M}}+1\,m},\ldots,V_{K+K_{\text{3M}}+K_{\text{6M}}\,m}\right]^\intercal\\
%&\bm{w}^{(\text{1Y})}_m=\left[V_{K+K_{\text{3M}}+K_{\text{6M}}+1\,m},\ldots,V_{K+K_{\text{3M}}+K_{\text{6M}}+K_{\text{1Y}}\,m}\right]^\intercal
%\end{split}
%\ee
\bes
\begin{split}
  &\bm{w}^{(f)}_m=\sqrt{\gamma}_m\,\left[O_{1m},\ldots,O_{Km}\right]^\intercal\,,\\ &\bm{w}^{(\text{3M})}_m=\sqrt{\gamma}_m\,\left[O_{K+1\,m},\ldots,O_{K+K_{\text{3M}}\,m}\right]^\intercal\,,\\
\end{split}
\ees
\begin{center}
  ~\\
  FIGURE~\ref{fig:modvolwema} SHOULD BE HERE\\
  ~\\
\end{center}
Both the EONIA and EUR3M curves exhibit the same behaviour in terms of factors, confirming that the first three eigenvectors could properly be interpreted as the level, the slope and the curvature. Thus, this evidence may support the extension of factor model approach to the description of multiple yield curves.

\begin{figure}[h]
  \begin{minipage}[t]{1\textwidth}
    \centering
    \includegraphics[width=0.9\textwidth]{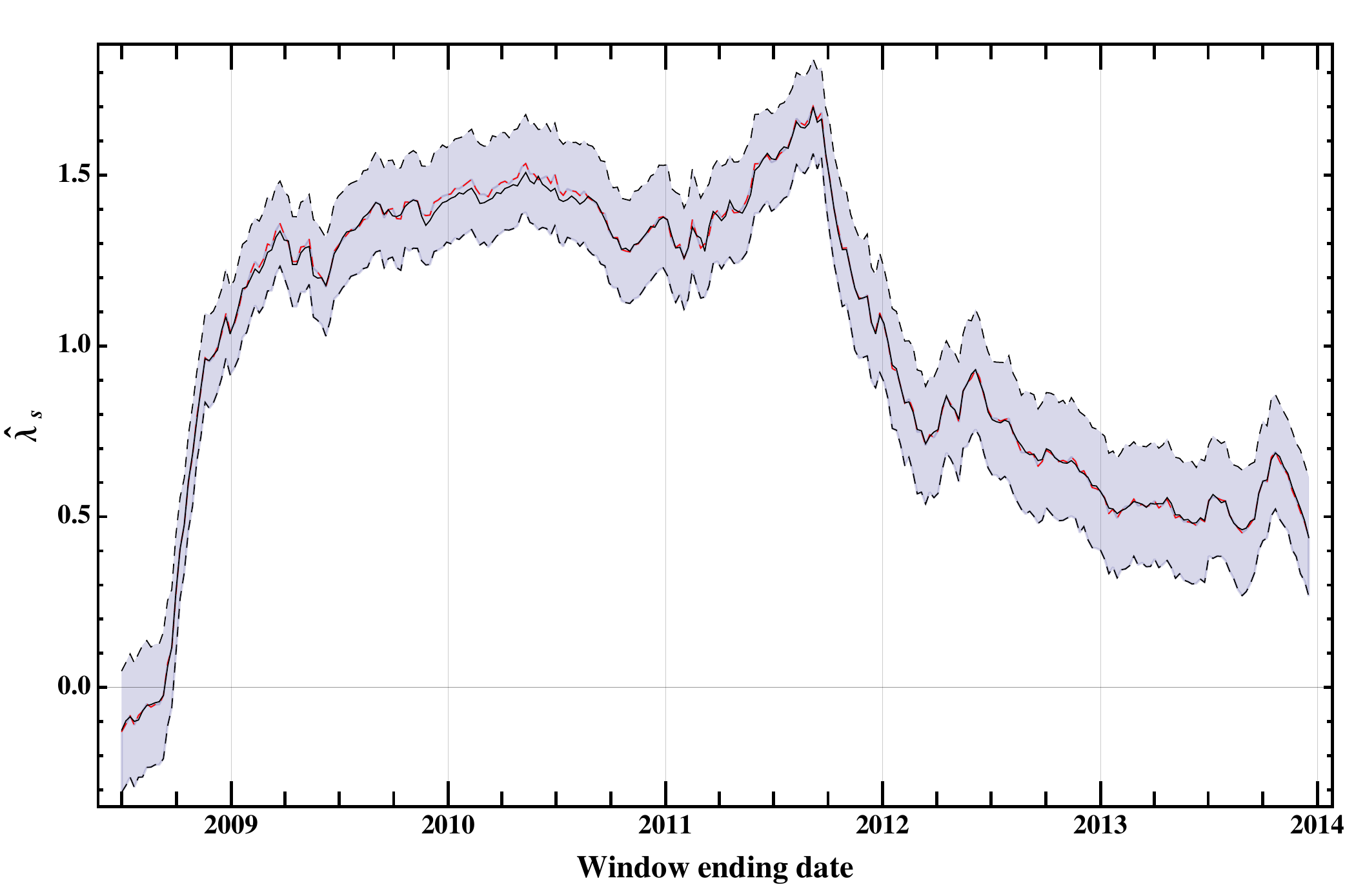}
    \caption{Black line: Value of $\lambda_s$ obtained estimating the model on three year samples of weekly spaced overlapping returns. Red line: Central value of the bootstrap estimator. Black dashed lines: Black line $\pm$ one standard deviation computed with bootstrap.}
    \label{fig:lambdashort}
  \end{minipage}
  \begin{minipage}[t]{1\textwidth}
    \centering
    \includegraphics[width=0.9\textwidth]{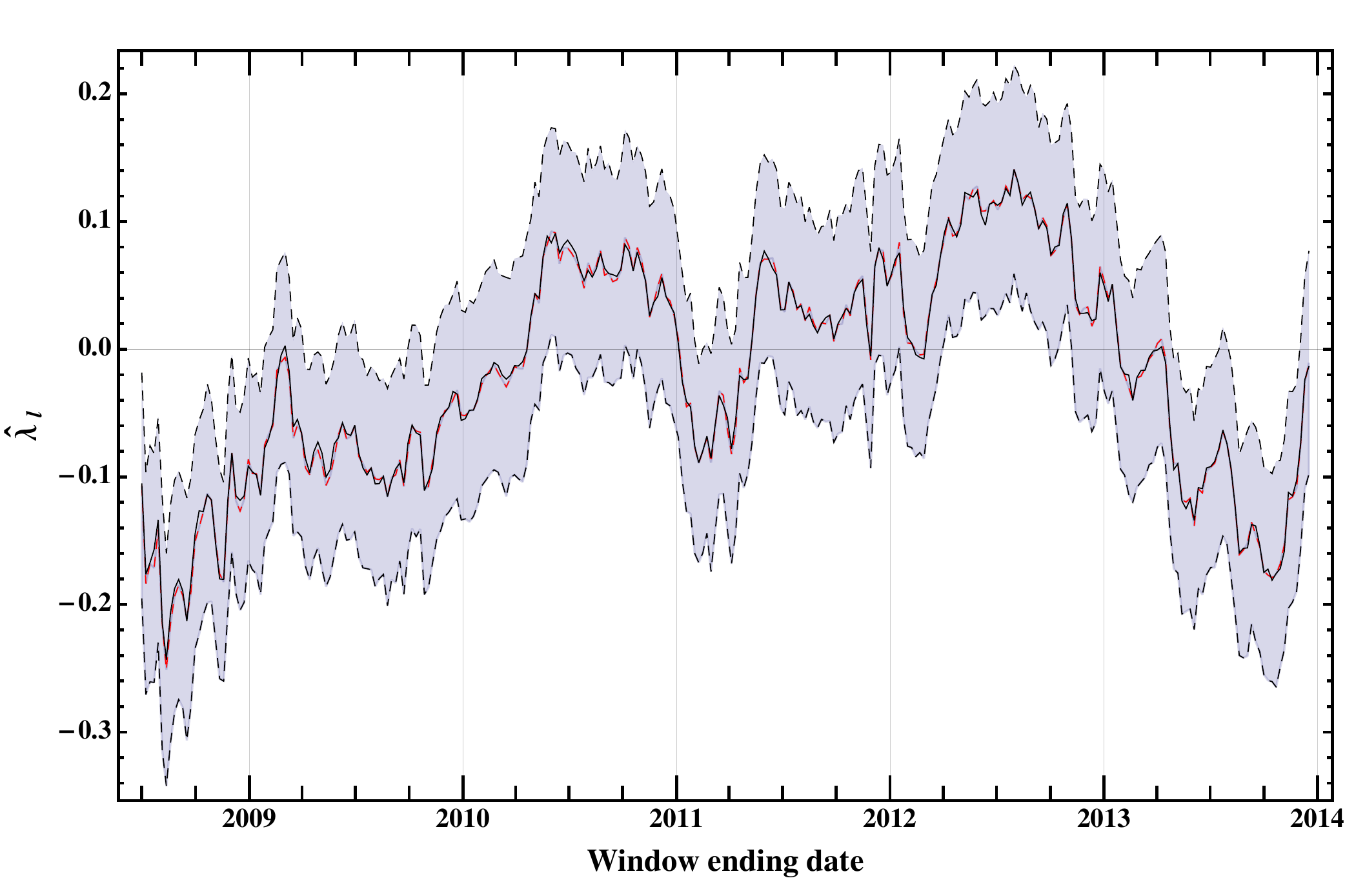}
    \caption{Black line: Value of $\lambda_l$ obtained estimating the model on three year samples of weekly spaced overlapping returns. Red and black dashed lines as in caption of Figure~\ref{fig:lambdashort}.}
    \label{fig:lambdalong}
  \end{minipage}
\end{figure}

\begin{figure}[h]
  \centering
  \includegraphics[width=0.9\textwidth]{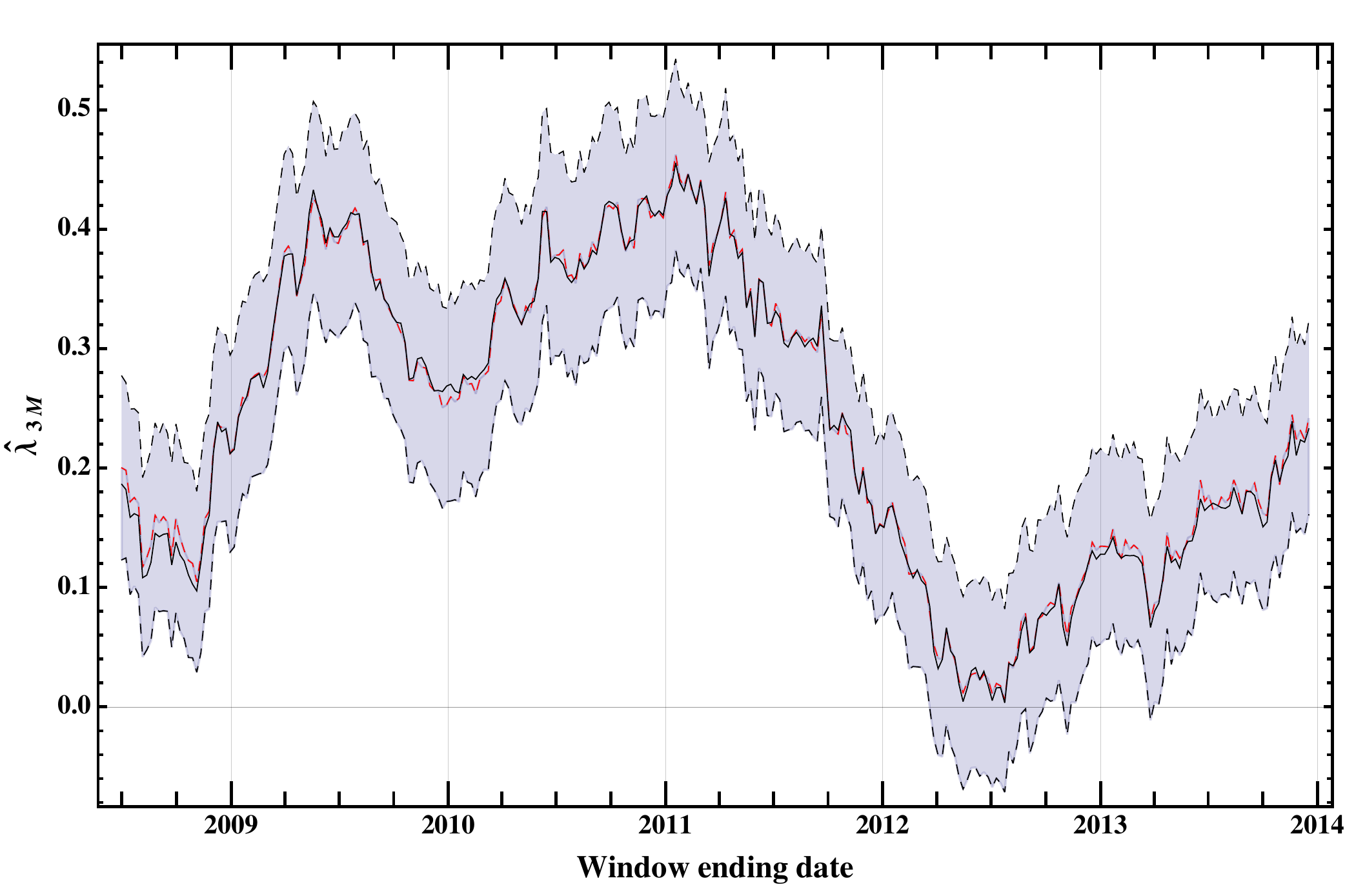}
  \caption{Black line: Value of $\lambda_{\text{3m}}$ obtained estimating the model on three year samples of weekly spaced overlapping returns. Red and black dashed lines as in caption of Figure~\ref{fig:lambdashort}.}
  \label{fig:lambda3m}
\end{figure}

We now move to the results of the maximum likelihood procedure.
Let us start from the vector of risk-premia, $\bm{\lambda}$, which is parametrised in terms of three components $\lambda_s$, $\lambda_l$ and $\lambda_{\text{3M}}$, as stated in Equation~(\ref{eq:riskpremia}).
In Figures~\ref{fig:lambdashort}, \ref{fig:lambdalong}, and~\ref{fig:lambda3m} we report the results of the estimation performed on three year rolling samples with weekly overlapping returns. Let us focus on the first component, $\lambda_s$, the one associated to the one month and two month buckets of the EONIA instantaneous forward curve. It starts from almost zero and rises up to very high positive values as soon as the credit crisis period is included in the calibration sample. Equation~(\ref{eq:driftpar}) allows a very neat interpretation of this result. The $-\bm{\omega}\circ R\bm{\lambda}$ term in the drift component gives a negative contribution for positive values of $R\bm{\lambda}$ and vice-versa for negative values. Let us consider the three years sample starting in september 2008, which gives rise to highest value $\lambda_s \sim 1.5$. In the period 2008-2009 the EONIA term structure changed sensibly its shape, going from an inverted curve towards a steep upward sloping one, with short rates around 0.5\% and long-term rates at nearly 4\%. If at $t^\star$, e.g. a day in September 2008, we were asked to forecast the one month interest rate observed in one year, our best guess would have been the forward rate implied by the yield curve at $t^\star$ valid between $t^\star+1y$ and $t^\star+1y+1m$. In September 2008 the curve showed an inverted shape, with the ten year rate around $4.2\%$, the one month rate at $4.3\%$, and the one month rate implied in one year around 4.9\% whereas in the aftermath of the crisis the shortest rates cooled down and reached the level of $0.35\%$. Thus our forecast would have needed a substantial downward correction, which is provided by the $-\omega_1\,\lambda_s$ with $\lambda_s >0$. The same argument holds true for the period at the onset of the crisis, namely the years 2006-2007, during which the term structure transformed its shape, going from a normal upward sloping curve towards a slightly inverted one. Also in this case our {\it naive} forecast based on the forward rate would need a downward correction, provided again by positive values of $\lambda_s$. However, since in this period the short rates exhibited a smaller variation, from 3.5\% to 4.2\%, the magnitude of this correction turns out to be smaller than the 2008-2009 one, and so it is our estimate of $\lambda_s$.

% In the period 2007-2008 the EONIA term structure changed sensibly its shape, going from a normal upward sloping curve towards an inverted curve in the August of 2007. If at $t^\star$, {\it e.g.} a day in September 2006, we were asked to forecast the one month interest rate observed in one year, our best guess would have been the rate observed at $t^\star$ with time-to-maturity one year plus one month. In September 2006 the curve showed a normal upward sloping shape, with the ten year rate around $3.6\%$ and the one month rate at $2.7\%$. As the crisis evolved the shortest rates reached the level of $4.1\%$. Thus our forecast would have needed a substantial downward correction, which is provided by the $-\bm{\omega}\circ R\bm{\lambda}$ with $\lambda_s >0$. 

The misalignment between the forward and the realised rates explains the behaviour observed in Figure~\ref{fig:lambdashort} where $\lambda_s$ increases to large positive values. This effect is even more evident when including the sovereign crisis of the EURO area in 2010-2011.
As soon as the credit crisis and the sovereign crisis are over, the level of the market price of risk associated with the shortest maturities cooled down, going back to low values. On the other hand the risk premia associated to the EONIA rates expiring in more than three months can be neglected since not statistically significant, as one can see from Figure~\ref{fig:lambdalong}.

Finally Figure~\ref{fig:lambda3m} shows that the risk-premium associated to the EUR3M FRA curve has a similar behaviour to the shortest maturity premium of the EONIA curve. Again, this is due to the fact that this component of the risk premia vector must account for the change observed between 2009 and 2013 in the shape -- from normal to inverted and back -- of the EUR3M term structure.

\begin{figure}[h]
  \begin{minipage}[t]{1\textwidth}
    \centering
    \includegraphics[width=0.9\textwidth]{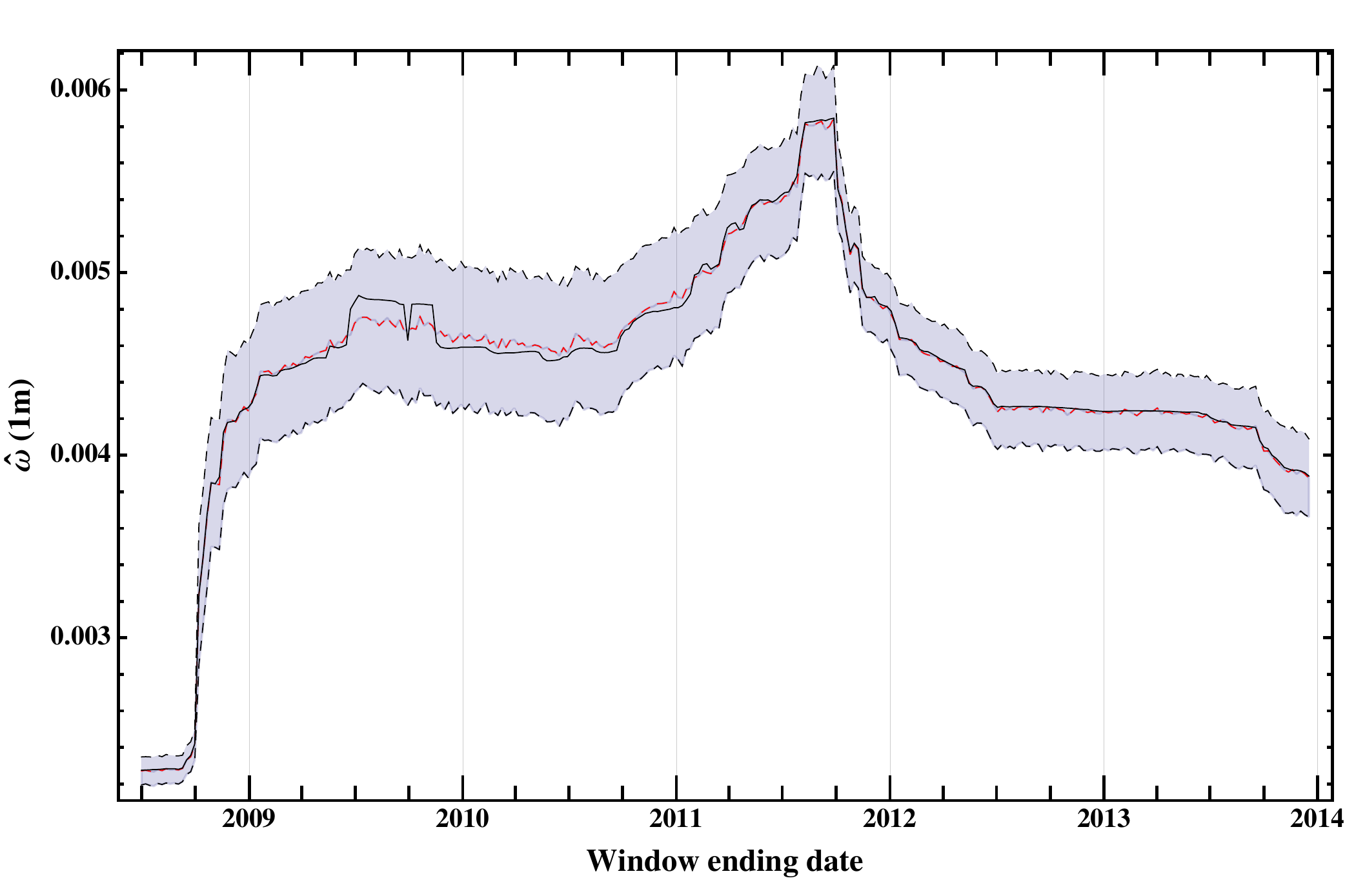}
    \caption{Black line: Value of $\omega(\text{1m})$ for the EONIA curve obtained estimating the model on three year samples of weekly spaced overlapping returns. Red and black dashed lines as in caption of Figure~\ref{fig:lambdashort}.}
    \label{fig:omegaeonia1m}
  \end{minipage}
  \begin{minipage}[t]{1\textwidth}
    \centering
    \includegraphics[width=0.9\textwidth]{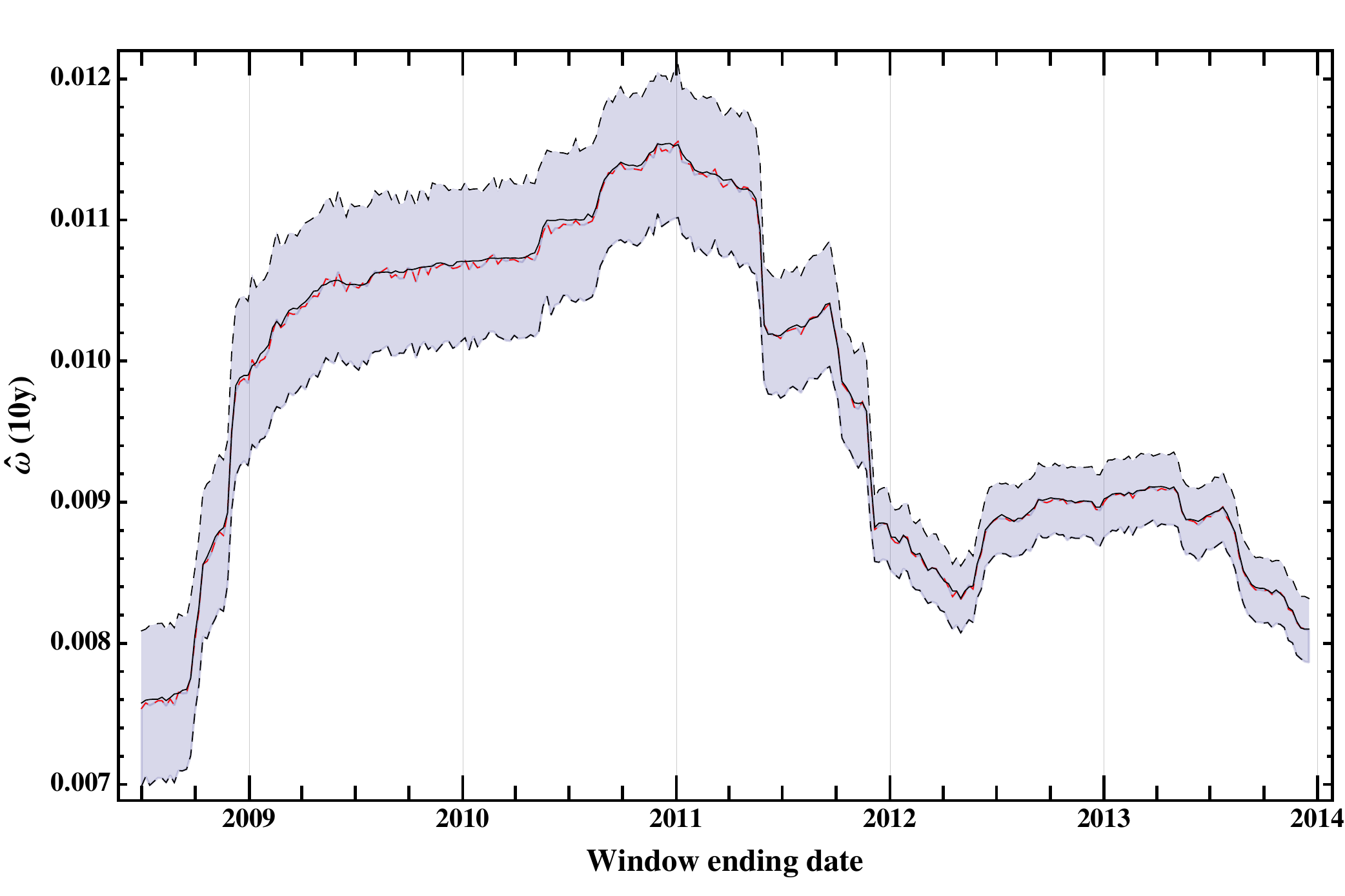}
    \caption{Black line: Value of $\omega(\text{10y})$ for the EONIA curve obtained estimating the model on three year samples of weekly spaced overlapping returns. Red and black dashed lines as in caption of Figure~\ref{fig:lambdashort}.}
    \label{fig:omegaeonia10y}
  \end{minipage}
\end{figure}

\begin{figure}[h]
  \begin{minipage}[t]{1\textwidth}
    \centering
    \includegraphics[width=0.9\textwidth]{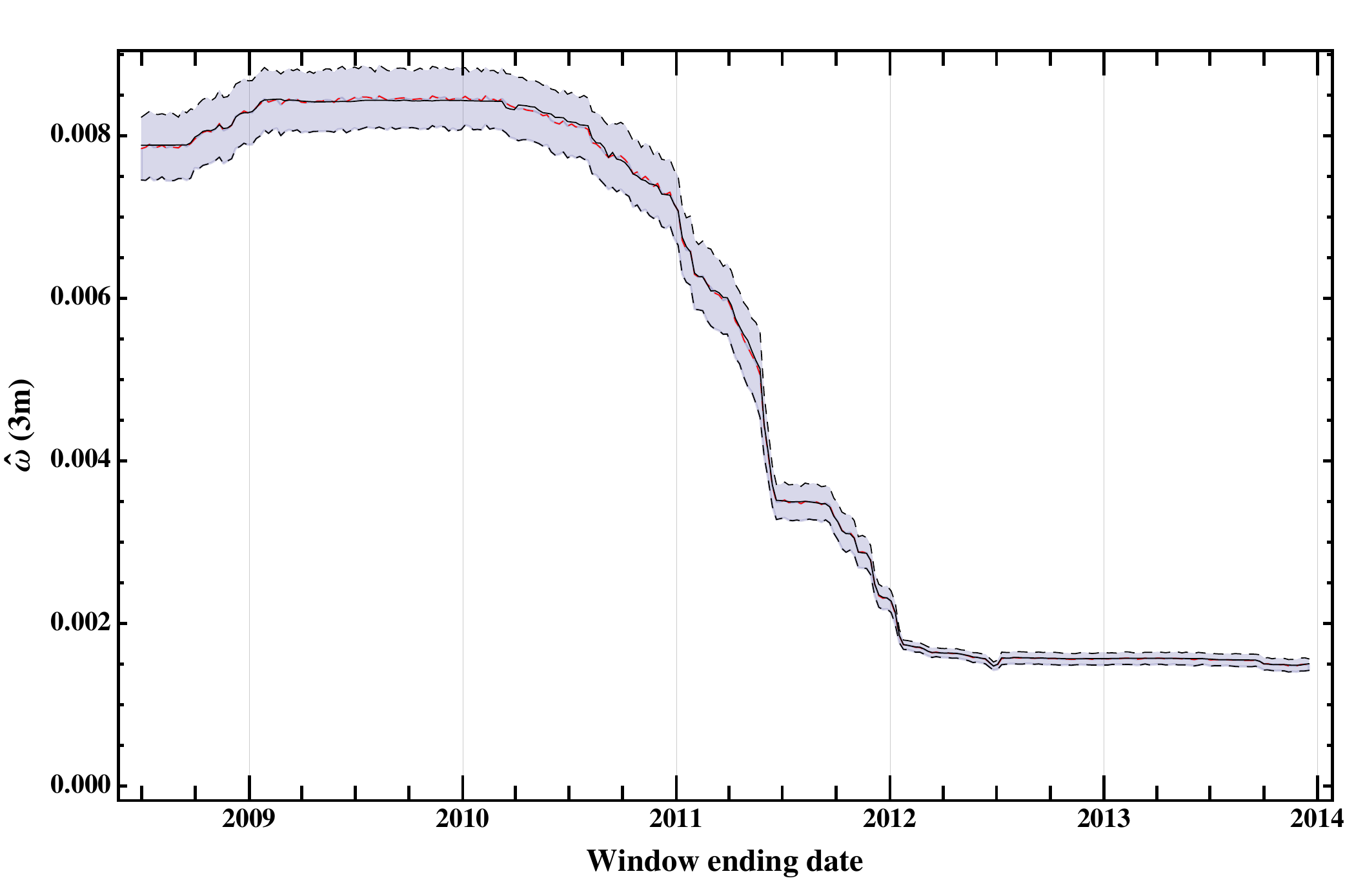}
    \caption{Black line: Value of $\omega(\text{3m})$ for the EUR3M curve obtained estimating the model on three year samples of weekly spaced overlapping returns. Red and black dashed lines as in caption of Figure~\ref{fig:lambdashort}.}
    \label{fig:omegaeur3m3m}
  \end{minipage}
  \begin{minipage}[t]{1\textwidth}
    \centering
    \includegraphics[width=0.9\textwidth]{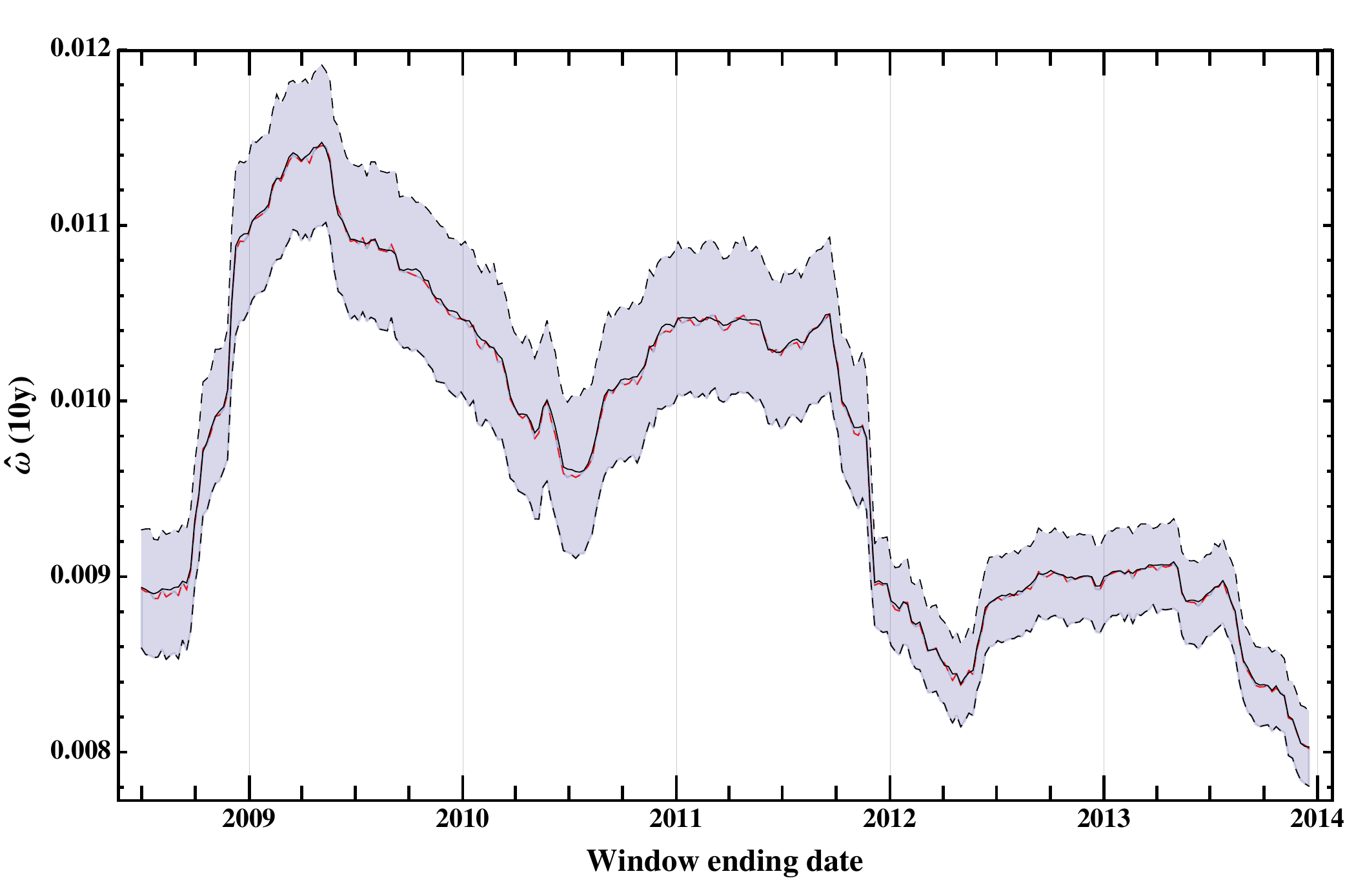}
    \caption{Black line: Value of $\omega(\text{10y})$ for the EUR3M curve obtained estimating the model on three year samples of weekly spaced overlapping returns. Red and black dashed lines as in caption of Figure~\ref{fig:lambdashort}.}
    \label{fig:omegaeur3m10y}
  \end{minipage}
\end{figure}

\begin{figure}[h]
  \begin{minipage}[t]{1\textwidth}
    \centering
    \includegraphics[width=0.9\textwidth]{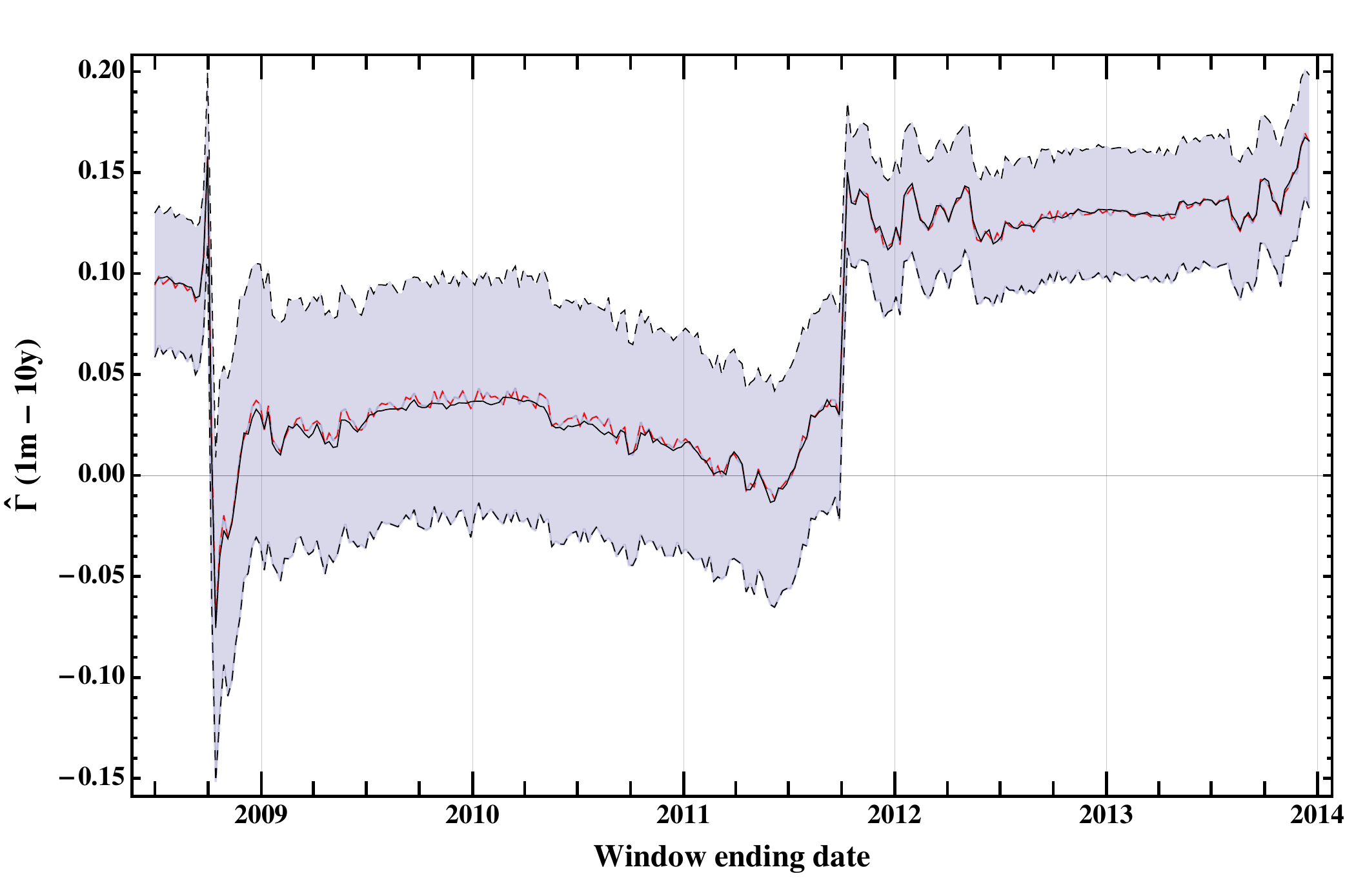}
    \caption{Black line: Value of $\Gamma(\text{1m}-\text{10y})$ for the EONIA curve obtained estimating the model on three year samples of weekly spaced overlapping returns. Red and black dashed lines as in caption of Figure~\ref{fig:lambdashort}.}
    \label{fig:rhoeonia1m10y}
  \end{minipage}
  \begin{minipage}[t]{1\textwidth}
    \centering
    \includegraphics[width=0.9\textwidth]{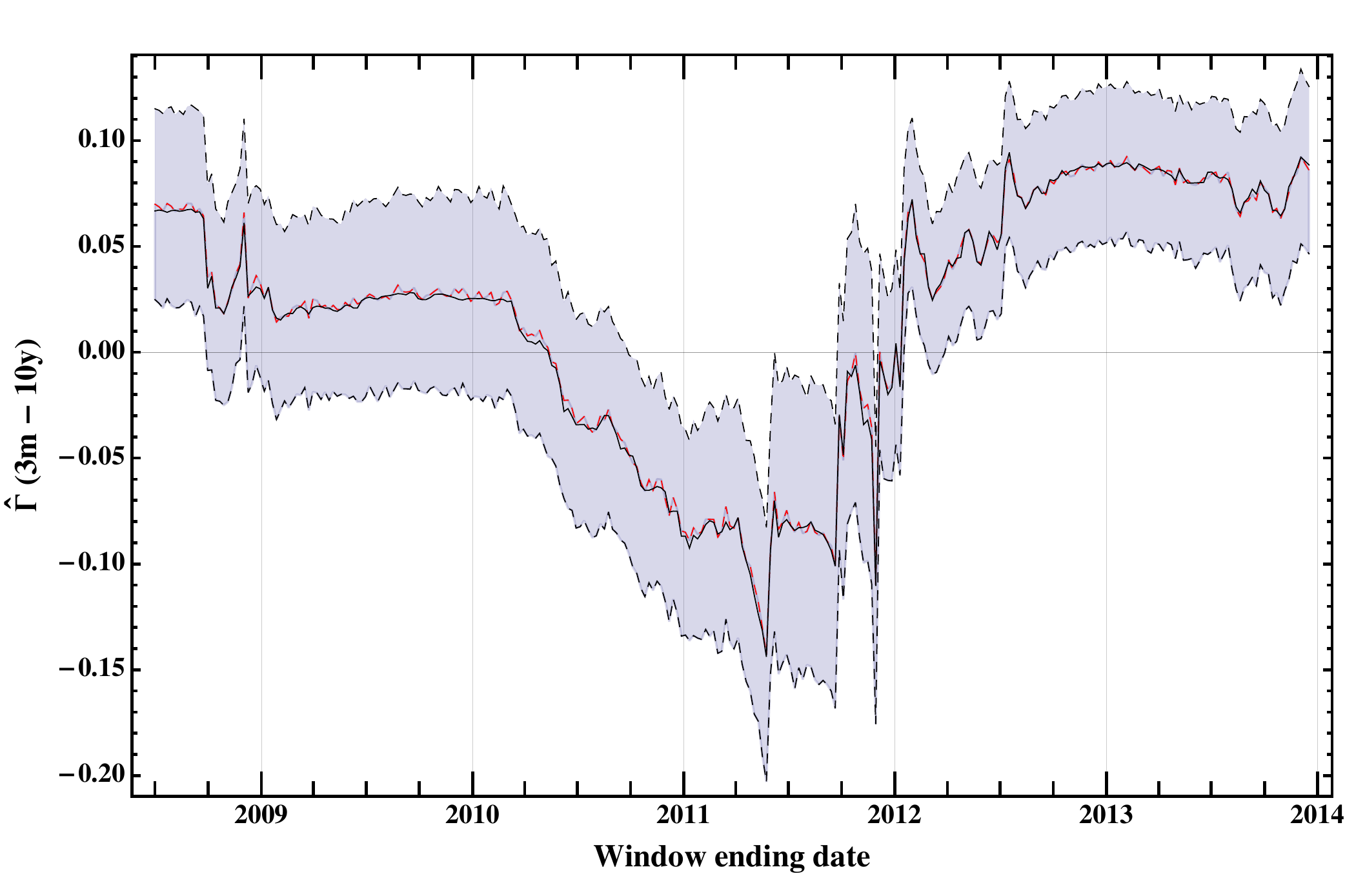}
    \caption{Black line: Value of $\Gamma(\text{3m}-\text{10 y})$ for the EUR3M curve obtained estimating the model on three year samples of weekly spaced overlapping returns. Red and black dashed lines as in caption of Figure~\ref{fig:lambdashort}.}
    \label{fig:rhoeur3m3m10y}
  \end{minipage}
\end{figure}

Beside the components of the market price of risk, estimation provides also the values for volatilities and correlation coefficients. We report the values of $\hat{\omega}_i$ and $\hat{\Gamma}_{ij}$ for a short and a long maturity rate of the EONIA and EUR3M term structure in Figures~\ref{fig:omegaeonia1m}, \ref{fig:omegaeonia10y}, and \ref{fig:rhoeonia1m10y} and Figures~\ref{fig:omegaeur3m3m}, \ref{fig:omegaeur3m10y}, and \ref{fig:rhoeur3m3m10y}, respectively.

\subsection{Forecasting the yield curves: Out-of-sample test}

We now investigate the predictive ability of our model performing an out-of-sample test. We compare the confidence intervals for the yield curves predicted by our model with the realised rates. In particular we put forward an overall frequency test, as described in~\cite{kupiec1995,christoffersen1998}.
For this analysis we consider a data set which is comprehensive of the EONIA yield curve and the EUR3M FRA term structure recorded daily for the period February 8 2005 - December 27 2013. Starting from the beginning of the time series, we estimate the parameters of our model on three year samples ranging from $t_k-3\text{ years}$ and $t_k$, with $t_k$ weekly spaced. For each of these samples we compute by Monte Carlo simulation confidence envelopes for a forecasting horizon of one week, three month and one year. For the EONIA curve, we denote the confidence intervals by
\bes
\left\{l_{t_k+\delta|t_k}(s_i;p),u_{t_k+\delta|t_k}(s_i;p)\right\},\quad k=1,\ldots,n_{\text{obs}}\,,\quad i=1,\ldots,K\,,\quad\text{and}\quad \delta=\text{1w},\text{3m},\text{1y}\,,
\ees
where $l_{t_k+\delta|t_k}(s_i;p)$ and $u_{t_k+\delta|t_k}(s_i;p)$ are the lower and upper bounds of the interval $\delta$-periods ahead forecast for the $s_i$ bucket computed at time $t_k$ for time $t_k+\delta$ with coverage probability $p$. For the longer tenor curve we compute confidence envelopes on the FRA term structure
\bes
\left\{l^{(\text{3M})}_{t_k+\delta|t_k}(s_{{\text{3M},i}};p),u^{(\text{3M})}_{t_k+\delta|t_k}(s_{{\text{3M},i}};p)\right\}\,,\quad k=1,\ldots,n_{\text{obs}}\,,\quad\text{and}\quad i=1,\ldots,K_{_\Delta}\,,\quad\text{and}\quad \delta=\text{1w},\text{3m},\text{1y}\,,
\ees
where $l^{(\text{3M})}_{t_k+\delta|t_k}(s_{{\text{3M},i}};p)$ and $u^{(\text{3M})}_{t_k+\delta|t_k}(s_{{\text{3M},i}};p)$ represent the lower and upper limits of the $\delta$-periods ahead interval forecast for the bucket $s_{\text{3M},i}$ with coverage probability $p$.
In the following analyses we consider $p=0.95$, $0.99$. We remind that $n_{\text{obs}}=290$.

For each time $t_k$ we compare the observed rates (instantaneous forward or FRA) with our forecasted interval and count the exceedances.
In other words, we introduce the following indicator variables
\bes
I^{(\delta)}_{k}(s_i;p)=
\begin{cases}
  1&\text{if } f(t_{k}+\delta,s_i)\notin \left\{l_{t_k+\delta|t_k}(s_i;p),u_{t_k+\delta|t_k}(s_i;p)\right\}\,,\\
  0&\text{otherwise}\,,
\end{cases}
\ees
and
\bes
I^{(\delta)}_{\text{3M},k}(s_{_{\text{3M},i}};p)=
\begin{cases}
  1&\text{if } \text{FRA}_{\text{3M}}(t_k+\delta,s_i)\notin \left\{l^{(\text{3M})}_{t_k+\delta|t_k}(s_{{\text{3M},i}};p),u^{(\text{3M})}_{t_k+\delta|t_k}(s_{{\text{3M},i}};p)\right\}\,,\\
  0&\text{otherwise}\,.
\end{cases}
\ees
Assuming independence among the $\left\{I^{(\delta)}_{k}(s_{i};p)\right\}_{k=1}^{n_\text{obs}}$, we want to test whether $\e\left[I^{(\delta)}_{k}(s_i;p)\right]=1-p$ against the alternative $\e\left[I^{\delta}_{k}(s_i;p)\right]\neq 1-p$, for each $i=1,\ldots,K$. We do the same for $\left\{I^{(\delta)}_{\text{3M},k}(s_{\text{3M} ,i};p)\right\}_{k=1}^{n_\text{obs}}$.
This type of test is often referred to as {\it unconditional coverage} test. The likelihood under the null hypothesis is given by
\bes
\begin{split}
&\mathcal{L}_{\text{UC}}(p;I^{(\delta)}_{1}(s_i;p),\ldots,I^{(\delta)}_{n_{\text{obs}}}(s_i;p))=(1-p)^{n_1}\,p^{n_0},\quad i=1,\ldots,K,\\
%&\mathcal{L}_{\text{UC}}(p;I_{\Delta,1}(s_{\Delta,i};\delta,p),\ldots,I_{\Delta,n_{\text{obs}}}(s_{\Delta,i};\delta,p))=(1-p)^{n_1}\,p^{n_0},\quad i=1,\ldots,K_\Delta,\quad \Delta=\text{3M},\text{6M},\text{1Y}\\
\end{split}
\ees
and analogously for the 3M-tenor FRA curve, where $n_0$ and $n_1$ are the number of occurrences of $0$ and $1$ in the sequence $\left\{I^{(\delta)}_{k}(s_i;p)\right\}_{k=1}^{n_{\text{obs}}}$, respectively. The likelihood under the alternative is instead
\bes
\mathcal{L}_{\text{UC}}(\pi;I^{(\delta)}_{1}(s_i;p),\ldots,I^{(\delta)}_{n_{\text{obs}}}(s_i;p))=(1-\pi)^{n_1}\,\pi^{n_0},
\ees
with $\pi=n_0/(n_0+n_1)$. 
This test can be formulated as a standard likelihood ratio test, where the log-likelihood ratio is asymptotically distributed as $\chi^2(1)$
\bes
\text{LR}_{\text{UC}}=-2\ln\left[\mathcal{L}_{\text{UC}}(p;I^{(\delta)}_{1}(s_i;p),\ldots,I^{(\delta)}_{n_{\text{obs}}}(s_i;p))/\mathcal{L}_{\text{UC}}(\pi;I^{(\delta)}_{1}(s_i;p),\ldots,I^{(\delta)}_{n_\text{obs}}(s_i;p))\right]\sim \chi^2(1).
\ees
We perform the same test on the EUR3M curve. 

In order to produce the confidence intervals required by the out-of-sample testing procedure, we use two methods: The Gaussian diffusive model outlined in Section~\ref{sec:gaussian} and the bootstrap methodology explained in \ref{sec:bootstrap}.
We then compare the outcomes of the test from these two settings.

\begin{figure}[H]
\centering
\begin{minipage}[t]{7.0truecm}
\centering
\includegraphics[width=7.0truecm]{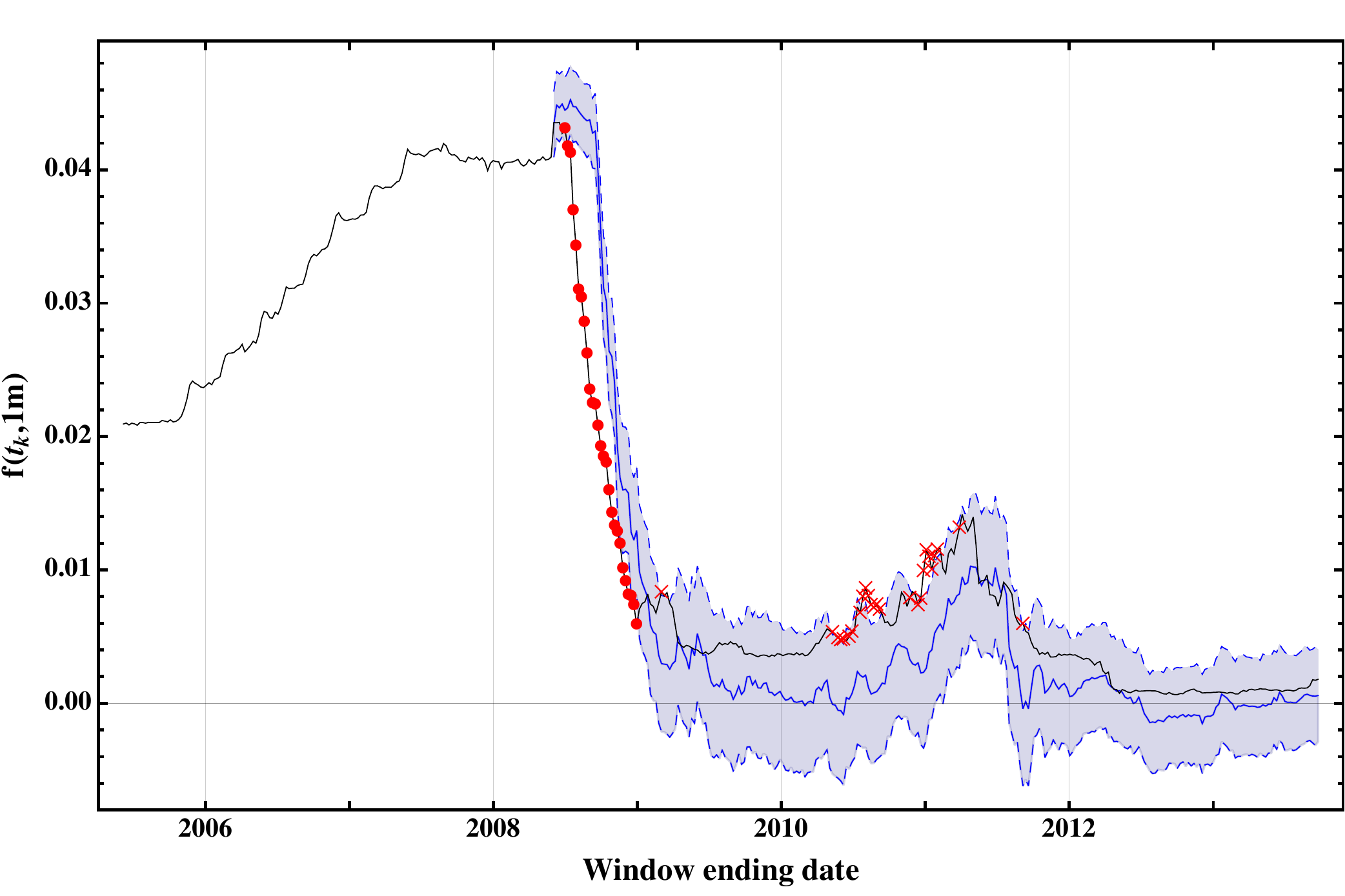}
\end{minipage}
\hspace{1truecm}
\begin{minipage}[t]{7.0truecm}
\centering
\includegraphics[width=7.0truecm]{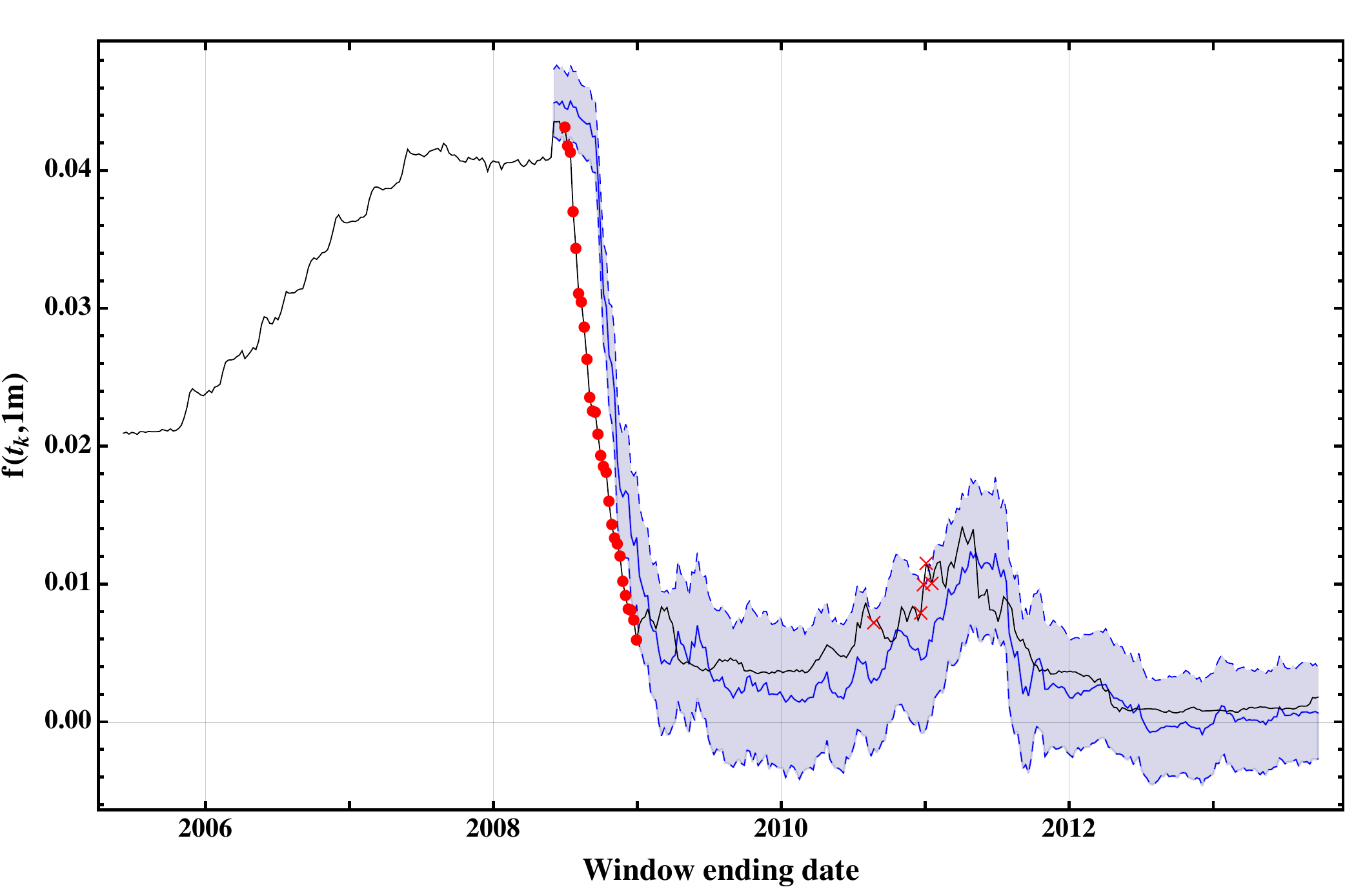}
\end{minipage}
\\
\vspace{1truecm}
\begin{minipage}[t]{7.0truecm}
\centering
\includegraphics[width=7.0truecm]{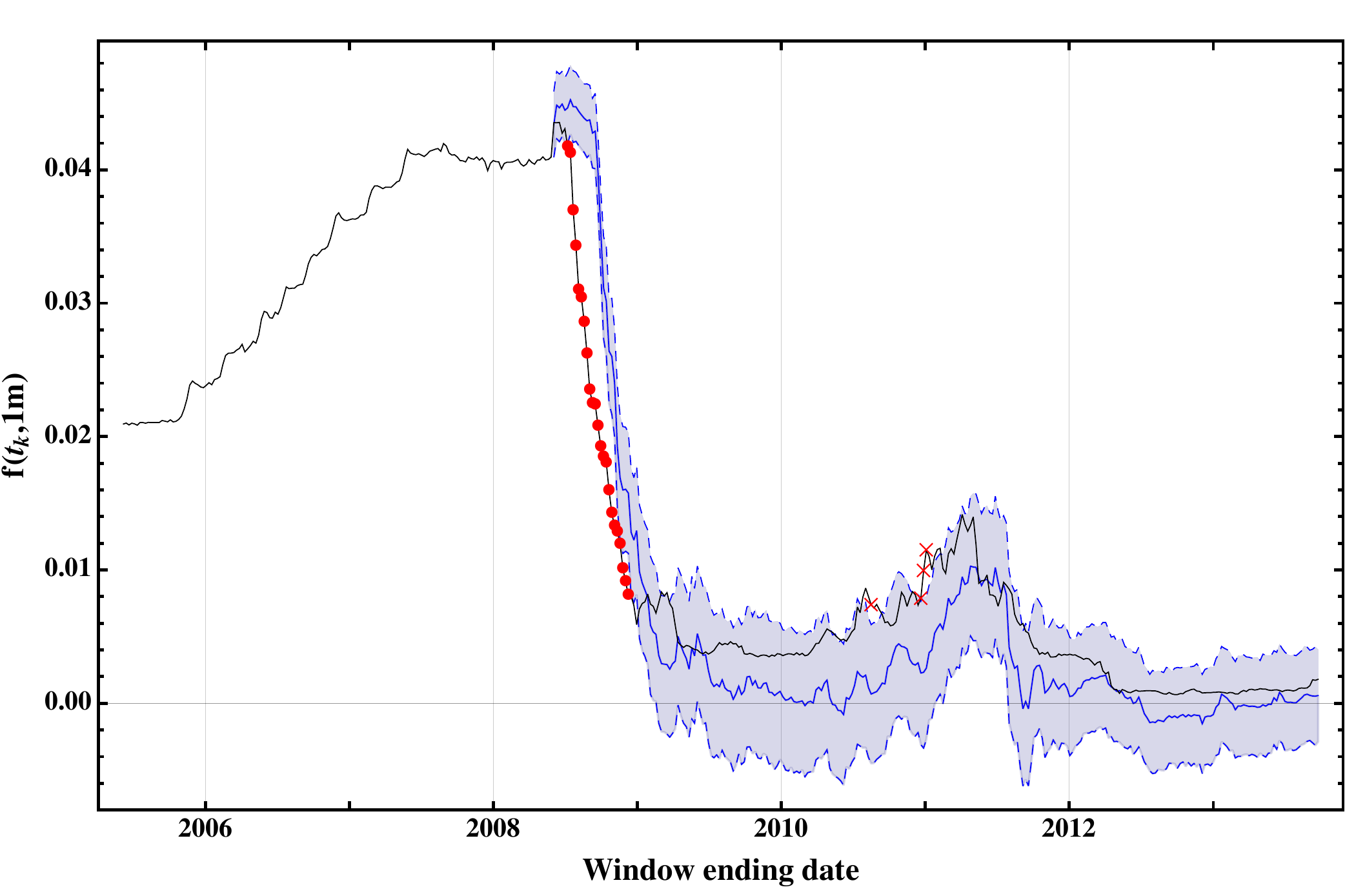}
\end{minipage}
\hspace{1truecm}
\begin{minipage}[t]{7.0truecm}
\centering
\includegraphics[width=7.0truecm]{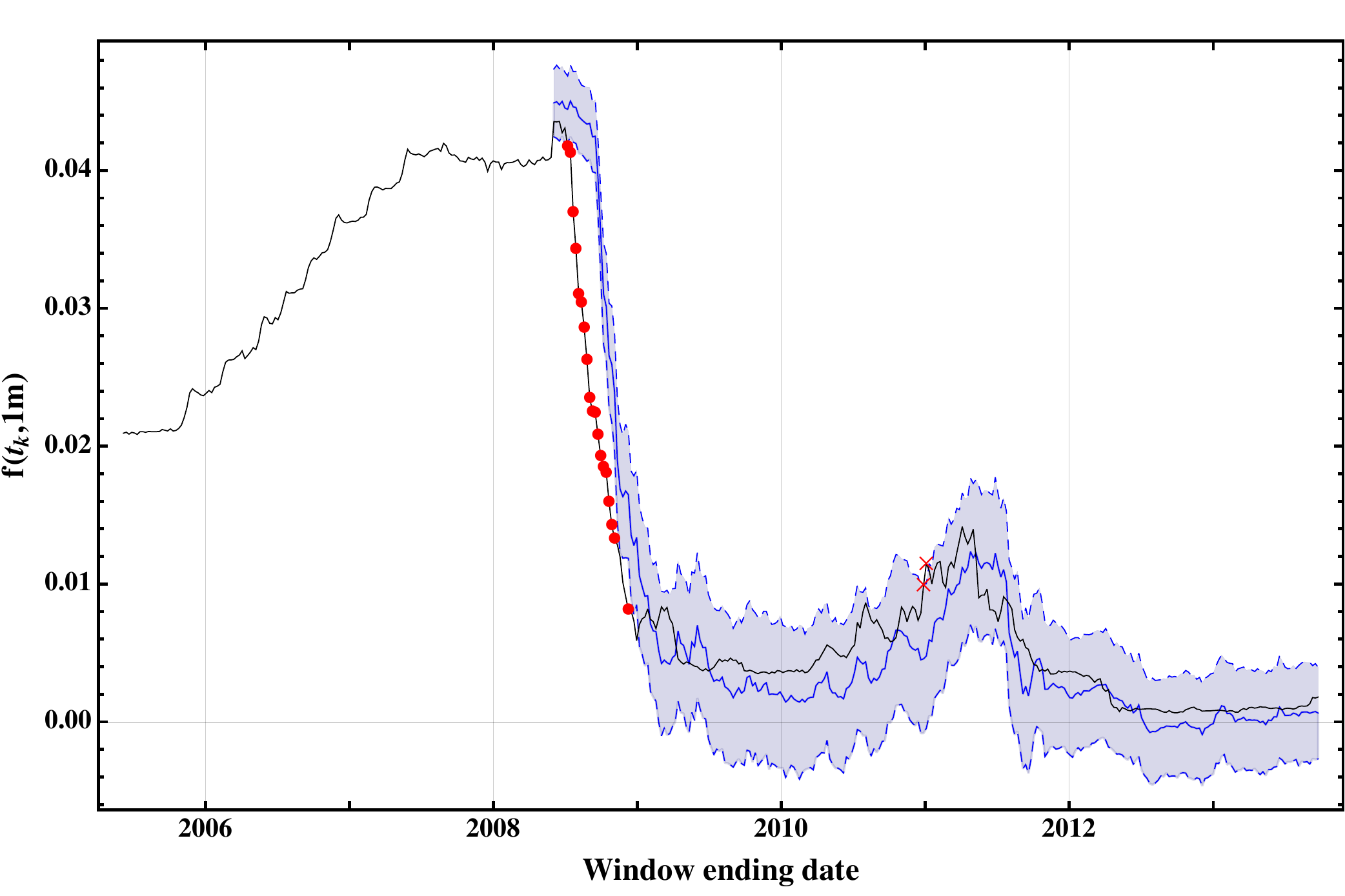}
\end{minipage}
\caption{Solid blue line: Mean of the three month ahead forecast distribution for the $f(t,1m)$ rate. Blue dashed lines: Mean $\pm$ two standard deviations of the three month ahead forecasted distribution for the $f(t,1m)$ rate. Black solid line: Realised $f(t,1m)$ rate. Red dots (crosses): Negative (positive) exceptions with coverage probability $95\%$ and $99\%$ on the first and second row respectively. The plots in the left column refer to the Gaussian diffusive model, whereas the ones in the right column to the bootstrap forecasting methodology.}
\label{fig:forvsrealEONIA1m}
\end{figure}

\begin{figure}[H]
\centering
\begin{minipage}[t]{7.0truecm}
\centering
\includegraphics[width=7.0truecm]{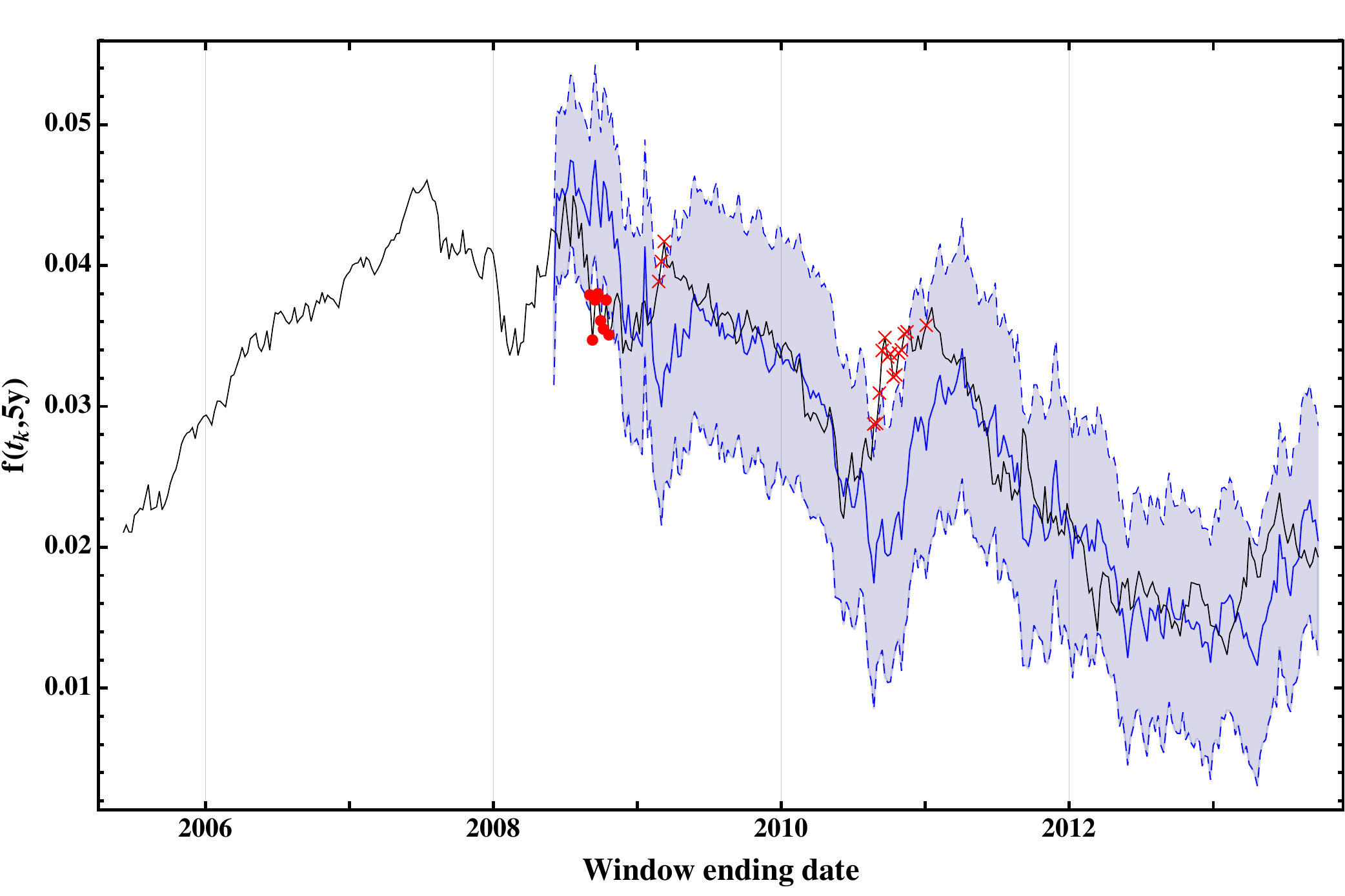}
\end{minipage}
\hspace{1truecm}
\begin{minipage}[t]{7.0truecm}
\centering
\includegraphics[width=7.0truecm]{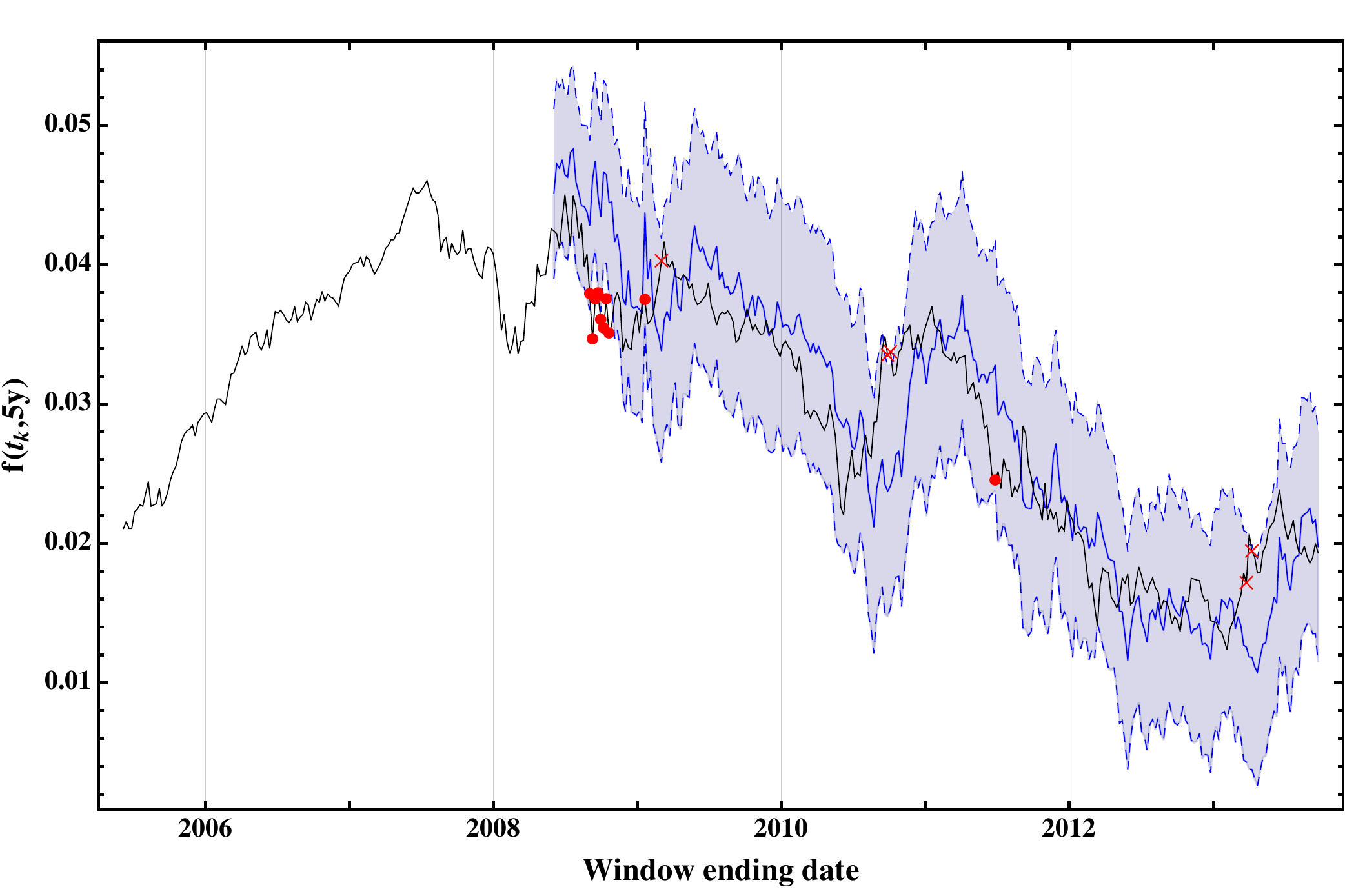}
\end{minipage}
\\
\vspace{1truecm}
\begin{minipage}[t]{7.0truecm}
\centering
\includegraphics[width=7.0truecm]{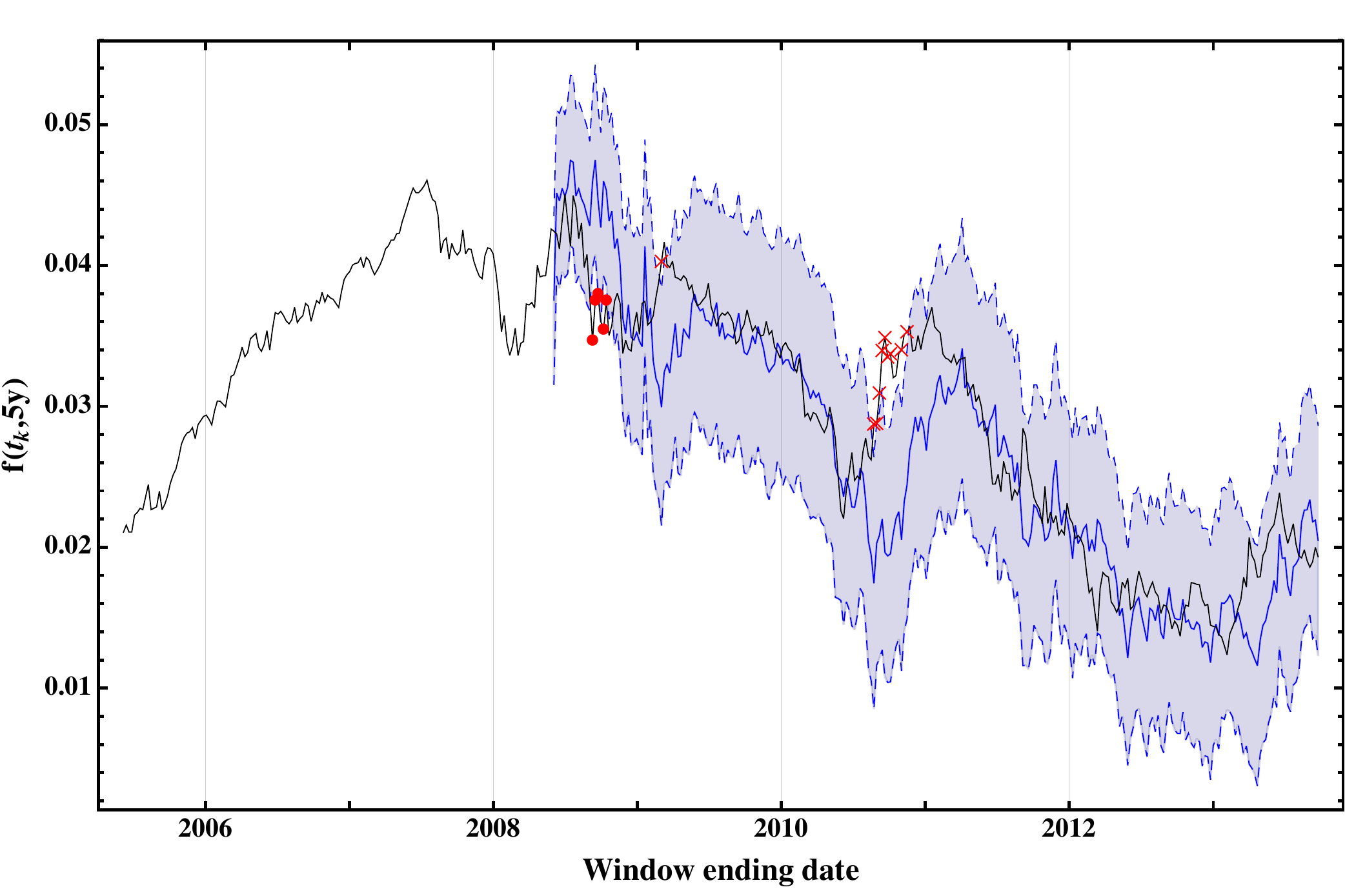}
\end{minipage}
\hspace{1truecm}
\begin{minipage}[t]{7.0truecm}
\centering
\includegraphics[width=7.0truecm]{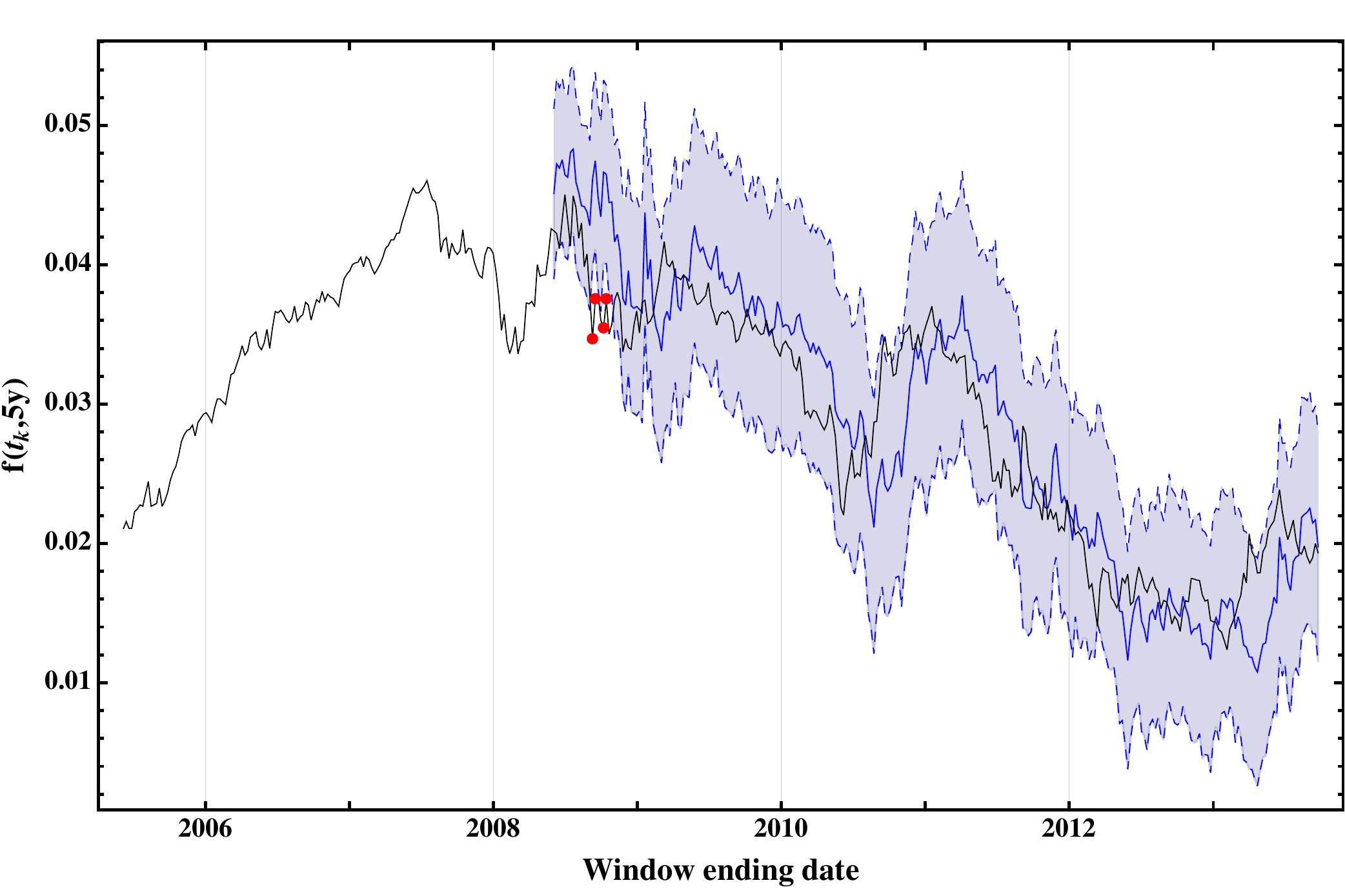}
\end{minipage}
\caption{Solid blue line: Mean of the three month ahead forecast distribution for the $f(t,5y)$ rate. Blue dashed lines: Mean $\pm$ two standard deviations of the  three month ahead forecasted distribution for the $f(t,5y)$ rate. Black solid line: Realised $f(t,5y)$ rate. Red dots (crosses): Negative (positive) exceptions with coverage probability $95\%$ and $99\%$ on the first and second row, respectively. Left and right columns as in Figure~\ref{fig:forvsrealEONIA1m}.}
\label{fig:forvsrealEONIA5y}
\end{figure}

The results of the unconditional coverage tests for the EONIA instantaneous forward curve are reported in Tables~\ref{tab:testEONIA1wgauss}, \ref{tab:testEONIA12wgauss}, and~\ref{tab:testEONIA52wgauss} for the Gaussian diffusive model, whereas Tables~\ref{tab:testEONIA1wboot}, \ref{tab:testEONIA12wboot}, and ~\ref{tab:testEONIA52wboot} contains the results obtained through the bootstrap procedure. The symbols $^{(*)}$ and $^{(**)}$ correspond to statistical significance at 95\% and 99\%, respectively. In Figures~\ref{fig:forvsrealEONIA1m}, and \ref{fig:forvsrealEONIA5y} we show the time evolution of the three month forecast confidence intervals (blue lines) together with the realised rates (black line), in order to localise negative and positive exceptions (red dots and crosses respectively). We select a short time-to-maturity rate, $f(t,1m)$ and a long time-to-maturity one, $f(t,5y)$. In the first row we use a coverage probability of $95\%$ comparing the Gaussian diffusive model (left panel) to the bootstrap method (right panel), whereas the second row show the results for the $99\%$ coverage probability, comparing again Gaussian diffusive (left panel) with bootstrap (right panel).

\begin{figure}[H]
\centering
\begin{minipage}[t]{7.0truecm}
\centering
\includegraphics[width=7.0truecm]{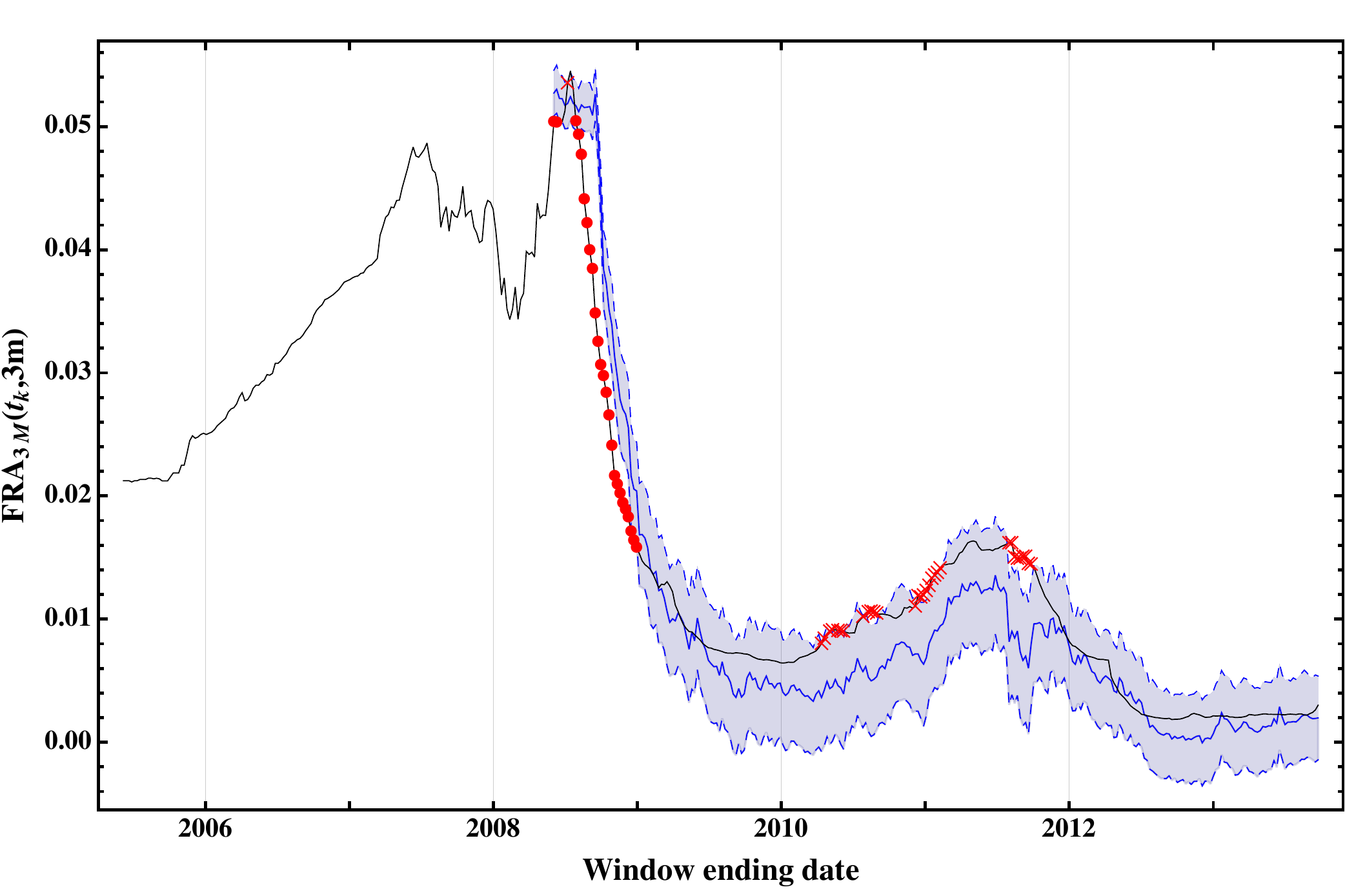}
\end{minipage}
\hspace{1truecm}
\begin{minipage}[t]{7.0truecm}
\centering
\includegraphics[width=7.0truecm]{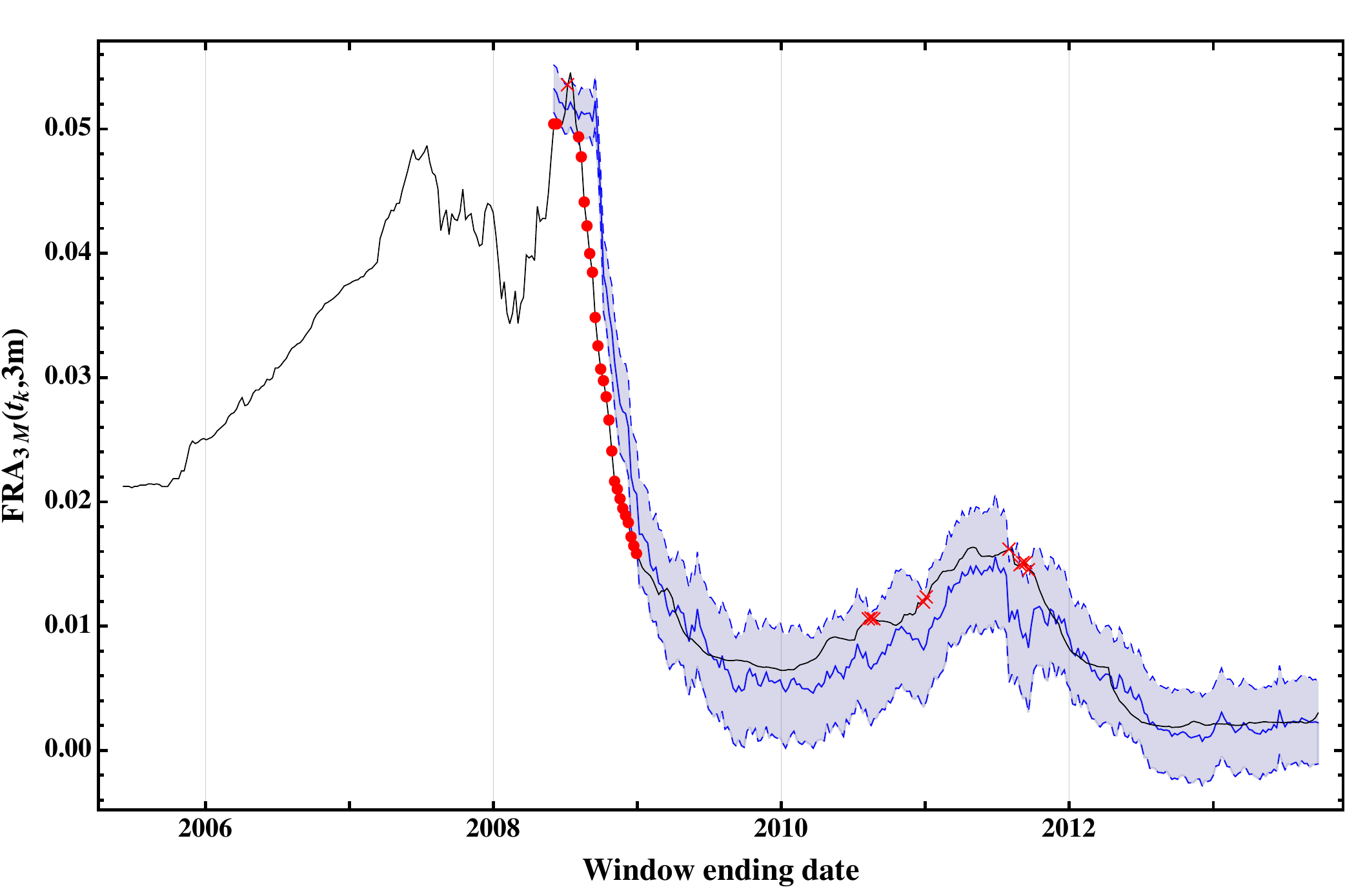}
\end{minipage}
\\
\vspace{1truecm}
\begin{minipage}[t]{7.0truecm}
\centering
\includegraphics[width=7.0truecm]{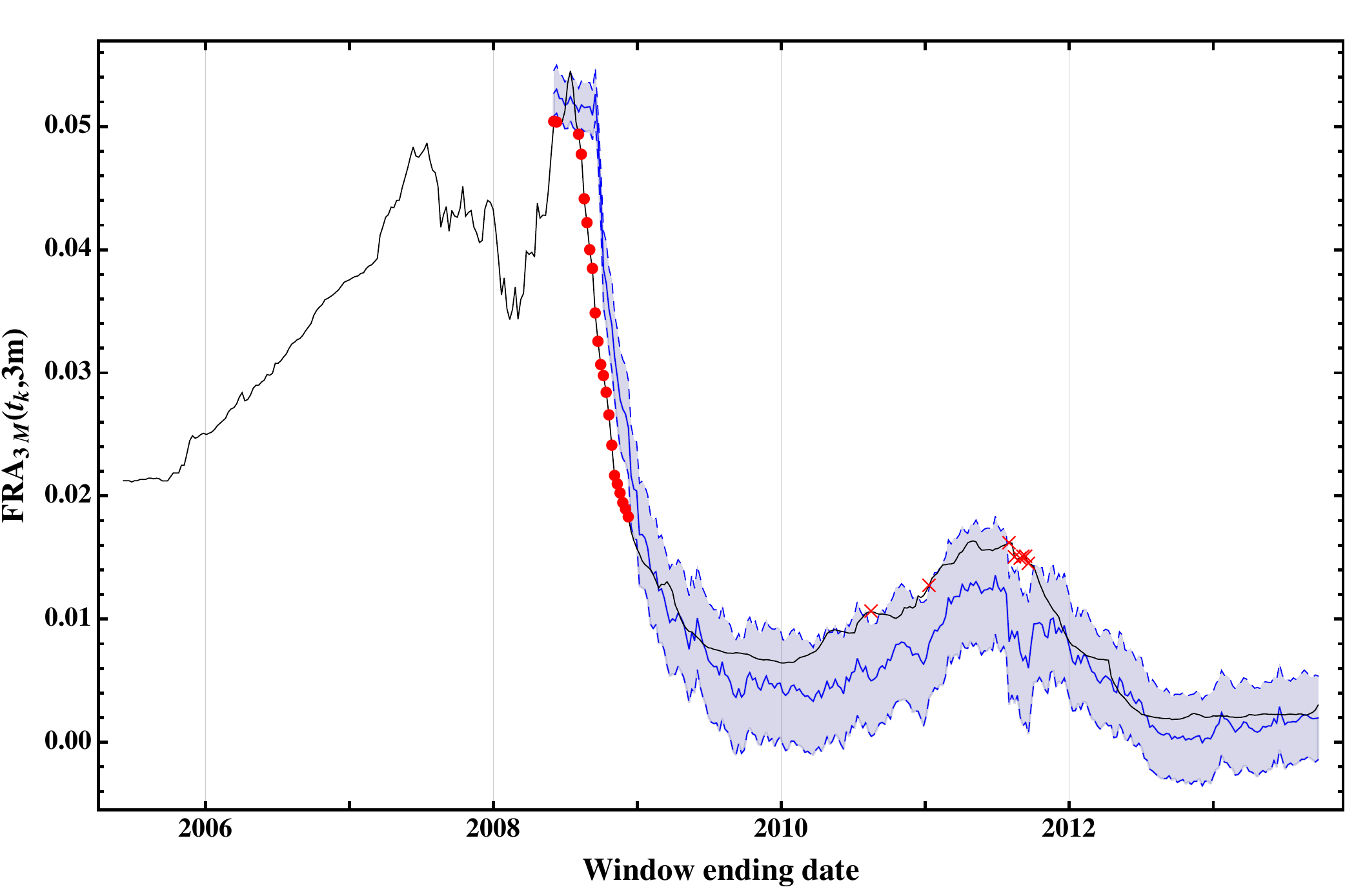}
\end{minipage}
\hspace{1truecm}
\begin{minipage}[t]{7.0truecm}
\centering
\includegraphics[width=7.0truecm]{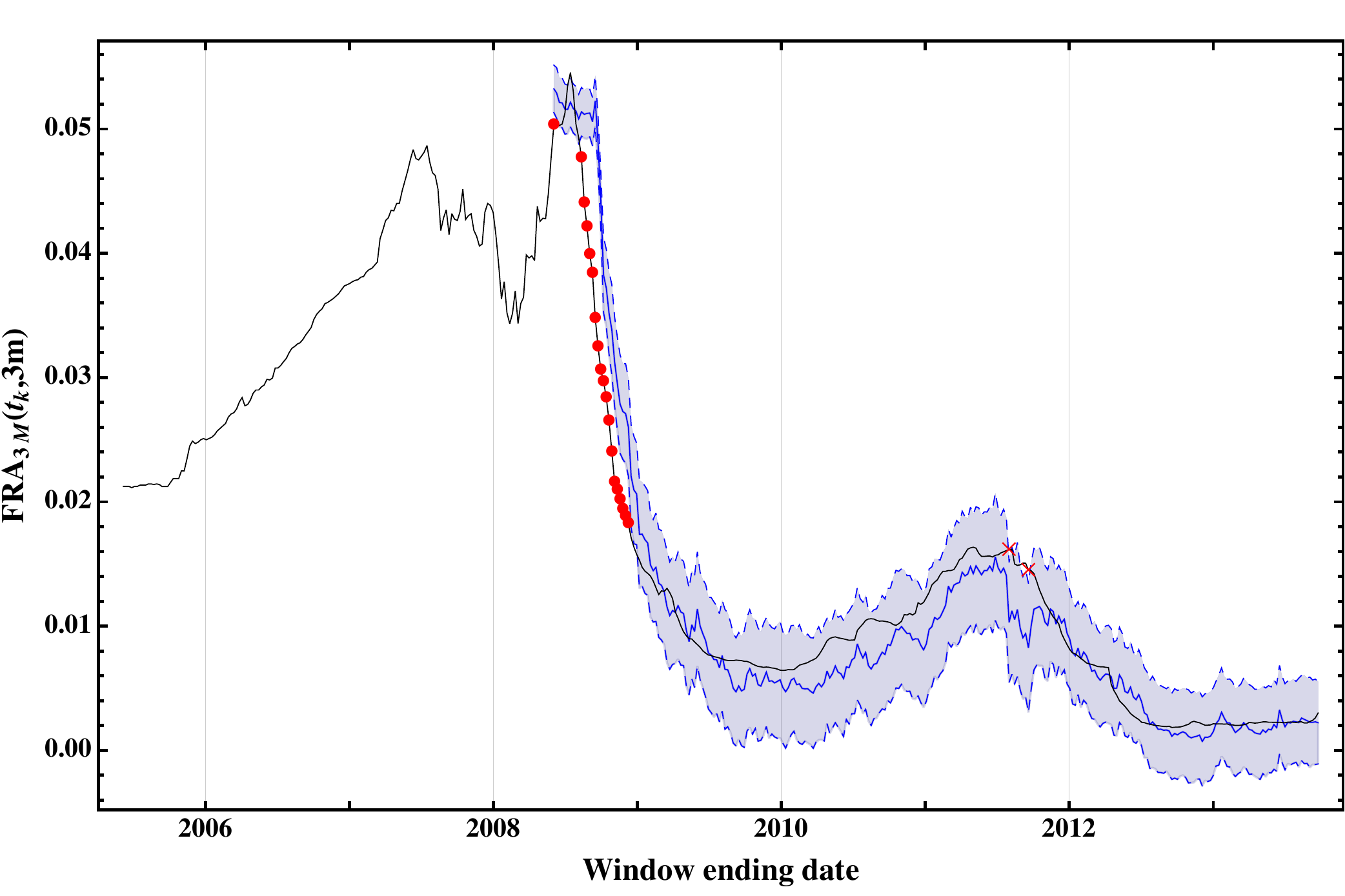}
\end{minipage}
\caption{Solid blue line: Mean of the three month ahead forecast distribution for the $\text{FRA}_{\text{3M}}(t,3m)$ rate. Blue dashed lines: Mean $\pm$ two standard deviations of the three month ahead forecasted distribution for the $\text{FRA}_{\text{3M}}(t,3m)$ rate. Black solid line: Realised $\text{FRA}_{\text{3M}}(t,3m)$ rate. Red dots (crosses): Negative (positive) exceptions with coverage probability $95\%$ and $99\%$ on the first and second row respectively.  Left and right columns as in Figure~\ref{fig:forvsrealEONIA1m}.}
\label{fig:forvsrealEUR3M3m}
\end{figure}

\begin{figure}[H]
\centering
\begin{minipage}[t]{7.0truecm}
\centering
\includegraphics[width=7.0truecm]{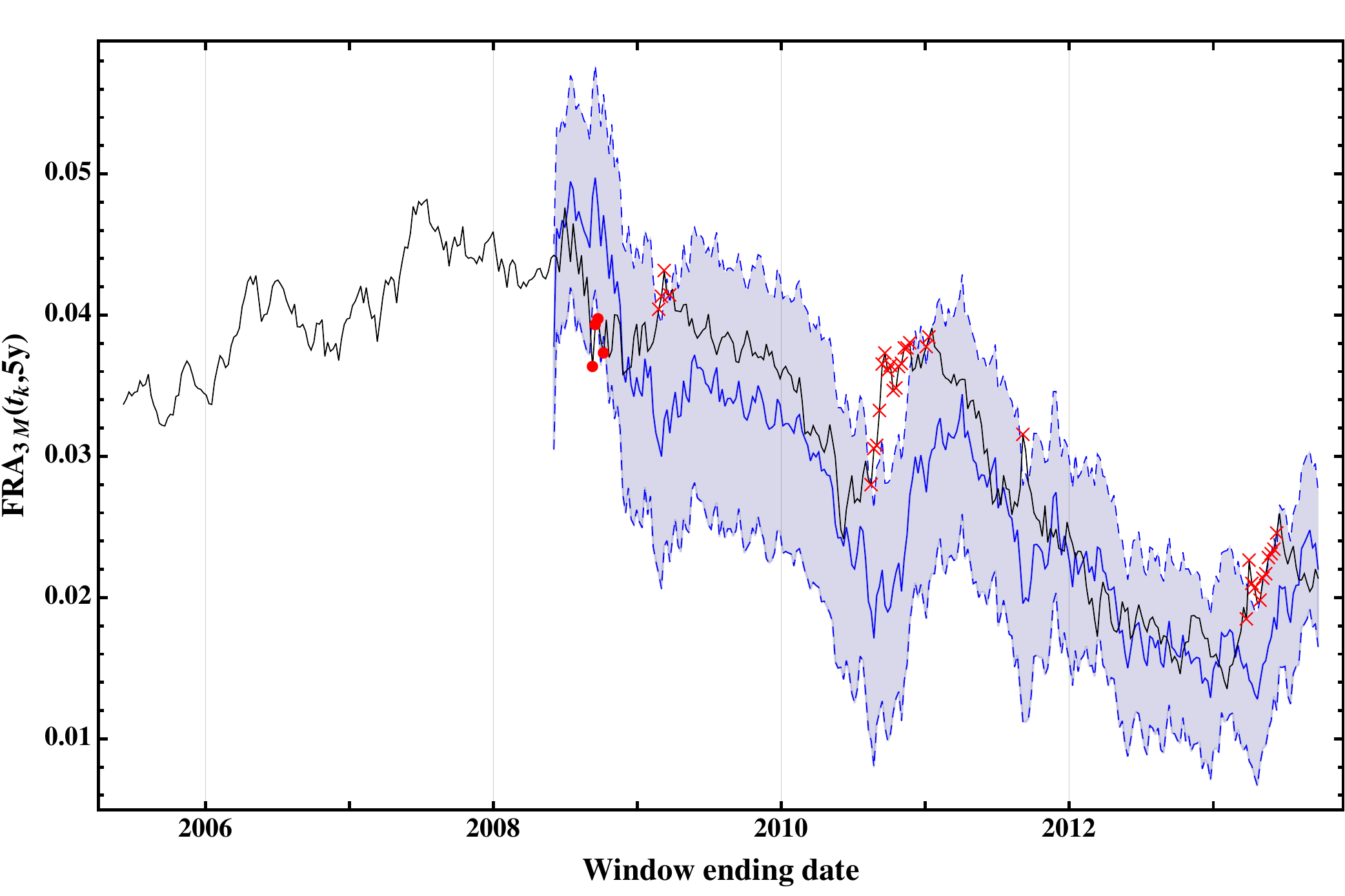}
\end{minipage}
\hspace{1truecm}
\begin{minipage}[t]{7.0truecm}
\centering
\includegraphics[width=7.0truecm]{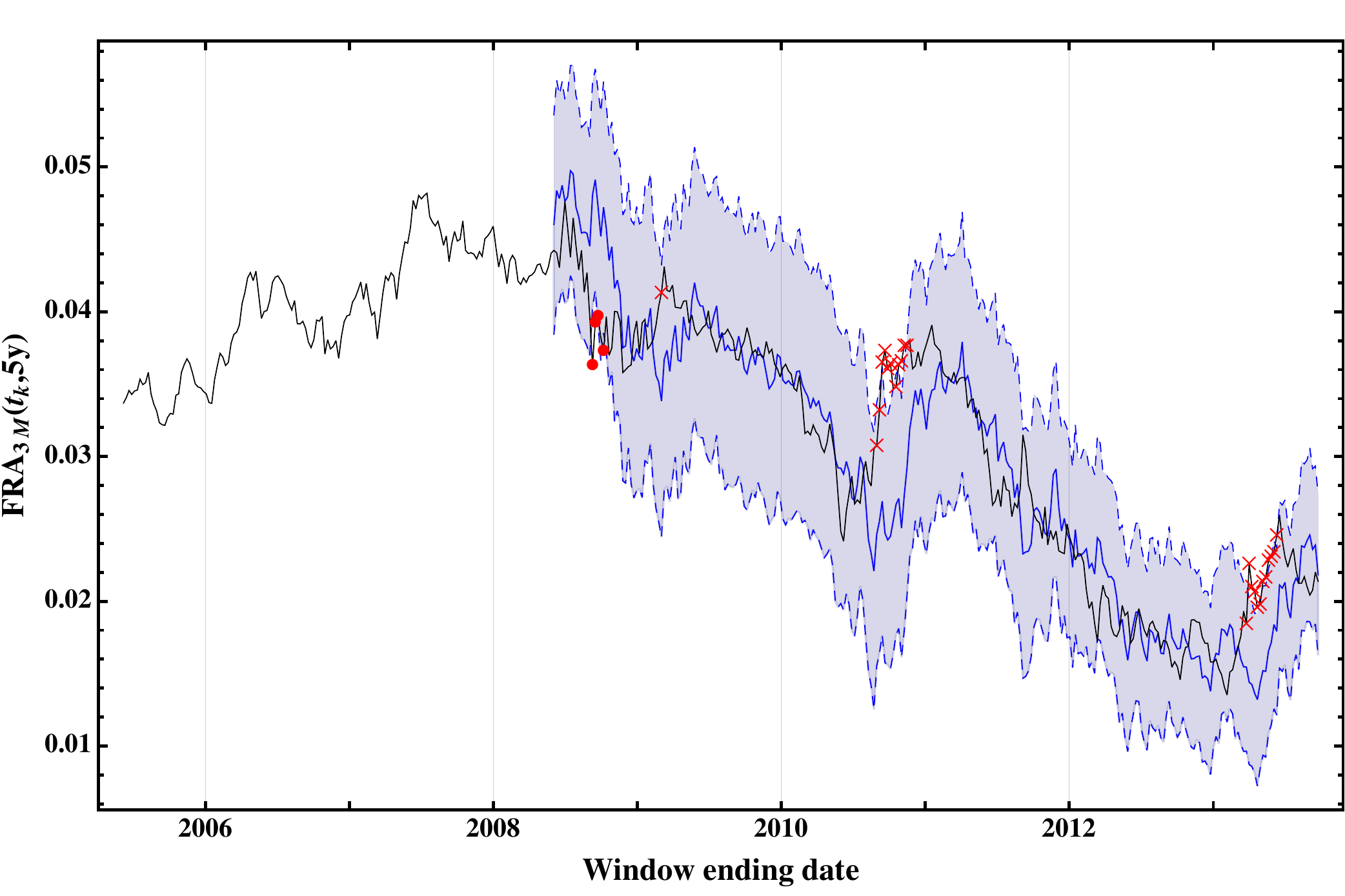}
\end{minipage}
\\
\vspace{1truecm}
\begin{minipage}[t]{7.0truecm}
\centering
\includegraphics[width=7.0truecm]{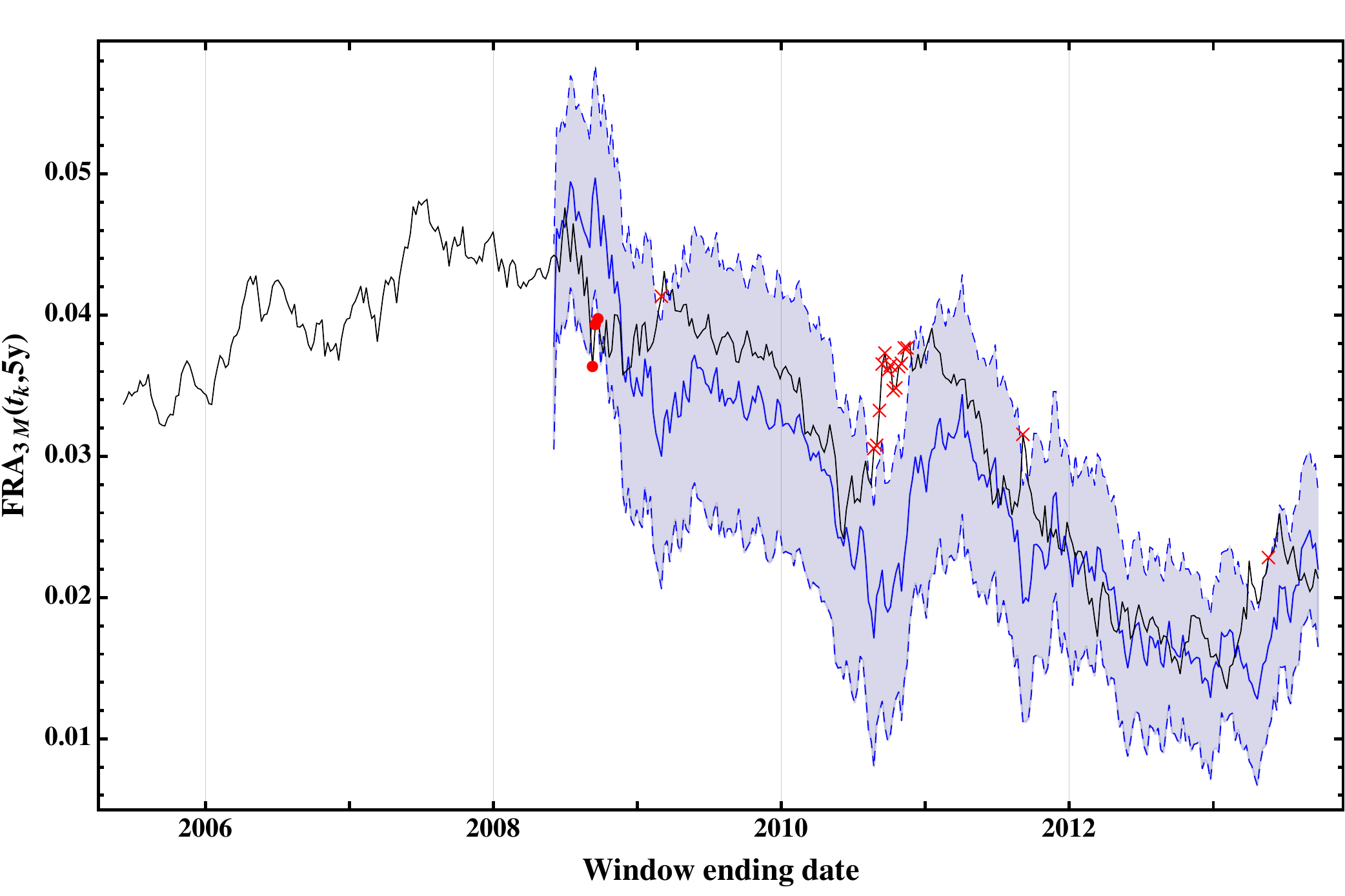}
\end{minipage}
\hspace{1truecm}
\begin{minipage}[t]{7.0truecm}
\centering
\includegraphics[width=7.0truecm]{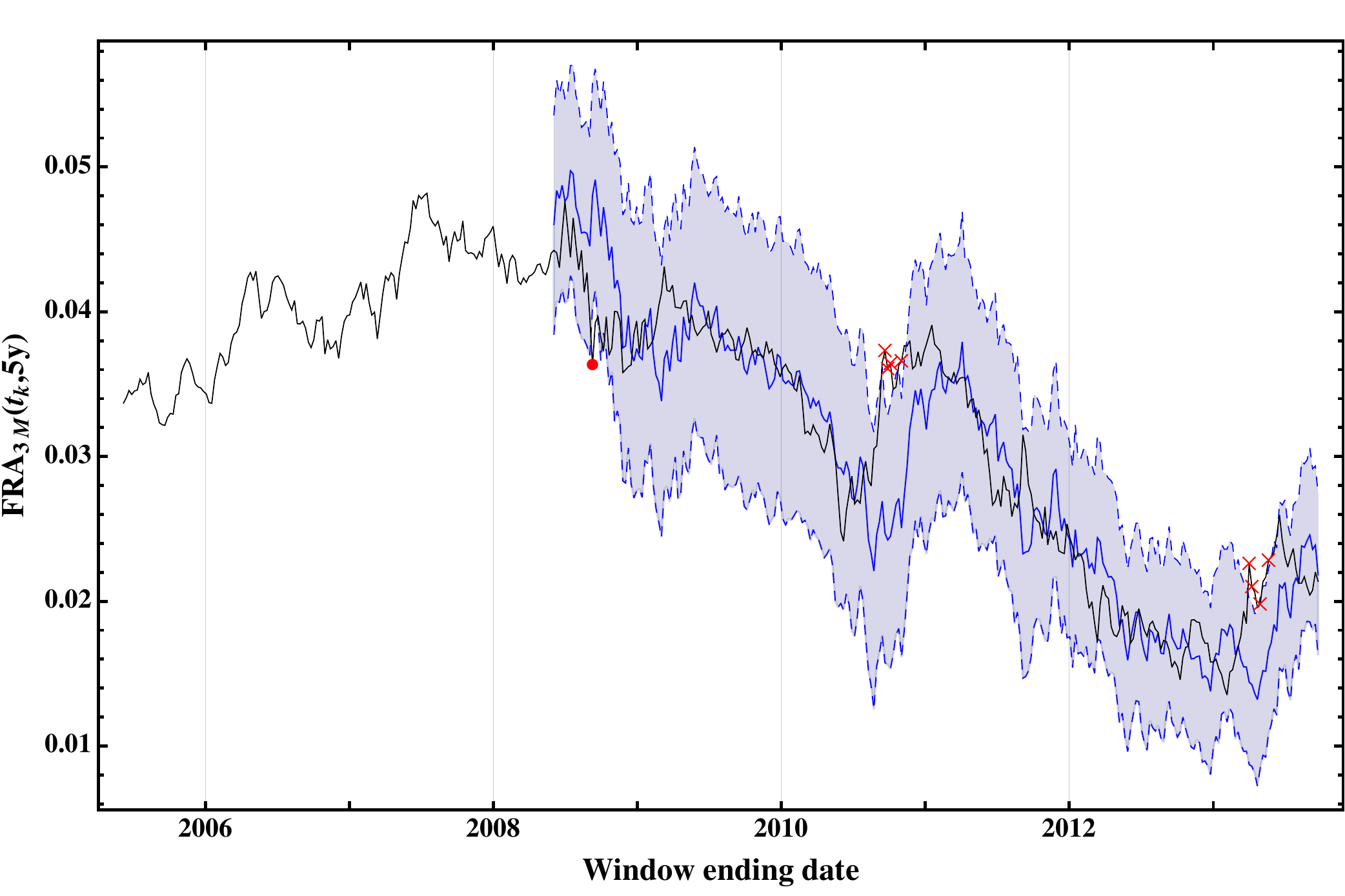}
\end{minipage}
\caption{Solid blue line: Mean of the three month ahead forecast distribution for the $\text{FRA}_{\text{3M}}(t,5y)$ rate. Blue dashed lines: Mean $\pm$ two standard deviations of the three month ahead forecasted distribution for the $\text{FRA}_{\text{3M}}(t,5y)$ rate. Black solid line: Realised $\text{FRA}_{\text{3M}}(t,5y)$ rate. Red dots (crosses): Negative (positive) exceptions with coverage probability $95\%$ and $99\%$ on the first and second row respectively.  Left and right columns as in Figure~\ref{fig:forvsrealEONIA1m}.}
\label{fig:forvsrealEUR3M5y}
\end{figure}

As for the EUR3M FRA term structure, the results of the unconditional coverage tests are reported in Tables~\ref{tab:testEUR3M1wgauss}, \ref{tab:testEUR3M12wgauss}, and~\ref{tab:testEUR3M52wgauss} for the Gaussian diffusive model, whereas Tables~\ref{tab:testEUR3M1wboot}, \ref{tab:testEUR3M12wboot}, and ~\ref{tab:testEUR3M52wboot} contains the results obtained through the bootstrap procedure. Finally in Figures~\ref{fig:forvsrealEUR3M3m}, and \ref{fig:forvsrealEUR3M5y} we report the time evolution of the three month ahead forecast confidence intervals for a short and a long rate of the EUR3M curve, i.e. $\text{FRA}_{\text{3M}}(t,3m)$ $\text{FRA}_{\text{3M}}(t,5y)$.

As far as the EONIA curve is concerned, results of the test are very satisfactory at short forecasting horizons (one week) with $95\%$ of coverage probability for both methods (multivariate Normal and bootstrap). When considering a higher value for the coverage probability , i.e. $p=99\%$, the bootstrap methodology further improves the quality of the test. This fact remains true also for longer forecasting horizons -- three month and one year -- even though the results overall deteriorate. The reason behind such behaviour can be guessed looking at the relation between the realised time series and the mean of the forecast distribution. Refer, for instance, to the top-left panel of Figure~\ref{fig:forvsrealEONIA1m}. From October 2008 to April 2009 the ECB cuts five times the reference short-term rate. Consistently, the short-term component of the EONIA curve follows the rapid decline in the level of rates. However, whereas the historical time series switches abruptly from a markedly upward trend to a markedly downward one, the forecast rate does not follow immediately the same behaviour. It takes a while to the forecast to revert the drift, i.e. to the risk premium to rise to large positive values and provide a strong negative correction. Indeed, from Figure~~\ref{fig:lambdashort} we see that the short term risk premium does not increase before the final quarter of 2009. Such a delayed adjustment of the risk premium in correspondence of abrupt changes of the rate dynamics is responsible of the majority of forecast failures. As can be readily understood the longer is the forecast horizon the more severe will be the mismatch between forecast and realised rates. Thus, we conclude that a model which does not take into account explicitly the possibility of abrupt rate movements due to the ECB monetary policy cannot succeed in producing reliable long-run forecast. Quite consistently with the behaviour observed for the EONIA curve, forecast results for the EUR3M curve are satisfactory at short forecasting horizons (one week) with $95\%$. For the three month tenor curve, however, results based on the Gaussian model performs slightly worse than those obtained with the bootstrap forecasting methodology. Also in this case, when considering a higher value for the coverage probability, the bootstrap methodology improves the quality of the test. Nonetheless, the same conclusions drawn for the EONIA case for longer forecasting horizon apply also for the three month tenor term structure. 

%%%%%%%%%%%%%%%%%%%%%%%%%%%%%%%%%%%%%%%%%%%%%%%%%%%%%%%%%%%%%%%%%%%%%%%%%%%%%%%%%%%%%%%%%%%%%%%%%%%%%%%%%%%%%%%%%%%%%%%%%%%%%%%%%%%%%%%%%%%%%%%%%%
\section{Conclusions}
\label{sec:conclusions}
In this paper we propose a novel approach to the projection of interest rate term structures which is tailored to the post-crisis world of multiple yield curves. As opposed to single curve scenarios, where one relies on generating techniques based on factor analysis, filtered historical simulation, or the popular RiskMetrics$^\text{TM}$ methodology, currently the quality and effectiveness of interest rate risk management depends on the ability to describe properly both the EONIA term structure and the three month, six month, and one year curves. To the best of our knowledge this is the first attempt to capture the volatility-correlation structure of the historical time series which is natively designed for the multiple yield curves. We present a HJM modelling framework where we describe the discounting curve in terms of instantaneous forward rates while the dynamics of FRA rates determines the evolution of the longer tenor term structures. We show how to approximate the continuous time infinite-dimensional SDE's in the HJM setting by a Vector Autoregressive process of finite dimension. The reduction to a discrete time model significantly eases the estimation of the model parameters.  Through an iterative maximum-likelihood procedure we estimate volatilities, correlation coefficients and risk-premia. In particular we display the evolution of the risk premia during the investigated time period (2005-2013) which accounts for changes observed in the interest rate term structure. A significantly positive risk premium indicates that the forward rates implied by the market term structures require a strong negative correction in order to describe the realised evolution of the spot rates. We then perform numerical out-of-sample tests which finally prove the reliability of our approach over a forecasting horizon of one week. Our Gaussian model can be further improved to capture tail events, if one employs a simple bootstrap methodology in the forecasting procedure. As longer forecasting horizons are considered, the performances of the model deteriorate. The empirical analysis strongly supports the hypothesis that the large majority of failures accumulates in correspondence of abrupt changes in the rate dynamics. Such events -- plausibly driven by shifts in the monetary policy of the ECB -- cannot be easily accommodated for within the current modelling approach. As a future perspective of the current work, we plan to partially accommodate for regime shifts augmenting our estimation procedure including survey-based forecasts of short-term yields. As far as the evidence of stochasticity in the volatility time series is concerned, we are currently working on a modified version of our discrete time setting able to capture the heteroskedastic nature and possible asymmetries of the instantaneous forward and FRA rate volatilities.

\section*{Acknowledgments}
We thank Flavia Barsotti, Andrea Bertagna, Tommaso Colozza, Fulvio Corsi, Niccol\`o Cottini, Lorenzo Liesch, Stefano Marmi, Aldo Nassigh, Andrea Pallavicini and Roberto Ren\`o for many inspiring discussions. We also acknowledge Andrea Sillari for having provided the historical data. The research activity of CS is supported by UniCredit S.p.A. Grant n.1300240/2013 administered by the Scuola Normale Superiore. GB acknowledges research support from the Scuola Normale Superiore Grant SNS14\_B\_BORMETTI.

\appendix

%%%%%%%%%%%%%%%%%%%%%%%%%%%%%%%%%%%%%%%%%%%%%%%%%%%%%%%%%%%%%%%%%%%%%%%%%%%%%%%%%%%%%%%%%%%%%%%%%%%%%%%%%%%%%%%%%%%%%%%%%%%%%%%%%%%%%%%%%%%%%%%%%%
\section{Bessel cubic spline}\label{app:spline}
We use the Bessel cubic spline method every time we need to interpolate a curve defined on a finite set of points.
Let us consider a generic curve $g(u)$ defined through the following $K$ points
\bd
\left\{(s_1,g(s_1)),\ldots, ,(s_K,g(s_K))\right\}.
\ed
The cubic spline interpolating function is defined by a set of $K-1$ third order polynomials
\bes
g_{{\rm spline}}(x)=a_{h}+b_{h}\,(x-s_{h})+c_{h}\,(x-s_{h})^2+d_{h}\,(x-s_{h})^3,\quad {\rm for}\;\;s_{h}\leq x\leq s_{h+1}\;\;{\rm and}\;\; h=1,\ldots,K-1.
\ees
For each polynomial there are 4 coefficients to be determined, so that the total number of constraints one needs to impose is $4K-4$.
In~\cite{hagan2006} the authors discuss the constraints and detail the derivation of the results which follow.
%One requires that:
%\begin{itemize}
%\item[\textopenbullet] the interpolating function indeed meets the given data $a_{k,h}=f(t_k,s_h)$ for $h=1,\ldots K-1$ and 
%$a_{k,K-1}+b_{k,K-1}\,(s_{K}-s_{K-1})+c_{k,K-1}\,(s_{K}-s_{K-1})^2+d_{k,K-1}\,(s_K-s_{K-1})^3=f(t_k,s_K)$;
%\item[\textopenbullet] the entire interpolating function is continuous, thus $a_{k,h}+b_{k,h}\,(s_{h+1}-s_{h})+c_{k,h}\,(s_{h+1}-s_{h})^2+d_{k,h}\,(s_{h+1}-s_{h})^3=f(t_k,s_{h+1})$ for $h=1,\ldots,K-2$;
%\item[\textopenbullet] the entire interpolating function is differentiable, thus $b_{k,h}+2c_{k,h}\,(s_{h+1}-s_{h})+3d_{k,h}\,(s_{h+1}-s_{h})^2=b_{k,h+1}$ for $h=1,\ldots,K-2$.
%\end{itemize}
%The three conditions above give $3K-4$ constraints. The missing $K$ conditions in the Bessel method for cubic splines are
%~\footnote{We define $b_{k,K}$ as
%\bd
%b_{k,K}=b_{k,K-1}+2c_{k,K}(s_K-s_{K-1})+3d_{k,K}(s_K-s_{K-1})^2.
%\ed
%}
%\bd
%\begin{split}
%&b_{k,h}=\text{ the slope of the quadratic that passes through}\;\;(s_{h-1},f(t_k,x_{h-1})), (s_{h},f(t_k,s_{h})), (s_{h+1},f(t_k,s_{h+1}))\\
%&\qquad\qquad\qquad{\rm for}\;\;h=2,\ldots,K-1\\
%&b_{k,1}=\text{ the slope of the quadratic that passes through}\;\;(s_{1},f(t_k,s_{1})), (s_{2},f(t_k,s_{2})), (s_{3},f(t_k,s_{3}))\\
%&b_{k,K}=\text{ the slope of the quadratic that passes through}\;\;(s_{K-2},f(t_k,s_{K-2})), (s_{K-1},f(t_k,s_{K-1})), (s_{K},f(t_k,s_{K})).
%\end{split}
%\ed
%
While $a_{i}$ are simply equal to the values of the interpolated function at $s_i$, i.e. $a_{i}=g(s_i)$, the analytic expressions for the coefficients $b_{i}$ correspond to  
\bd
\begin{split}
&b_{1}=\frac{1}{s_3-s_1}\left[\frac{s_3+s_2-2s_1}{s_2-s_1}(g(s_2)-g(s_1))-\frac{s_2-s_1}{s_3-s_2}(g(s_3)-g(s_2))\right],\\
&b_{i}=\frac{1}{s_{i+1}-s_{i-1}}\left[\frac{s_{i+1}-s_{i}}{s_{i}-s_{i-1}}(g(s_i)-g(s_{i-1}))+\frac{s_{i}-s_{i-1}}{s_{i+1}-s_{i}}(g(s_{i+1})-g(s_{i}))\right]\quad {\rm for}\;\;i=2,\ldots,K-1,\\
&b_{K}=-\frac{1}{s_{K}-s_{K-2}}\left[\frac{s_{K}-s_{K-1}}{s_{K-1}-s_{K-2}}(g(s_{K-1})-g(s_{K-2}))-\frac{2s_{K}-s_{K-1}-s_{K-2}}{s_{K}-s_{K-1}}(f(t_k,s_{K})-f(t_k,s_{K-1}))\right],
\end{split}
\ed
The equations above can be rewritten as linear superpositions of instantaneous forward rates for three adjacent buckets at $t_k$ with coefficients depending only on the times to maturity of these three buckets
\bd
\begin{split}
&b_{1}=A_{1}(s_1,s_2,s_3)\,g(s_1)+B_{1}(s_1,s_2,s_3)\,g(s_2)+C_{1}(s_1,s_2,s_3)\,g(s_3),\\
&b_{i}=A_{i}(s_{i-1},s_i,s_{i+1})\,g(s_{i-1})+B_{i}(s_{i-1},s_i,s_{i+1})\,g(s_i)+C_{i}(s_{i-1},s_i,s_{i+1})\,g(s_{i+1}),\quad {\rm for}\;\;i=2,\ldots,K-1,\\
&b_{K}=A_{K}(s_{K-2},s_{K-1},s_{K})\,g(s_{K-2})+B_{K}(s_{K-2},s_{K-1},s_{K})\,g(s_{K-1})+C_{K}(s_{K-2},s_{K-1},s_{K})\,g(s_{K}),
\end{split}
\ed
where
\bd
\begin{cases}
&A_1(s_1,s_2,s_3)=\frac{2s_1-s_3-s_2}{(s_2-s_1)(s_3-s_1)},\\
&B_1(s_1,s_2,s_3)=\frac{s_3-s_1}{(s_3-s_2)(s_2-s_1)},\\
&C_1(s_1,s_2,s_3)=\frac{s_1-s_2}{(s_3-s_2)(s_3-s_1)},
\end{cases}
\ed
\bd
\begin{cases}
&A_i(s_{i-1},s_{i},s_{i+1})=\frac{s_{i}-s_{i+1}}{(s_{i}-s_{i-1})(s_{i+1}-s_{i-1})},\\
&B_i(s_{i-1},s_{i},s_{i+1})=\frac{s_{i-1}-2s_{i}+s_{i+1}}{(s_{i}-s_{i-1})(s_{i+1}-s_{i-1})},\\
&C_i(s_{i-1},s_{i},s_{i+1})=\frac{s_{i}-s_{i-1}}{(s_{i+1}-s_{i})(s_{i+1}-s_{i-1})},
\end{cases}\qquad i=2,\ldots,K-1,
\ed
\bd
\begin{cases}
&A_K(s_{K-2},s_{K-1},s_{K})=\frac{s_{K}-s_{K-1}}{(s_{K-1}-s_{K-2})(s_{K}-s_{K-2})},\\
&B_K(s_{K-2},s_{K-1},s_{K})=\frac{s_{K-2}-s_{K}}{(s_{K-1}-s_{K-2})(s_{K}-s_{K-1})},\\
&C_K(s_{K-2},s_{K-1},s_{K})=\frac{2s_{K}-s_{K-1}-s_{K-2}}{(s_{K}-s_{K-1})(s_{K}-s_{K-2})},
\end{cases}
\ed
The same reasoning applies for the coefficients $c_{h}$ and $d_{h}$
\bd
\begin{split}
  &c_{1}=\frac{1}{s_1-s_2}\left[\frac{g(s_3)-g(s_1)}{s_3-s_1}-\frac{g(s_3)-g(s_2)}{s_3-s_2}\right],\\
  &d_{1}=0\\
  &c_{i}=\frac{1}{s_{i}-s_{i+1}}\left[\frac{2(g(s_{i})-g(s_{i-1}))}{(s_i-s_{i-1})}-\frac{2(g(s_{i+1})-g(s_{i-1}))}{s_{i+1}-s_{i-1}}-\frac{g(s_{i+2})-g(s_{i})}{s_{i+2}-s_{i}}+\frac{g(s_{i+2})-g(s_{i+1})}{s_{i+2}-s_{i+1}}\right],\\
  &d_{i}=\frac{1}{(s_{i}-s_{i+1})^2}\left[\frac{g(s_{i})-g(s_{i-1})}{(s_i-s_{i-1})}-\frac{g(s_{i+1})-g(s_{i-1})}{s_{i+1}-s_{i-1}}-\frac{g(s_{i+2})-g(s_{i})}{s_{i+2}-s_{i}}+\frac{g(s_{i+2})-g(s_{i+1})}{s_{i+2}-s_{i+1}}\right],\\
\end{split}
\ed

\noindent for $i=2,\ldots,K-1$. Thus, both $c_{h}$ and $d_{h}$ can be written as linear superposition of the values of four adjacent instantaneous forward rates with coefficients which depends exclusively on the time-to-maturity buckets vector
\bd
\begin{split}
&c_{1}=A^\prime_{1}(s_1,s_2,s_3)\,g(s_1)+B^\prime_{1}(s_1,s_2,s_3)\,g(s_2)+C^\prime_{1}(s_1,s_2,s_3)\,g(s_3),\\
&c_{i}=A^\prime_{i}(s_{i-1},s_i,s_{i+1},s_{i+2})\,g(s_{i-1})+B^\prime_{i}(s_{i-1},s_i,s_{i+1},s_{i+2})\,g(s_i)+C^\prime_{i}(s_{i-1},s_i,s_{i+1},s_{i+2})\,g(s_{i+1}),\\
&\quad\quad\quad +D^\prime_{i}(s_{i-1},s_i,s_{i+1},s_{i+2})\,g(s_{i+2}),\hspace{200pt} {\rm for}\;\;i=2,\ldots,K-2,\\
&c_{K-1}=A^\prime_{K-1}(s_{K-2},s_{K-1},s_K)\,g(s_{K-2})+B^\prime_{K-1}(s_{K-2},s_{K-1},s_K)\,g(s_{K-1})+C^\prime_{K-1}(s_{K-2},s_{K-1},s_{K})\,g(s_K),\\
\end{split}
\ed
where
\bd
\begin{cases}
&A^\prime_1(s_1,s_2,s_3)=\frac{1}{(s_2-s_1)(s_3-s_1)},\\
&B^\prime_1(s_1,s_2,s_3)=-\frac{1}{(s_3-s_2)(s_2-s_1)},\\
&C^\prime_1(s_1,s_2,s_3)=\frac{1}{(s_3-s_2)(s_3-s_1)},
\end{cases}
\ed
\bd
\begin{cases}
&A^\prime_i(s_{i-1},s_{i},s_{i+1},s_{i+2})=\frac{2}{(s_{i}-s_{i-1})(s_{i+1}-s_{i-1})},\\
&B^\prime_i(s_{i-1},s_{i},s_{i+1},s_{i+2})=\frac{s_{i-1}+s_{i}-2s_{i+2}}{(s_{i}-s_{i-1})(s_{i+1}-s_{i})(s_{i+2}-s_i)},\\
&C^\prime_i(s_{i-1},s_{i},s_{i+1},s_{i+2})=-\frac{s_{i-1}+s_{i+1}-2s_{i+2}}{(s_{i+1}-s_{i-1})(s_{i+1}-s_{i})(s_{i+2}-s_{i+1})},\\
&D^\prime_i(s_{i-1},s_{i},s_{i+1})=-\frac{1}{(s_{i+2}-s_{i})(s_{i+2}-s_{i+1})},
\end{cases}\qquad i=2,\ldots,K-2,
\ed
\bd
\begin{cases}
&A^\prime_{K-1}(s_{K-2},s_{K-1},s_K)=\frac{1}{(s_{K-1}-s_{K-2})(s_{K}-s_{K-2})},\\
&B^\prime_{K-1}(s_{K-2},s_{K-1},s_K)=-\frac{1}{(s_{K-1}-s_{K-2})(s_{K}-s_{K-1})},\\
&C^\prime_{K-1}(s_{K-2},s_{K-1},s_K)=\frac{1}{(s_{K}-s_{K-2})(s_{K}-s_{K-1})},
\end{cases}
\ed
and
\bd
\begin{split}
&d_{1}=0,\\
&d_{i}=A^{\prime\prime}_{i}(s_{i-1},s_i,s_{i+1},s_{i+2})\,g(s_{i-1})+B^{\prime\prime}_{i}(s_{i-1},s_i,s_{i+1},s_{i+2})\,g(s_i)+C^{\prime\prime}_{i}(s_{i-1},s_i,s_{i+1},s_{i+2})\,g(s_{i+1}),\\
&\quad\quad\quad +D^{\prime\prime}_{i}(s_{i-1},s_i,s_{i+1},s_{i+2})\,g(s_{i+2}),\hspace{175pt} {\rm for}\;\;i=2,\ldots,K-2\\
&d_{k,K-1}=0,\\
\end{split}
\ed
where
\bd
\begin{cases}
&A^{\prime\prime}_i(s_{i-1},s_{i},s_{i+1},s_{i+2})=\frac{1}{(s_{i}-s_{i-1})(s_{i+1}-s_{i-1})(s_{i+1}-s_i)},\\
&B^{\prime\prime}_i(s_{i-1},s_{i},s_{i+1},s_{i+2})=\frac{s_{i+2}-s_{i-1}}{(s_{i}-s_{i-1})(s_{i+2}-s_{i})(s_{i+1}-s_{i})^2},\\
&C^{\prime\prime}_i(s_{i-1},s_{i},s_{i+1},s_{i+2})=\frac{s_{i+2}-s_{i-1}}{(s_{i+1}-s_{i-1})(s_{i+2}-s_{i+1})(s_{i+1}-s_{i})^2},\\
&D^{\prime\prime}_i(s_{i-1},s_{i},s_{i+1})=\frac{1}{(s_{i+1}-s_{i})(s_{i+2}-s_{i})(s_{i+2}-s_{i+1})},
\end{cases}\qquad i=2,\ldots,K-2,
\ed
In conclusion, all coefficients can be expressed as a matrix-vector product
\be
\label{eq:splinecoeff}
\begin{split}
  &a_{h}=\left[\bm{g}\right]_h,\quad h=1,\ldots, K-1,\\
  &b_{h}=\left[M(\bm{s})\,\bm{g}\right]_h,\quad h=1,\ldots, K,\\
  &c_{h}=\left[M^\prime(\bm{s})\,\bm{g}\right]_h,\quad h=1,\ldots, K-1,\\
  &d_{h}=\left[M^{\prime\prime}(\bm{s})\,\bm{g}\right]_h,\quad h=1,\ldots, K-1,\\
\end{split}
\ee
where~\footnote{We have dropped the dependence of $A_i, B_i,\ldots$ on $\bm{s}$ for ease of notation.}
\begin{eqnarray*}
\bm{g}&=&\left[g(s_1),\ldots,g(s_K)\right]^\intercal,\\
M(\bm{s})&=&\left[\begin{array}{cccccccc}
A_1 & B_1 & C_1 & 0 &0&0&\dots & 0\\ 
A_2 & B_2 & C_2 & 0 &0&0&\dots & 0 \\ 
0&A_3&B_3& C_3 &0&0& \dots &0\\
\vdots&&&&&&\vdots\\
0&0&0&\dots &0&A_{K-1}&B_{K-1}& C_{K-1}\\
0&0&0&\dots &0&A_{K}&B_{K}& C_{K}\\
\end{array}\right],\\
M^\prime(\bm{s})&=&\left[\begin{array}{cccccccc}
A^\prime_1 & B^\prime_1 & C^\prime_1 & 0 &0& 0&\dots & 0\\ 
A^\prime_2 & B^\prime_2 & C^\prime_2 & D^\prime_2 &0& 0&\dots & 0 \\ 
0&A^\prime_3&B^\prime_3& C^\prime_3 &D^\prime_3& 0&\dots &0\\
\vdots&&&&&&&\vdots\\
0&0&\dots &0&A^\prime_{K-2}&B^\prime_{K-2}&C^\prime_{K-2}& D^\prime_{K-2}\\
0&0&\dots &0&0&A^\prime_{K-1}&B^\prime_{K-1}& C^\prime_{K-1}\\
\end{array}\right],\\
M^{\prime\prime}(\bm{s})&=&\left[\begin{array}{cccccccc}
0 & 0 & 0 & 0 &0& 0&\dots & 0\\ 
A^{\prime\prime}_2 & B^{\prime\prime}_2 & C^{\prime\prime}_2 & D^{\prime\prime}_2 &0& 0&\dots & 0 \\ 
0&A^{\prime\prime}_3&B^{\prime\prime}_3& C^{\prime\prime}_3 &D^{\prime\prime}_3& 0&\dots &0\\
\vdots&&&&&&&\vdots\\
0&0&\dots &0&A^{\prime\prime}_{K-2}&B^{\prime\prime}_{K-2}&C^{\prime\prime}_{K-2}& D^{\prime\prime}_{K-2}\\
0&0&\dots &0&0&0&0& 0\\
\end{array}\right].
\end{eqnarray*}
The derivative of the spline interpolated version of the function $g(u)$ at the interpolation points reduces to the coefficients $b_{i}$
\bd
\partial_x\, g_{{\rm spline}}(x)\Big|_{x=x_i}=b_{i}=\left[M(\bm{s})\,\bm{g}\right]_i,\qquad i=1,\ldots,K.
\ed
As for the integral, we need to compute the following quantity
\bd
\int_0^{s_i}\,du\,g_{{\rm spline}}(u),
\ed
where $s_i$ is one of the interpolation points.
We assume that $s_1>0$, as it is the case in our analysis, and chose to extrapolate the function flat from $s_1$ down to 0 so that
\bd
\begin{split}
\int_0^{s_i}\,du\,g_{{\rm spline}}(u)&=\int_0^{s_1}du\,a_1+\sum_{\substack{h=1}}^{i-1}\int_{s_h}^{s_{h+1}}du\,\left[a_h+b_h(u-s_h)+c_h(u-s_h)^2+d_h(u-s_h)^3\right]\\
&=a_1\,s_1+\sum_{\substack{h=1}}^{i-1}\left[a_h\,(s_{h+1}-s_h)+\frac{b_h}{2}\,(s_{h+1}-s_h)^2+\frac{c_h}{3}\,(s_{h+1}-s_h)^3+\frac{d_h}{4}\,(s_{h+1}-s_h)^4\right].
\end{split}
\ed
Now we can use the explicit expression of the coefficients $a_h$, $b_h$, $c_h$ and $d_h$ as reported in Equation~(\ref{eq:splinecoeff})~\footnote{We drop the dependence of $M$, $M^\prime$ and $M^{\prime\prime}$ on $\bm{s}$ for ease of notation.}
\begin{eqnarray*}
&&\int_0^{s_i}\,du\,g_{{\rm spline}}(u)\\
&&=g(s_1)s_1+\sum_{\substack{h=1}}^{i-1}\left[g_h(s_{h+1}-s_h)+\frac{\left[M\bm{g}\right]_h}{2}(s_{h+1}-s_h)^2+\frac{\left[M^\prime \bm{g}\right]_h}{3}(s_{h+1}-s_h)^3+\frac{\left[M^{\prime\prime}\bm{g}\right]_h}{4}(s_{h+1}-s_h)^4\right]\\
&&=g(s_1)s_1+\sum_{\substack{j=1}}^{K}\sum_{\substack{h=1}}^{i-1}\left[\delta_{jh}(s_{h+1}-s_h)+\frac{M_{hj}}{2}(s_{h+1}-s_h)^2+\frac{M^{\prime}_{hj}}{3}(s_{h+1}-s_h)^3+\frac{M^{\prime\prime}_{hj}}{4}(s_{h+1}-s_h)^4\right]g(s_j)\\
&&\equiv\sum_{\substack{j=1}}^{K}P(\bm{s})_{ij}\,g(s_j)=\left[P(\bm{s})\,\bm{g}\right]_i\,,
\end{eqnarray*}
with
\bes
P(\bm{s})_{ij}=s_1\delta_{j1}+\sum_{\substack{h=1}}^{i-1}\left[\delta_{jh}(s_{h+1}-s_h)+\frac{M_{hj}}{2}(s_{h+1}-s_h)^2+\frac{M^\prime_{hj}}{3}(s_{h+1}-s_h)^3+\frac{M^{\prime\prime}_{hj}}{4}(s_{h+1}-s_h)^4\right]\,,
\ees
so that
\bd
\begin{split}
&P(\bm{s})_{11}=s_1\quad \text{and}\quad P_{1j}=0,\quad j>1,\\
&P(\bm{s})_{2j}=s_2\,\delta_{j1}+\frac{1}{2}\,M_{1j}\,(s_{2}-s_1)^2+\frac{1}{3}M^\prime_{1j}\,(s_{2}-s_1)^3+\frac{1}{4}\,M^{\prime\prime}_{1j}\,(s_{2}-s_1)^4\quad \Rightarrow\quad  P_{2j}(\bm{s})=0\quad j>3,\\
\end{split}
\ed
and so on and so forth for larger $i$ and $j$.
Finally, the form of the matrix $P$ is given by 
\bes
P(\bm{s})=\left[\begin{array}{ccccccc}
P(\bm{s})_{11} & 0 & 0 & 0 &0&\dots & 0\\ 
P(\bm{s})_{21} & P(\bm{s})_{22} & P(\bm{s})_{23} & 0 &0&\dots & 0 \\ 
P(\bm{s})_{31} & P(\bm{s})_{32} & P(\bm{s})_{33} & P(\bm{s})_{34} &0& \dots &0\\
\\
\vdots&&&&&&\vdots\\
%P_{K-1\,1}& P_{K-1\,1} & P_{K-1\,2} & \dots & P_{K-1\,K-3} &P_{K-1\,K-2} & P_{K-1,K-1}& P_{K-1\,K}\\
\\
P(\bm{s})_{K\,1}& P(\bm{s})_{K\,1} & P(\bm{s})_{K\,2} & \dots &  \dots &  \dots  & P(\bm{s})_{K\,K}\\
\end{array}\right]\,.\\
\ees

We obtain the results presented in Equations~(\ref{eq:derf}) and~(\ref{eq:derfra}) replacing $g_{{\rm spline}}(x)$ with the time $t_k$ observation of the EONIA instantaneous forward curve $f_{{\rm int}}(t_k,x)$ and the FRA term structure $FRA_{\Delta,{\rm int}}(t_k,x)$ respectively. Equation~(\ref{eq:intsigma}) is readily derived substituting $g_{{\rm spline}}(x)$ with the time $t_k$ observation of the components of $\bm{\sigma}_{f,\text{int}}(t_k,x)$.
%%%%%%%%%%%%%%%%%%%%%%%%%%%%%%%%%%%%%%%%%%%%%%%%%%%%%%%%%%%%%%%%%%%%%%%%%%%%%%%%%%%%%%%%%%%%%%%%%%%%%%%%%%%%%%%%%%%%%%%%%%%%%%%%%%%%%%%%%%%%%%%%%%

\bibliography{HJM_for_Risk_Management_new}

\begin{thebibliography}{10}

\bibitem{litterman_scheinkman1991}
R.~B. Litterman and J.~Scheinkman, ``Common factors affecting bond returns,''
  {\em The Journal of Fixed Income}, vol.~1, no.~1, pp.~54--61, 1991.

\bibitem{knez1994}
P.~J. Knez, R.~Litterman, and J.~Scheinkman, ``Explorations into factors
  explaining money market returns,'' {\em The Journal of Finance}, vol.~49,
  no.~5, pp.~1861--1882, 1994.

\bibitem{jamshidian_zhu1996}
F.~Jamshidian and Y.~Zhu, ``Scenario simulation: Theory and methodology,'' {\em
  Finance and stochastics}, vol.~1, no.~1, pp.~43--67, 1996.

\bibitem{rebonato1998}
R.~Rebonato, {\em Interest Rate Option Models}.
\newblock Wiley Series in Financial Engineering, Wiley, second~ed., May 1998.

\bibitem{scherer2002}
K.~P. Scherer and M.~Avellaneda, ``All for one ... one for all? {A} principal
  component analysis of {L}atin {A}merican brady bond debt from 1994 to 2000,''
  {\em International Journal of Theoretical and Applied Finance}, vol.~05,
  no.~01, pp.~79--106, 2002.

\bibitem{driessen2003}
J.~Driessen, P.~Klaassen, and B.~Melenberg, ``The performance of multi-factor
  term structure models for pricing and hedging caps and swaptions,'' {\em
  Journal of Financial and Quantitative Analysis}, vol.~38, pp.~635--672, 9
  2003.

\bibitem{nelson1987}
C.~R. Nelson and A.~F. Siegel, ``Parsimonious modeling of yield curves,'' {\em
  The Journal of Business}, vol.~60, no.~4, pp.~473 -- 489, 1987.

\bibitem{diebold_li2006}
F.~X. Diebold and C.~Li, ``Forecasting the term structure of government bond
  yields,'' {\em Journal of econometrics}, vol.~130, no.~2, pp.~337--364, 2006.

\bibitem{bliss1997}
R.~R. Bliss, ``{Movements in the term structure of interest rates},'' {\em
  Economic Review}, no.~Q 4, pp.~16--33, 1997.

\bibitem{dai2000}
Q.~Dai and K.~J. Singleton, ``{Specification Analysis of Affine Term Structure
  Models},'' {\em Journal of Finance}, vol.~55, pp.~1943--1978, October 2000.

\bibitem{dejong1999}
F.~de~Jong and P.~Santa-Clara, ``{The Dynamics of the Forward Interest Rate
  Curve: A Formulation with State Variables},'' {\em Journal of Financial and
  Quantitative Analysis}, vol.~34, pp.~131--157, March 1999.

\bibitem{dejong2000}
F.~de~Jong, ``{Time-series and Cross-section Information in Affine Term
  Structure Models},'' CEPR Discussion Papers 2065, C.E.P.R. Discussion Papers,
  Feb. 1999.

\bibitem{duffee2002}
G.~R. Duffee, ``{Term Premia and Interest Rate Forecasts in Affine Models},''
  {\em Journal of Finance}, vol.~57, pp.~405--443, 02 2002.

\bibitem{bernadell2005}
C.~Bernadell, J.~Coche, and K.~Nyholm, ``Yield curve prediction for the
  strategic investor,'' {\em ECB Working Paper Series no 472, April 2005},
  2005.

\bibitem{Coroneo2011393}
L.~Coroneo, K.~Nyholm, and R.~Vidova-Koleva, ``How arbitrage-free is the
  {N}elson - {S}iegel model?,'' {\em Journal of Empirical Finance}, vol.~18,
  no.~3, pp.~393 -- 407, 2011.

\bibitem{monfort2007}
{Monfort, Alain and Pegoraro, Fulvio}, ``{Switching {VARMA} Term Structure
  Models - Extended Version},'' {\em Banque de France Working Paper No. 191.},
  2007.

\bibitem{gourieroux2013}
C.~Gourieroux, A.~Monfort, F.~Pegoraro, and J.-P. Renne, ``Regime switching and
  bond pricing,'' {\em Banque de France Working Paper No. 456.}, 2013.

\bibitem{barone1999}
G.~Barone-Adesi, K.~Giannopoulos, and L.~Vosper, ``{V}a{R} without correlations
  for portfolios of derivative securities,'' {\em Journal of Futures Markets},
  vol.~19, no.~5, pp.~583--602, 1999.

\bibitem{barone2002}
G.~Barone-Adesi, K.~Giannopoulos, and L.~Vosper, ``Backtesting derivative
  portfolios with filtered historical simulation ({FHS}),'' {\em European
  Financial Management}, vol.~8, no.~1, pp.~31--58, 2002.

\bibitem{audrino2004}
F.~Audrino and F.~Trojani, ``{Accurate Yield Curve Scenarios Generation using
  Functional Gradient Descent},'' {\em Banque de France Working Paper No.
  456.}, 2004.

\bibitem{teichmann_wuthrich2013}
J.~Teichmann and M.~V. W{\"u}thrich, ``Consistent yield curve prediction,''
  {\em Preprint, ETH Zurich}, 2013.

\bibitem{Vasicek1977}
O.~Vasicek, ``An equilibrium characterization of the term structure,'' {\em
  Journal of Financial Economics}, vol.~5, no.~2, pp.~177 -- 188, 1977.

\bibitem{heath1992}
D.~Heath, R.~Jarrow, and A.~Morton, ``Bond pricing and the term structure of
  interest rates: A new methodology for contingent claims valuation,'' {\em
  Econometrica}, vol.~60, no.~1, pp.~77--105, 1992.

\bibitem{sahalia2010}
Y.~A{\"\i}t-Sahalia and R.~L. Kimmel, ``Estimating affine multifactor term
  structure models using closed-form likelihood expansions,'' {\em Journal of
  Financial Economics}, vol.~98, no.~1, pp.~113 -- 144, 2010.

\bibitem{henrard2007}
M.~Henrard, ``The irony in derivatives discounting,'' {\em Wilmott Magazine,
  July 2007}, 2007.

\bibitem{morini2009}
M.~Morini, ``Solving the puzzle in the interest rate market,'' {\em
  http://ssrn.com/abstract=1506046}, 2009.

\bibitem{mercurio2009}
F.~Mercurio, ``Interest rates and the credit crunch: New formulas and market
  models,'' {\em Bloomberg Portfolio Research Paper No.2010-01-FRONTIERS.},
  2009.

\bibitem{mercurio2010}
F.~Mercurio, ``{LIBOR} market models with stochastic basis,'' {\em Risk
  Magazine, December 2010}, pp.~84--89, 2010.

\bibitem{mercurio2010b}
F.~Mercurio, ``Modern {LIBOR} market models: Using different curves for
  projecting rates and for discounting,'' {\em International Journal of
  Theoretical and Applied Finance}, vol.~13, no.~01, pp.~113--137, 2010.

\bibitem{pallavicini2010}
A.~Pallavicini and M.~Tarenghi, ``Interest-rate modeling with multiple yield
  curves,'' {\em arXiv:1006.4767v1 [q-fin.PR]}, 2010.

\bibitem{fujii2011}
M.~Fujii, Y.~Shimada, and A.~Takahashi, ``A market model of interest rates with
  dynamic basis spreads in the presence of collateral and multiple
  currencies,'' {\em Wilmott}, vol.~2011, no.~54, pp.~61--73, 2011.

\bibitem{crepey2012}
S.~Cr\'epey, Z.~Grbac, and H.-N. Nguyen, ``A multiple-curve {HJM} model of
  interbank risk,'' {\em Mathematics and Financial Economics}, vol.~6, no.~3,
  pp.~155--190, 2012.

\bibitem{crepey2013}
S.~Crépey, Z.~Grbac, N.~Ngor, and D.~Skovmand, ``A l\`evy hjm multiple-curve
  model with application to cva computation,'' {\em
  http://ssrn.com/abstract=2334865}, 2013.

\bibitem{pallavicini2014}
N.~Moreni and A.~Pallavicini, ``Parsimonious {HJM} modelling for multiple yield
  curve dynamics,'' {\em Quantitative Finance}, vol.~14, no.~2, pp.~199--210,
  2014.

\bibitem{cuchiero2014}
C.~Cuchiero, C.~Fontana, and A.~Gnoatto, ``A general hjm framework for multiple
  yield curve modeling,'' {\em http://ssrn.com/abstract=2454003}, 2014.

\bibitem{henrard2010}
M.~Henrard, ``The irony in derivatives discounting part {II}: The crisis,''
  {\em Wilmott Journal}, vol.~2, no.~6, pp.~301--316, 2010.

\bibitem{henrard2013}
M.~P.~A. Henrard, ``Multi-curves framework with stochastic spread: A coherent
  approach to {STIR} futures and their options,'' {\em OpenGamma Quantitative
  Research, No. 11, March 2013.}, 2013.

\bibitem{bianchetti2010}
M.~Bianchetti, ``Two curves, one price,'' {\em Risk magazine}, vol.~23, no.~8,
  pp.~66--72, 2010.

\bibitem{bianchetti2011}
M.~Bianchetti and M.~Carlicchi, ``Interest rates after the credit crunch:
  Multiple-curve vanilla derivatives and {SABR},'' {\em arXiv preprint
  arXiv:1103.2567}, 2011.

\bibitem{kijima2009}
M.~Kijima, K.~Tanaka, and T.~Wong, ``A multi-quality model of interest rates,''
  {\em Quantitative Finance}, vol.~9, no.~2, pp.~133--145, 2009.

\bibitem{kenyon2010}
C.~Kenyon, ``Post-shock short-rate pricing,'' {\em Risk Magazine, November
  2010}, pp.~83--87, 2010.

\bibitem{grasselli2014}
M.~Grasselli and G.~Miglietta, ``A flexible spot multiple-curve model,'' {\em
  http://ssrn.com/abstract=2424242}, 2014.

\bibitem{morino2014}
L.~Morino and W.~J. Ruggaldier, ``On multicurve models for the term
  structure,'' {\em http://arxiv.org/abs/1401.5431}, 2014.

\bibitem{henrard2014}
M.~Henrard, {\em {Interest Rate Modelling in the Multi-Curve Framework}}.
\newblock Applied Quantitative Finance, Palgrave Macmillan, 2014 edition~ed.,
  May 2014.

\bibitem{kupiec1995}
P.~H. Kupiec, ``Technique for verifying the accuracy of risk measurement
  models,'' {\em The Journal of Derivatives}, vol.~3, no.~2, pp.~pp. 73--84,
  1995.

\bibitem{christoffersen1998}
P.~F. Christoffersen, ``Evaluating interval forecasts,'' {\em International
  Economic Review}, vol.~39, no.~4, pp.~pp. 841--862, 1998.

\bibitem{renne2014}
J.-P. Renne, ``A model of the {E}uro-area yield-curve with discrete policy
  rates,'' {\em Available at SSRN 2015236}, 2014.

\bibitem{kimorphanides2012}
D.~H. Kim and A.~Orphanides, ``Term structure estimation with survey data on
  interest rate forecasts,'' {\em Journal of Financial and Quantitative
  Analysis}, vol.~47, no.~01, pp.~241--272, 2012.

\bibitem{musiela1993}
M.~Musiela, ``Stochastic {PDE}s and term structure models,'' {\em Preprint},
  1993.

\bibitem{brace_musiela1994}
A.~Brace and M.~Musiela, ``A multifactor {G}auss {M}arkov implementation of
  {H}eath, {J}arrow, and {M}orton,'' {\em Mathematical Finance}, vol.~4, no.~3,
  pp.~259--283, 1994.

\bibitem{bjork2004}
T.~Bj{\"o}rk, {\em Arbitrage Theory in Continuous Time}.
\newblock Oxford University press, 2004.

\bibitem{ametrano2009}
F.~M. Ametrano and M.~Bianchetti, ``Bootstrapping the illiquidity: Multiple
  yield curves construction for market coherent forward rates estimations,''
  {\em Modeling Interest Rates, Fabio Mercurio, ed., Risk Books, Incisive
  Media}, 2009.

\bibitem{hagan2006}
P.~S. Hagan and G.~West, ``Interpolation methods for curve construction,'' {\em
  Applied Mathematical Finance}, vol.~13, no.~2, pp.~89--129, 2006.

\bibitem{anderson1963}
T.~W. Anderson, ``Asymptotic theory for principal component analysis,'' {\em
  The Annals of Mathematical Statistics}, vol.~34, pp.~122--148, 03 1963.

\bibitem{flury1997}
B.~Flury, {\em {A First Course in Multivariate Statistics}}.
\newblock Springer Texts in Statistics, Springer, 1997 edition~ed., Aug 1997.

\bibitem{efron1979}
B.~Efron, ``Bootstrap methods: Another look at the jackknife,'' {\em Ann.
  Statist.}, vol.~7, pp.~1--26, 01 1979.

\end{thebibliography}
\bibliographystyle{ieeetr}

\newpage 

% \section*{Figures}

\afterpage{\clearpage}

\afterpage{\clearpage}

\afterpage{\clearpage}

\afterpage{\clearpage}

\afterpage{\clearpage}

\afterpage{\clearpage}

\afterpage{\clearpage}

\afterpage{\clearpage}

\afterpage{\clearpage}

\afterpage{\clearpage}

\afterpage{\clearpage}

\afterpage{\clearpage}

\afterpage{\clearpage}

\afterpage{\clearpage}

\newpage

% \section*{Tables}

\begin{center}
  \begin{table}[H!]
    \centering
    \begin{tabular}{lll}
      \hline
      {\bf \text{Tenor}} & {\bf Starting Date} & {\bf Ending Date} \\
      \hline
      \hline
      EONIA & \text{02/08/2005} & \text{12/27/2013} \\
      EUR3M & \text{02/08/2005} & \text{12/27/2013} \\
      % EUR6M & \text{04/22/1999} & \text{12/27/2013} \\
      % EUR1Y & \text{02/08/2005} & \text{12/27/2013}\\
      \hline
    \end{tabular}
    \caption{Starting and ending dates of the historical time series.}
    \label{tab:dataset2dates}
  \end{table}
\end{center}
\afterpage{\clearpage}

\begin{center}
  \begin{table}[H!]
    \centering
    \begin{tabular}{llllllllllllllllllllllllllllllllll}
      \hline
      {\bf  \text{Tenor}} & \multicolumn{12}{c}{{\bf \text{Time to Maturity}}}\\
      \hline
      \hline
      EONIA & \text{1d} & \text{7d} & \text{1m} & \text{2m} & \text{3m} & \text{6m} & \text{9m} & \text{1y} & \text{1y 6m} & \text{2y} & \text{3y} & \text{4y} \\
      & \text{5y} &   \text{6y} & \text{7y} & \text{8y} & \text{9y} & \text{10y} & \text{12y} & \text{15y} & \text{20y} & \text{25y}  & \text{30y}\\
      \hline
      EUR3M & \text{1d} & & \text{1m} & \text{2m} & \text{3m} &\text{6m} & \text{9m} & \text{1y} & \text{1y 6m} &\text{2y} & \text{3y} & \text{4y}\\
      & \text{5y} & \text{6y} & \text{7y} & \text{8y} & \text{9y} & \text{10y} & \text{12y} & \text{15y} & \text{20y} & \text{25y} & \text{30y}\\
      \hline
      % EUR6M & \text{1d} & & \text{1m} & \text{2m} & \text{3m} &\text{6m}  & \text{9m} & \text{1y} & \text{1y 6m} & \text{2y} & \text{3y} & \text{4y}\\
      % & \text{5y} & \text{6y} & \text{7y} & \text{8y} & \text{9y} & \text{10y} & \text{12y} & \text{15y} & \text{20y} & \text{25y} & \text{30y}\\
      % \hline
      % EUR1Y & \text{1d} & & \text{1m} & \text{2m} & \text{3m} & \text{6m} & \text{9m} & \text{1y} && \text{2y} & \text{3y} & \text{4y}\\
      % & \text{5y} & \text{6y} &\text{7y} & \text{8y} & \text{9y} & \text{10y} & \text{12y} & \text{15y} & \text{20y} & \text{25y} & \text{30y}\\
      \hline
    \end{tabular}
    \caption{Time to maturity grids.}
    \label{tab:dataset2mat}
  \end{table}
\end{center}
\afterpage{\clearpage}

\begin{center}
  \begin{table}[H!]
    \centering
    \begin{tabular}{llllllllllllllllllllllllllllllll}
      \hline
      {\bf  \text{Tenor}} & \multicolumn{12}{c}{{\bf \text{Time to Maturity}}}\\
      \hline
      \hline
      EONIA &\text{1m} &\text{2m} &\text{3m} &\text{6m} &\text{9m} &\text{1y} &\text{5y} &\text{10y} & \text{15y} &\text{20y} &\text{25y} &\text{30y} \\
      EUR3M &&&\text{3m} &\text{6m} &\text{9m} &\text{1y} &\text{5y} &\text{10y} & \text{15y} &\text{20y} &\text{25y} &\text{30y} \\
      \hline
    \end{tabular}
    \caption{Time to maturity grid used in the empirical analysis for the EONIA and the EUR3M curves.}
    \label{tab:dataset2matspread}
  \end{table}
\end{center}
\afterpage{\clearpage}

%%%%%%%%%%%%%%%%%%%%%%%%%%%%%%%%%

\begin{table}
\begin{center}
\begin{tabular}{l|lll|lll}
\multirow{2}{*}{$s_i$}&\multicolumn{3}{c}{$p=0.95$}&
\multicolumn{3}{|c}{$p=0.99$} \\
\cline{2-7}
& $n_1$& $\text{LR}_{\text{UC}}$& $\text{p-value}$& $n_1$& $\text{LR}_{\text{UC}}$& $\text{p-value}$\\
 \hline
\text{1m} & 22 & 3.6 & 5.76 & 12 & \text{16.24 (**)} & 0.01 \\
 \text{2m} & 20 & 2.01 & 15.58 & 10 & \text{10.78 (**)} & 0.1 \\
 \text{3m} & 15 & 0.02 & 88.27 & 8 & \text{6.16 (*)} & 1.3 \\
 \text{6m} & 13 & 0.16 & 69.08 & 7 & \text{4.22 (*)} & 3.98 \\
 \text{9m} & 11 & 0.94 & 33.19 & 6 & 2.58 & 10.82 \\
 \text{1y} & 14 & 0.01 & 90.29 & 5 & 1.28 & 25.84 \\
 \text{5y} & 23 & \text{4.55 (*)} & 3.29 & 11 & \text{13.42 (**)} & 0.02 \\
 \text{10y} & 17 & 0.45 & 50.26 & 8 & \text{6.16 (*)} & 1.3 \\
 \text{15y} & 22 & 3.6 & 5.76 & 8 & \text{6.16 (*)} & 1.3 \\
 \text{20y} & 16 & 0.17 & 68.07 & 7 & \text{4.22 (*)} & 3.98 \\
 \text{25y} & 19 & 1.38 & 24.04 & 7 & \text{4.22 (*)} & 3.98 \\
 \text{30y} & 17 & 0.45 & 50.26 & 12 & \text{16.24 (**)} & 0.01 \\
\end{tabular}
\end{center}
\caption{Unconditional coverage test results for the one week ahead forecast for the EONIA curve obtained with the Gaussian diffusive model.}
\label{tab:testEONIA1wgauss}
\end{table}
\afterpage{\clearpage}

\begin{table}
\begin{center}
\begin{tabular}{l|lll|lll}
\multirow{2}{*}{$s_i$}&\multicolumn{3}{c}{$p=0.95$}&
\multicolumn{3}{|c}{$p=0.99$} \\
\cline{2-7}
& $n_1$& $\text{LR}_{\text{UC}}$& $\text{p-value}$& $n_1$& $\text{LR}_{\text{UC}}$& $\text{p-value}$\\
 \hline
\text{1m} & 19 & 1.38 & 24.04 & 7 & \text{4.22 (*)} & 3.98 \\
 \text{2m} & 17 & 0.45 & 50.26 & 4 & 0.38 & 53.51 \\
 \text{3m} & 11 & 0.94 & 33.19 & 3 & 0. & 94.85 \\
 \text{6m} & 12 & 0.46 & 49.63 & 3 & 0. & 94.85 \\
 \text{9m} & 9 & 2.49 & 11.49 & 2 & 0.31 & 57.75 \\
 \text{1y} & 11 & 0.94 & 33.19 & 3 & 0. & 94.85 \\
 \text{5y} & 18 & 0.85 & 35.53 & 7 & \text{4.22 (*)} & 3.98 \\
 \text{10y} & 18 & 0.85 & 35.53 & 2 & 0.31 & 57.75 \\
 \text{15y} & 16 & 0.17 & 68.07 & 2 & 0.31 & 57.75 \\
 \text{20y} & 14 & 0.01 & 90.29 & 3 & 0. & 94.85 \\
 \text{25y} & 14 & 0.01 & 90.29 & 5 & 1.28 & 25.84 \\
 \text{30y} & 18 & 0.85 & 35.53 & 7 & \text{4.22 (*)} & 3.98 \\
\end{tabular}
\end{center}
\caption{Unconditional coverage test results for the one week ahead forecast for the EONIA curve obtained with the bootstrap methodology.}
\label{tab:testEONIA1wboot}
\end{table}
\afterpage{\clearpage}

\begin{table}
\begin{center}
\begin{tabular}{l|lll|lll}
\multirow{2}{*}{$s_i$}&\multicolumn{3}{c}{$p=0.95$}&
\multicolumn{3}{|c}{$p=0.99$} \\
\cline{2-7}
& $n_1$& $\text{LR}_{\text{UC}}$& $\text{p-value}$& $n_1$& $\text{LR}_{\text{UC}}$& $\text{p-value}$\\
 \hline
\text{1m} & 51 & \text{63.87 (**)} & 0. & 25 & \text{67.23 (**)} & 0. \\
 \text{2m} & 53 & \text{69.77 (**)} & 0. & 25 & \text{67.23 (**)} & 0. \\
 \text{3m} & 48 & \text{55.38 (**)} & 0. & 25 & \text{67.23 (**)} & 0. \\
 \text{6m} & 45 & \text{47.34 (**)} & 0. & 23 & \text{58.28 (**)} & 0. \\
 \text{9m} & 40 & \text{35.03 (**)} & 0. & 22 & \text{53.95 (**)} & 0. \\
 \text{1y} & 37 & \text{28.33 (**)} & 0. & 20 & \text{45.59 (**)} & 0. \\
 \text{5y} & 25 & \text{7.62 (**)} & 0.58 & 13 & \text{20.05 (**)} & 0. \\
 \text{10y} & 17 & 0.68 & 40.9 & 4 & 0.48 & 49.02 \\
 \text{15y} & 22 & \text{4.25 (*)} & 3.92 & 7 & \text{4.55 (*)} & 3.29 \\
 \text{20y} & 35 & \text{24.17 (**)} & 0. & 12 & \text{16.97 (**)} & 0. \\
 \text{25y} & 34 & \text{22.19 (**)} & 0. & 14 & \text{23.29 (**)} & 0. \\
 \text{30y} & 21 & 3.32 & 6.83 & 15 & \text{26.68 (**)} & 0. \\
\end{tabular}
\end{center}
\caption{Unconditional coverage test results for three month ahead forecast for the EONIA curve obtained with the Gaussian diffusive model.}
\label{tab:testEONIA12wgauss}
\end{table}
\afterpage{\clearpage}

\begin{table}
\begin{center}
\begin{tabular}{l|lll|lll}
\multirow{2}{*}{$s_i$}&\multicolumn{3}{c}{$p=0.95$}&
\multicolumn{3}{|c}{$p=0.99$} \\
\cline{2-7}
& $n_1$& $\text{LR}_{\text{UC}}$& $\text{p-value}$& $n_1$& $\text{LR}_{\text{UC}}$& $\text{p-value}$\\
 \hline
\text{1m} & 32 & \text{18.44 (**)} & 0. & 21 & \text{49.72 (**)} & 0. \\
 \text{2m} & 29 & \text{13.33 (**)} & 0.03 & 19 & \text{41.57 (**)} & 0. \\
 \text{3m} & 27 & \text{10.31 (**)} & 0.13 & 19 & \text{41.57 (**)} & 0. \\
 \text{6m} & 28 & \text{11.78 (**)} & 0.06 & 18 & \text{37.66 (**)} & 0. \\
 \text{9m} & 26 & \text{8.93 (**)} & 0.28 & 18 & \text{37.66 (**)} & 0. \\
 \text{1y} & 25 & \text{7.62 (**)} & 0.58 & 6 & 2.83 & 9.25 \\
 \text{5y} & 15 & 0.09 & 76.49 & 4 & 0.48 & 49.02 \\
 \text{10y} & 6 & \text{5.95 (*)} & 1.47 & 1 & 1.53 & 21.66 \\
 \text{15y} & 12 & 0.29 & 59.28 & 4 & 0.48 & 49.02 \\
 \text{20y} & 13 & 0.06 & 80.24 & 12 & \text{16.97 (**)} & 0. \\
 \text{25y} & 17 & 0.68 & 40.9 & 12 & \text{16.97 (**)} & 0. \\
 \text{30y} & 15 & 0.09 & 76.49 & 12 & \text{16.97 (**)} & 0. \\
\end{tabular}
\end{center}
\caption{Unconditional coverage test results for three month ahead forecast for the EONIA curve obtained with the bootstrap methodology.}
\label{tab:testEONIA12wboot}
\end{table}
\afterpage{\clearpage}

\begin{table}
\begin{center}
\begin{tabular}{l|lll|lll}
\multirow{2}{*}{$s_i$}&\multicolumn{3}{c}{$p=0.95$}&
\multicolumn{3}{|c}{$p=0.99$} \\
\cline{2-7}
& $n_1$& $\text{LR}_{\text{UC}}$& $\text{p-value}$& $n_1$& $\text{LR}_{\text{UC}}$& $\text{p-value}$\\
 \hline
\text{1m} & 93 & \text{253.6 (**)} & 0. & 70 & \text{359.74 (**)} & 0. \\
 \text{2m} & 88 & \text{229.04 (**)} & 0. & 69 & \text{352.32 (**)} & 0. \\
 \text{3m} & 88 & \text{229.04 (**)} & 0. & 63 & \text{308.68 (**)} & 0. \\
 \text{6m} & 78 & \text{182.65 (**)} & 0. & 51 & \text{226.17 (**)} & 0. \\
 \text{9m} & 64 & \text{124.19 (**)} & 0. & 20 & \text{51.26 (**)} & 0. \\
 \text{1y} & 36 & \text{34.17 (**)} & 0. & 7 & \text{5.95 (*)} & 1.47 \\
 \text{5y} & 30 & \text{20.77 (**)} & 0. & 9 & \text{10.89 (**)} & 0.1 \\
 \text{10y} & 27 & \text{15.07 (**)} & 0.01 & 3 & 0.15 & 69.79 \\
 \text{15y} & 46 & \text{61.62 (**)} & 0. & 20 & \text{51.26 (**)} & 0. \\
 \text{20y} & 71 & \text{152.44 (**)} & 0. & 27 & \text{84.58 (**)} & 0. \\
 \text{25y} & 67 & \text{136.05 (**)} & 0. & 31 & \text{105.53 (**)} & 0. \\
 \text{30y} & 55 & \text{90.98 (**)} & 0. & 21 & \text{55.72 (**)} & 0. \\
\end{tabular}
\end{center}
\caption{Unconditional coverage test results for the 1 year ahead forecast for the EONIA curve obtained with the Gaussian diffusive model.}
\label{tab:testEONIA52wgauss}
\end{table}
\afterpage{\clearpage}

\begin{table}
\begin{center}
\begin{tabular}{l|lll|lll}
\multirow{2}{*}{$s_i$}&\multicolumn{3}{c}{$p=0.95$}&
\multicolumn{3}{|c}{$p=0.99$} \\
\cline{2-7}
& $n_1$& $\text{LR}_{\text{UC}}$& $\text{p-value}$& $n_1$& $\text{LR}_{\text{UC}}$& $\text{p-value}$\\
 \hline
\text{1m} & 35 & \text{31.76 (**)} & 0. & 20 & \text{51.26 (**)} & 0. \\
 \text{2m} & 32 & \text{24.95 (**)} & 0. & 20 & \text{51.26 (**)} & 0. \\
 \text{3m} & 30 & \text{20.77 (**)} & 0. & 20 & \text{51.26 (**)} & 0. \\
 \text{6m} & 25 & \text{11.69 (**)} & 0.06 & 12 & \text{19.99 (**)} & 0. \\
 \text{9m} & 9 & 0.81 & 36.83 & 0 & \text{4.78 (*)} & 2.87 \\
 \text{1y} & 0 & \text{24.42 (**)} & 0. & 0 & \text{4.78 (*)} & 2.87 \\
 \text{5y} & 17 & 2.04 & 15.29 & 0 & \text{4.78 (*)} & 2.87 \\
 \text{10y} & 0 & \text{24.42 (**)} & 0. & 0 & \text{4.78 (*)} & 2.87 \\
 \text{15y} & 2 & \text{13.09 (**)} & 0.03 & 0 & \text{4.78 (*)} & 2.87 \\
 \text{20y} & 3 & \text{9.88 (**)} & 0.17 & 0 & \text{4.78 (*)} & 2.87 \\
 \text{25y} & 12 & 0. & 97.63 & 1 & 1.03 & 30.93 \\
 \text{30y} & 15 & 0.79 & 37.47 & 7 & \text{5.95 (*)} & 1.47 \\
\end{tabular}
\end{center}
\caption{Unconditional coverage test results for the 1 year ahead forecast for the EONIA curve obtained with the bootstrap methodology.}
\label{tab:testEONIA52wboot}
\end{table}
\afterpage{\clearpage}

%%%%%%%%%%%%%%%%%%%%%%%%%%%%%%%%%

\begin{table}
\begin{center}
\begin{tabular}{l|lll|lll}
\multirow{2}{*}{$s_i$}&\multicolumn{3}{c}{$p=0.95$}&
\multicolumn{3}{|c}{$p=0.99$} \\
\cline{2-7}
& $n_1$& $\text{LR}_{\text{UC}}$& $\text{p-value}$& $n_1$& $\text{LR}_{\text{UC}}$& $\text{p-value}$\\
 \hline
\text{3m} & 10 & 1.61 & 20.46 & 7 & \text{4.22 (*)} & 3.98 \\
 \text{6m} & 26 & \text{7.94 (**)} & 0.48 & 13 & \text{19.24 (**)} & 0. \\
 \text{9m} & 35 & \text{22.4 (**)} & 0. & 19 & \text{40.27 (**)} & 0. \\
 \text{1y} & 26 & \text{7.94 (**)} & 0.48 & 11 & \text{13.42 (**)} & 0.02 \\
 \text{5y} & 25 & \text{6.72 (**)} & 0.95 & 7 & \text{4.22 (*)} & 3.98 \\
 \text{10y} & 24 & \text{5.59 (*)} & 1.81 & 6 & 2.58 & 10.82 \\
 \text{15y} & 33 & \text{18.69 (**)} & 0. & 17 & \text{32.73 (**)} & 0. \\
 \text{20y} & 19 & 1.38 & 24.04 & 8 & \text{6.16 (*)} & 1.3 \\
 \text{25y} & 18 & 0.85 & 35.53 & 9 & \text{8.36 (**)} & 0.38 \\
 \text{30y} & 16 & 0.17 & 68.07 & 8 & \text{6.16 (*)} & 1.3 \\
\end{tabular}
\end{center}
\caption{Unconditional coverage test results for the one week ahead forecast for the EUR3M curve obtained with the Gaussian diffusive model.}
\label{tab:testEUR3M1wgauss}
\end{table}
\afterpage{\clearpage}

\begin{table}
\begin{center}
\begin{tabular}{l|lll|lll}
\multirow{2}{*}{$s_i$}&\multicolumn{3}{c}{$p=0.95$}&
\multicolumn{3}{|c}{$p=0.99$} \\
\cline{2-7}
& $n_1$& $\text{LR}_{\text{UC}}$& $\text{p-value}$& $n_1$& $\text{LR}_{\text{UC}}$& $\text{p-value}$\\
 \hline
\text{3m} & 6 & \text{6.61 (*)} & 1.01 & 4 & 0.38 & 53.51 \\
 \text{6m} & 22 & 3.6 & 5.76 & 8 & \text{6.16 (*)} & 1.3 \\
 \text{9m} & 34 & \text{20.51 (**)} & 0. & 14 & \text{22.4 (**)} & 0. \\
 \text{1y} & 19 & 1.38 & 24.04 & 8 & \text{6.16 (*)} & 1.3 \\
 \text{5y} & 21 & 2.76 & 9.68 & 3 & 0. & 94.85 \\
 \text{10y} & 17 & 0.45 & 50.26 & 3 & 0. & 94.85 \\
 \text{15y} & 25 & \text{6.72 (**)} & 0.95 & 15 & \text{25.7 (**)} & 0. \\
 \text{20y} & 18 & 0.85 & 35.53 & 6 & 2.58 & 10.82 \\
 \text{25y} & 18 & 0.85 & 35.53 & 7 & \text{4.22 (*)} & 3.98 \\
 \text{30y} & 15 & 0.02 & 88.27 & 6 & 2.58 & 10.82 \\
\end{tabular}
\end{center}
\caption{Unconditional coverage test results for the one week ahead forecast for the EUR3M curve obtained with the bootstrap methodology.}
\label{tab:testEUR3M1wboot}
\end{table}
\afterpage{\clearpage}

\begin{table}
\begin{center}
\begin{tabular}{l|lll|lll}
\multirow{2}{*}{$s_i$}&\multicolumn{3}{c}{$p=0.95$}&
\multicolumn{3}{|c}{$p=0.99$} \\
\cline{2-7}
& $n_1$& $\text{LR}_{\text{UC}}$& $\text{p-value}$& $n_1$& $\text{LR}_{\text{UC}}$& $\text{p-value}$\\
 \hline
\text{3m} & 57 & \text{82.12 (**)} & 0. & 29 & \text{86.14 (**)} & 0. \\
 \text{6m} & 58 & \text{85.32 (**)} & 0. & 33 & \text{106.29 (**)} & 0. \\
 \text{9m} & 64 & \text{105.42 (**)} & 0. & 36 & \text{122.14 (**)} & 0. \\
 \text{1y} & 45 & \text{47.34 (**)} & 0. & 29 & \text{86.14 (**)} & 0. \\
 \text{5y} & 37 & \text{28.33 (**)} & 0. & 19 & \text{41.57 (**)} & 0. \\
 \text{10y} & 21 & 3.32 & 6.83 & 10 & \text{11.35 (**)} & 0.08 \\
 \text{15y} & 35 & \text{24.17 (**)} & 0. & 21 & \text{49.72 (**)} & 0. \\
 \text{20y} & 32 & \text{18.44 (**)} & 0. & 12 & \text{16.97 (**)} & 0. \\
 \text{25y} & 39 & \text{32.74 (**)} & 0. & 16 & \text{30.21 (**)} & 0. \\
 \text{30y} & 38 & \text{30.5 (**)} & 0. & 16 & \text{30.21 (**)} & 0. \\
\end{tabular}
\end{center}
\caption{Unconditional coverage test results for three month ahead forecast for the EUR3M curve obtained with the Gaussian diffusive model.}
\label{tab:testEUR3M12wgauss}
\end{table}

\afterpage{\clearpage}

\begin{table}
\begin{center}
\begin{tabular}{l|lll|lll}
\multirow{2}{*}{$s_i$}&\multicolumn{3}{c}{$p=0.95$}&
\multicolumn{3}{|c}{$p=0.99$} \\
\cline{2-7}
& $n_1$& $\text{LR}_{\text{UC}}$& $\text{p-value}$& $n_1$& $\text{LR}_{\text{UC}}$& $\text{p-value}$\\
 \hline
\text{3m} & 35 & \text{24.17 (**)} & 0. & 21 & \text{49.72 (**)} & 0. \\
 \text{6m} & 38 & \text{30.5 (**)} & 0. & 20 & \text{45.59 (**)} & 0. \\
 \text{9m} & 40 & \text{35.03 (**)} & 0. & 26 & \text{71.83 (**)} & 0. \\
 \text{1y} & 26 & \text{8.93 (**)} & 0.28 & 20 & \text{45.59 (**)} & 0. \\
 \text{5y} & 28 & \text{11.78 (**)} & 0.06 & 9 & \text{8.85 (**)} & 0.29 \\
 \text{10y} & 6 & \text{5.95 (*)} & 1.47 & 0 & \text{5.59 (*)} & 1.81 \\
 \text{15y} & 23 & \text{5.28 (*)} & 2.15 & 5 & 1.45 & 22.89 \\
 \text{20y} & 15 & 0.09 & 76.49 & 13 & \text{20.05 (**)} & 0. \\
 \text{25y} & 19 & 1.78 & 18.26 & 12 & \text{16.97 (**)} & 0. \\
 \text{30y} & 16 & 0.32 & 57.21 & 12 & \text{16.97 (**)} & 0. \\
\end{tabular}
\end{center}
\caption{Unconditional coverage test results for three month ahead forecast for the EUR3M curve obtained with the bootstrap methodology.}
\label{tab:testEUR3M12wboot}
\end{table}
\afterpage{\clearpage}

\begin{table}
\begin{center}
\begin{tabular}{l|lll|lll}
\multirow{2}{*}{$s_i$}&\multicolumn{3}{c}{$p=0.95$}&
\multicolumn{3}{|c}{$p=0.99$} \\
\cline{2-7}
& $n_1$& $\text{LR}_{\text{UC}}$& $\text{p-value}$& $n_1$& $\text{LR}_{\text{UC}}$& $\text{p-value}$\\
 \hline
\text{3m} & 97 & \text{273.88 (**)} & 0. & 78 & \text{420.52 (**)} & 0. \\
 \text{6m} & 85 & \text{214.73 (**)} & 0. & 69 & \text{352.32 (**)} & 0. \\
 \text{9m} & 80 & \text{191.63 (**)} & 0. & 56 & \text{259.73 (**)} & 0. \\
 \text{1y} & 75 & \text{169.47 (**)} & 0. & 38 & \text{145. (**)} & 0. \\
 \text{5y} & 78 & \text{182.65 (**)} & 0. & 43 & \text{175.1 (**)} & 0. \\
 \text{10y} & 112 & \text{354.86 (**)} & 0. & 43 & \text{175.1 (**)} & 0. \\
 \text{15y} & 47 & \text{64.67 (**)} & 0. & 17 & \text{38.53 (**)} & 0. \\
 \text{20y} & 45 & \text{58.61 (**)} & 0. & 8 & \text{8.29 (**)} & 0.4 \\
 \text{25y} & 65 & \text{128.1 (**)} & 0. & 25 & \text{74.59 (**)} & 0. \\
 \text{30y} & 75 & \text{169.47 (**)} & 0. & 39 & \text{150.9 (**)} & 0. \\
\end{tabular}
\end{center}
\caption{Unconditional coverage test results for the 1 year ahead forecast for the EUR3M curve obtained with the Gaussian diffusive model.}
\label{tab:testEUR3M52wgauss}
\end{table}
\afterpage{\clearpage}

\begin{table}
\begin{center}
\begin{tabular}{l|lll|lll}
\multirow{2}{*}{$s_i$}&\multicolumn{3}{c}{$p=0.95$}&
\multicolumn{3}{|c}{$p=0.99$} \\
\cline{2-7}
& $n_1$& $\text{LR}_{\text{UC}}$& $\text{p-value}$& $n_1$& $\text{LR}_{\text{UC}}$& $\text{p-value}$\\
 \hline
\text{3m} & 57 & \text{98.04 (**)} & 0. & 35 & \text{127.68 (**)} & 0. \\
 \text{6m} & 35 & \text{31.76 (**)} & 0. & 19 & \text{46.9 (**)} & 0. \\
 \text{9m} & 21 & \text{6.03 (*)} & 1.41 & 17 & \text{38.53 (**)} & 0. \\
 \text{1y} & 17 & 2.04 & 15.29 & 13 & \text{23.39 (**)} & 0. \\
 \text{5y} & 2 & \text{13.09 (**)} & 0.03 & 0 & \text{4.78 (*)} & 2.87 \\
 \text{10y} & 1 & \text{17.36 (**)} & 0. & 0 & \text{4.78 (*)} & 2.87 \\
 \text{15y} & 9 & 0.81 & 36.83 & 1 & 1.03 & 30.93 \\
 \text{20y} & 3 & \text{9.88 (**)} & 0.17 & 0 & \text{4.78 (*)} & 2.87 \\
 \text{25y} & 13 & 0.1 & 74.7 & 2 & 0.06 & 79.91 \\
 \text{30y} & 16 & 1.35 & 24.56 & 7 & \text{5.95 (*)} & 1.47 \\
\end{tabular}
\end{center}
\caption{Unconditional coverage test results for the 1 year ahead forecast for the EUR3M curve obtained with the bootstrap methodology.}
\label{tab:testEUR3M52wboot}
\end{table}
\afterpage{\clearpage}

%%%%%%%%%%%%%%%%%%%%%%%%%%%%%%%%%

\end{document}